\documentclass[longauth]{aa}  

\usepackage{xcolor}
\usepackage{multirow}
\usepackage{graphicx}
\usepackage{lscape}
\usepackage{placeins}
\usepackage{amsmath,amssymb}
\usepackage{txfonts}
\usepackage{hyperref}

\begin{document}

   \title{X-Shooting ULLYSES: Massive stars at low metallicity}

%  \subtitle{XI. Clumping, X-rays, Mass-loss rates and Stellar Properties of B supergiants}
  \subtitle{VII. Stellar and wind properties of B supergiants in the Small Magellanic Cloud}

   \author{M. Bernini-Peron\inst{\ref{inst:ari}}
          \and 
          A.A.C.\ Sander\inst{\ref{inst:ari}}
          \and
          V.\ Ramachandran\inst{\ref{inst:ari}}
          \and
          L.M.\ Oskinova\inst{\ref{inst:up}}
          \and
          J.S.\ Vink\inst{\ref{inst:aop}}
          \and
          O.\ Verhamme\inst{\ref{inst:kul}}
          \and
          F.\ Najarro\inst{\ref{inst:cab}}
          \and
          J.\ Josiek\inst{\ref{inst:ari}}
          \and 
          S.A.\ Brands\inst{\ref{inst:api}}
          \and
          P.A.\ Crowther\inst{\ref{inst:uos}}
          \and
          V.M.A.\ G\'omez-Gonz\'alez\inst{\ref{inst:up}}
          \and
          A.C.\ Gormaz-Matamala\inst{\ref{inst:camk}}
          \and
          C.\ Hawcroft\inst{\ref{inst:stsci}}
          \and
          R.\ Kuiper\inst{\ref{inst:ude}}
          \and
          L.\ Mahy\inst{\ref{inst:rob}}
          \and
          W.L.F.\ Marcolino\inst{\ref{inst:obval}}
          \and
          L.P.\ Martins\inst{\ref{inst:nat}}
          \and
          A.\ Mehner\inst{\ref{inst:esoc}}
          \and
          T.N.\ Parsons\inst{\ref{inst:ucl}}
          \and
          D.\ Pauli\inst{\ref{inst:up}}
          \and
          T.\ Shenar\inst{\ref{inst:tau}}
          \and
          A.\ Schootemeijer\inst{\ref{inst:aifa}}
          \and
          H.\ Todt\inst{\ref{inst:up}}
          \and
          J.Th.\ van~Loon\inst{\ref{inst:keele}}
          \and
          the XShootU collaboration
          }

   \institute{
     {Zentrum f{\"u}r Astronomie der Universit{\"a}t Heidelberg, Astronomisches Rechen-Institut, M{\"o}nchhofstr. 12-14, 69120 Heidelberg \label{inst:ari}}\\
        \email{matheus.bernini@uni-heidelberg.de}
     \and
      {Institut f{\"u}r Physik und Astronomie, Universit{\"a}t Potsdam, Karl-Liebknecht-Str. 24/25, 14476 Potsdam, Germany\label{inst:up}}
     \and
       {Armagh Observatory and Planetarium, College Hill, BT61 9DG Armagh, Northern Ireland \label{inst:aop}}
     \and
      {Institute of Astronomy, KU Leuven, Celestijnenlaan 200D, B-3001 Leuven, Belgium\label{inst:kul}}
     \and 
      {Departamento de Astrof\'{\i}sica, Centro de Astrobiolog\'{\i}a, (CSIC-INTA), Ctra. Torrej\'on a Ajalvir, km 4,  28850 Torrej\'on de Ardoz, Madrid, Spain\label{inst:cab}}
     \and
      {Anton Pannekoek Institute for Astronomy, Universiteit van Amsterdam, Science Park 904, 1098 XH Amsterdam, The Netherlands\label{inst:api}}   
     \and
      {Department of Physics \& Astronomy, University of Sheffield, Hicks Building, Hounsfield Road, Sheffield S3 7RH, United Kingdom\label{inst:uos}}
     \and
      {Nicolaus Copernicus Astronomical Center, Polish Academy of Sciences, Bartycka 18, 00-716 Warsaw, Poland\label{inst:camk}}
     \and
      {Space Telescope Science Institute, 3700 San Martin Drive, Baltimore, MD 21218, USA\label{inst:stsci}}
      \and 
      {Faculty of Physics, University of Duisburg-Essen, Lotharstr. 1, D-47057 Duisburg, Germany\label{inst:ude}}
     \and
      {Royal Observatory of Belgium, Avenue Circulaire/Ringlaan 3, B-1180 Brussels, Belgium\label{inst:rob}}
     \and 
      {Observat\'{o}rio do Valongo, Universidade Federal do Rio de Janeiro, Ladeira Pedro Ant\^{o}nio, 43, CEP 20080-090, Rio de Janeiro, Brazil\label{inst:obval}}
     \and 
      {NAT - Universidade Cidade de S\~{a}o Paulo, Rua Galv\~{a}o Bueno, 868, S\~{a}o Paulo, Brazil\label{inst:nat}}
     \and
      {ESO - European Organisation for Astronomical Research in the Southern Hemisphere, Alonso de Cordova 3107, Vitacura, Santiago de Chile, Chile\label{inst:esoc}}
     \and
      {Department of Physics and Astronomy, University College London, Gower Street, London WC1E 6BT, UK\label{inst:ucl}}
     \and
      {The School of Physics and Astronomy, Tel Aviv University, Tel Aviv 6997801, Israel\label{inst:tau}}
     \and 
      {Argelander-Institut f{\"u}r Astronomie, Universit{\"a}t Bonn, Auf dem H{\"u}gel 71, 53121 Bonn, Germany\label{inst:aifa}}
     \and
      {Lennard-Jones Laboratories, Keele University, ST5 5BG, UK\label{inst:keele}}
  }

   \date{Received April 23, 2024; accepted July 05, 2024}

% \abstract{}{}{}{}{} 
% 5 {} token are mandatory
 
  \abstract
    {%
With the aim of  understanding massive stars and their feedback in the early epochs of our Universe, the ULLYSES and XShootU collaborations collected the biggest homogeneous dataset of high-quality hot star spectra at low metallicity. Within the rich ``zoo'' of massive star stellar types, B supergiants (BSGs) represent an important connection between the main sequence and more extreme evolutionary stages. Additionally, lying  toward the cool end of the hot star regime, determining their wind properties is crucial to gauging our expectations on the evolution and feedback of massive stars as, for instance, they are implicated in the bi-stability jump phenomenon.}
    {Here we undertake a detailed analysis of a representative sample of 18 Small Magellanic Cloud (SMC) BSGs within the  ULLYSES dataset. Our UV and optical analysis samples early- and late-type BSGs (from B0 to B8), covering the bi-stability jump region. Our aim is to evaluate their evolutionary status and verify what their wind properties say about the bi-stability jump in a low-metallicity environment.}  
    {We used the stellar atmosphere code CMFGEN to model the UV and optical spectra of the sample BSGs as well as photometry in different bands. The optical range encodes photospheric properties, while the wind information resides mostly in the UV. Further, we compare our results with different evolutionary models, with previous determinations in the literature of OB stars, and with diverging mass-loss prescriptions at the bi-stability jump. Additionally, for the first time we provide BSG models in the SMC including X-rays.}
    {Our analysis yielded the following main results: (i) From a single-stellar evolution perspective, the evolutionary status of early BSGs appear less clear than late BSGs, which agree reasonably well  with H-shell burning models. (ii)  Ultraviolet analysis shows evidence that the BSGs contain X-rays in their atmospheres, for which we provide constraints. In general, higher X-ray luminosity (close to the standard $\log (L_\mathrm{X}/L) \sim -7$) is favored for early BSGs, despite associated degeneracies. For later-type BSGs, lower values are preferred, $\log (L_\mathrm{X}/L) \sim -8.5$. (iii) The obtained mass-loss rates suggest neither a jump nor an unperturbed monotonic decrease with temperature. Instead, a rather constant trend appears to happen, which is at odds with the increase found for Galactic BSGs. (iv) The wind velocity behavior with temperature shows a sharp drop at $\sim$19~kK, very similar to the bi-stability jump observed for Galactic stars.}
    {}%

   
   \maketitle

\section{Introduction}
\label{sec:intro}

Massive stars ($M > 8\, \mathrm{M_\odot}$) are born as hot objects \citep[earlier than spectral type B2\,V; e.g.,][]{Harmanec+1988,Hillier2020} and are far outnumbered by  low-mass stars. However, they deeply impact their surroundings and shape the galaxies' dynamical and chemical evolution, both due to their usually violent deaths and  to their powerful winds, energetic radiation  largely emitted in the ultraviolet (UV),  and the metal-enriched material they depose in their vicinity. Thus, understanding how these objects behave throughout cosmic history is paramount to comprehending many aspects of the Universe since the birth of the first stars. 

Among the hot stars, B-type stars, especially supergiants, contribute significantly to that feedback in stellar populations younger than 50 Myr \citep{deMello+2000}. Additionally, most B stars observed in external galaxies are B supergiants (BSGs),\footnote{In general, BSG is used to refer to luminosity class I, Ib, Ia, and Iab. However, frequently stars of classes Ia+ (hypergiants) and II (bright giants) are conflated in the samples.} due to their usually high luminosity \citep[e.g.,][]{Kudritzki+2012}. Recently, this has reached even cosmological dimensions, as lensed BSG candidates were inferred from spectra taken of high-redshift objects (e.g., $z \sim 4.8$ and $z \sim 6.2$)  using JWST \citep[see, e.g.,][]{Welch+2022, Furtak+2024}. BSGs can even be progenitors of supernovae \citep[e.g., SN~1987A,][]{Walborn+1987}, resulting in compact objects such as neutron stars (NSs) and black holes (BHs). Their properties and evolution are therefore a major puzzle piece toward understanding high-mass stellar evolution and its endpoints, which are crucial in order to properly interpret gravitational wave events produced by BH and NS mergers \citep[e.g.,][]{Abbott+2016,Abbott+2017}.

B supergiants play a crucial role in our understanding of line-driven wind physics and mark an important stage among the puzzling diverse evolutionary paths of massive stars. BSGs lie at the cool end of the hot stars regime, with effective temperatures ranging from $\sim$30~kK down to $\sim$10~kK. Within this parameter range, important and drastic changes in the wind properties  (e.g., terminal wind velocities, clumping, and X-rays) appear to happen, especially at spectral types around B1, or temperatures of $\sim$22~kK \citep[e.g.,][]{Lamers+1995,Driessen+2019,Berghoefer+1997,Petrov+2014}. Traditionally, these changes are attributed to the   ``bi-stability phenomenon'' \citep[e.g.,][]{Pauldrach+1990, Lamers+1995, Vink+1999}, whose understanding is paramount to constraining the feedback and the evolution of these stars properly.

From a more observational perspective, such changes include a relatively sharp variation in terminal wind speed \citep[e.g.,][]{Lamers+1995, Markova+2008} and in the X-ray luminosity \citep{Berghoefer+1997} within this temperature range. Such drastic  changes align with theoretical findings that massive stars might also experience variations in their wind structure \citep[e.g.,][]{Driessen+2019}, ionization of elements \citep{Vink+1999,Petrov+2014,Petrov+2016}, and a steep jump in mass-loss rates \citep{Pauldrach+1990,Vink+1999, Vink+2000}. The last is particularly relevant for the evolution of high-mass stars, as different mass-loss rates can significantly change a star's evolutionary path \citep[see, e.g.,][]{Vink+2010,Renzo+2017}. However, recent theoretical works challenge the existence of a steep jump in the mass-loss rates in that region, either predicting a mild increase \citep{Krticka+2021} or even predicting a continuous decline \citep{Bjorklund+2023}.

At solar metallicity, recent empirical studies point toward an increase in mass-loss rates in the bi-stability region \citep[e.g.,][]{Bernini-Peron+2023}, albeit less pronounced than predicted by \citet{Vink+1999}. However, it is unclear whether and how the bi-stability phenomenon would manifest at lower metallicities. For instance, \citet{Evans+2004-vinf} shows that terminal wind velocities ($\varv_\infty$) of Small Magellanic Cloud (SMC) BSGs are similar to those from Galactic BSGs,  on the hot side and and on the cool side of the jump region, which resonates with theoretical predictions finding weak \citep[][]{Leitherer+1992} to almost no \citep{Vink-Sander2021} dependence of $\varv_\infty(Z)$, in particular for cooler BSGs.
When looking at mass-loss rates ($\dot{M}$), most studies in the literature invoke a power-law scaling of $\dot{M}$ with $(Z/Z_\odot)$ with the most commonly employed scaling being $\dot{M} \propto Z^{0.70}$ \citep{Vink+2001}. However, the actual $\dot{M}(Z)$ dependence could vary significantly depending on the temperature, luminosity, and proximity to the Eddington limit \citep[see, e.g.,][for recent studies]{Marcolino+2022, Krticka+2024}.

For this work we analyzed a representative sample of 18 SMC BSGs across the bi-stability jump region at $\sim$22~kK. To determine their stellar and wind properties, we used data from the ULLYSES initiative and the X-Shooting ULYSSES (XShootU) collaboration plus previous X-Shooter data from the ESO archive. We focused particularly on deriving the stars' terminal wind velocities, mass-loss rates, clumping factors, and X-ray properties. Additionally, with the stellar properties at hand, we discuss the evolutionary status of the BSGs. Since the stars are in the  SMC, the distance is well-constrained, which is advantageous for obtaining luminosities and masses.

For our analysis we used the non-LTE comoving-frame stellar atmosphere code CMFGEN \citep[][]{Hillier-Miller1998} to obtain the stellar and wind properties. A similar approach was previously taken by \citet{Evans+2004-cmfgen}, albeit for a smaller sample of objects and only up to early-type BSGs. In addition to extending the sample, in particular to the crucial regime of later-type BSGs, we also take into account X-rays in this regime for the first time. 
The outcomes of our study provide insights into the general behavior of BSGs at low metallicity (e.g., relations between wind and photospheric properties). Additionally, it will reduce the necessary parameter space for follow-up studies that will address the full ULLYSES BSGs sample. Specifically, we  provide constraints on the relative X-ray luminosity ($L_{\mathrm{X}}/L$), clumping, and evolutionary properties.

In Sect.\,\ref{sec:observ_data} we discuss the selection of our targets, the XShootU data acquisition, the selection of additional archival data, and interesting features we noticed in the spectra. We explain our analysis methods in Sect.\,\ref{sec:atmos_model} before discussing the results in Sect.\,\ref{sec:stellar-props} (stellar properties), Sect.\,\ref{sec:xrays} (X-rays), and Sect.\,\ref{sec:wind-props} (wind properties). The conclusions are drawn in Sect.\,\ref{sec:conclu}.

%%%%%%%%%%%%%%%%%%%%%%%%%%%%%%%%%%%%%%%%%%%%%%%%%%%%%%%%%%%%%%%%%%%%%%%%%%%%%%%%%%%%%%%%%%%%%%%%%%%%%%

\section{Observational data}
\label{sec:observ_data}
\subsection{Sample selection}

\begin{figure}
  \includegraphics[width=\columnwidth]{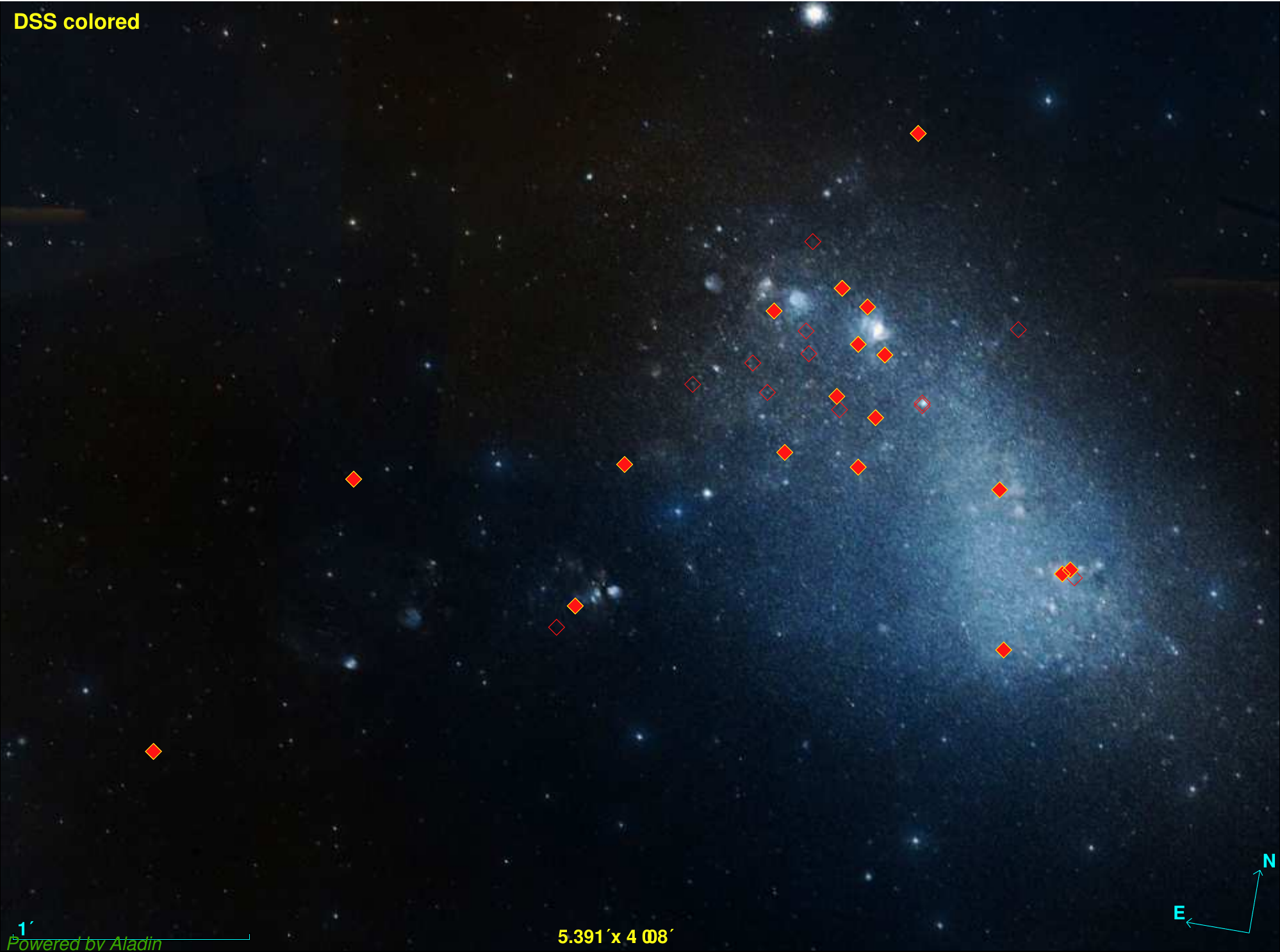}
  \caption{Positions of BSGs in the SMC from ULLYSES. The filled diamond symbols indicate the BSGs analyzed in this study, whereas the empty diamonds indicate the remaining BSGs of the total ULLYSES sample.}
  \label{fig:SMC-BSGs-positions}
\end{figure}
In the recent final ULLYSES data release DR7\footnote{The ULLYSES dataset can be retrieved via the ULLYSES search form: \url{https://ullyses.stsci.edu/ullyses-download.html}.} there are 30 SMC stars classified as BSGs. For this work, we performed a pilot study of the different spectral subtypes. Thus, we selected, whenever possible, two stars per spectral subtype as representatives of the full sample. {In the target selection process, the literature spectral classifications from Paper I \citep{Vink+2023} were used, some of which have been revised (cf.\ Table\,\ref{tab:sample}) following classification of new VLT/Xshooter observations following the SMC B supergiant scheme of Lennon (1997) with luminosity classes assigned using H-gamma morphologies
from Galactic B templates of \citet{Negueruela+2024}.}
In total, we cover 18 of the 30 ULLYSES targets.
A quick inspection of the optical and UV spectra indicated that some of our targets have similar spectral types and a similar spectral appearance in the optical while showing extreme differences in the UV.
This hints at important differences in wind properties despite similar photospheric properties, underlining the importance of picking more than one target per subtype where possible. We discuss our findings on this matter in Sect~\ref{sec:stellar-props}.

In this work, we focus only on apparently single stars. Therefore, we exclude objects such as the known high-mass X-ray binary AzV\,490 \citep[e.g.,][]{Dickey+2023} or Be/B[e] stars (namely, AzV\,261 and AzV\,16). 
While this does not rule out the presence of so far undetected companions, we checked the existing, partially multi-epoch, spectral data for signatures of multiplicity (e.g., radial velocity variations, steps in the UV absorption troughs, additional broad components in absorption lines) and found no clear evidence for them in any of the targets. The full list of our selected targets is shown in Table~\ref{tab:sample}.

\subsection{Sample spectra}

In Fig.~\ref{fig:SMC-BSGs-positions} we plot the positions of the ULLYSES BSGs in the SMC. 
We assume that all the stars are to be located at a distance of 62.45$\pm$2.61\,kpc \citep{Graczyk+2020} and correct the spectra for the radial velocities of each star in our analysis.

\begin{figure}
  \includegraphics[width=\columnwidth]{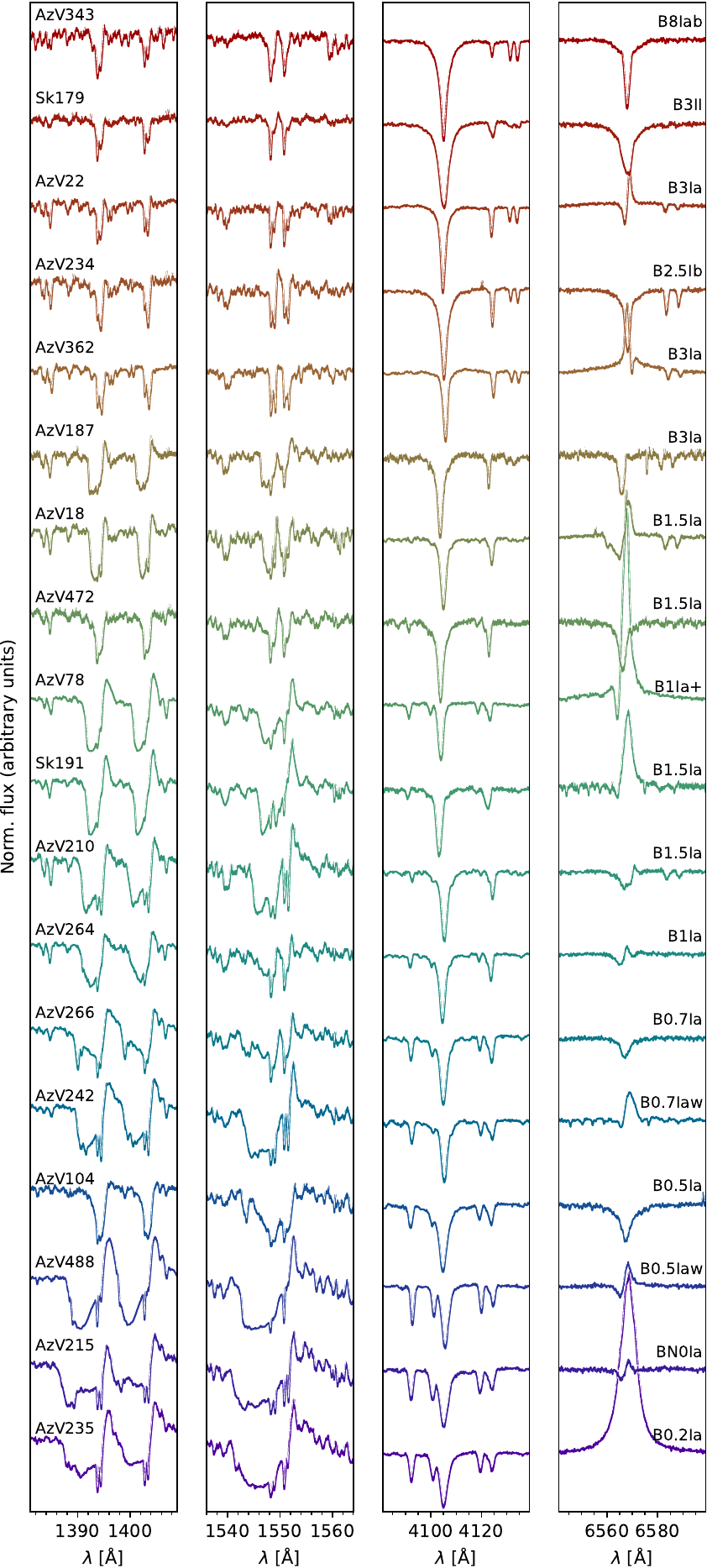}
  \caption{UV and optical spectra used to derive stellar parameters of the sample stars. This figure illustrates the variation in some spectral features with the spectral type. From left to right, the columns show spectral regions around \ion{Si}{IV}~$\lambda1400$, \ion{C}{IV}~$\lambda1550$, O+N+H$\delta$+\ion{He}{I}+Si lines, and H$\alpha$+\ion{C}{II}.}
  \label{fig:uv-op-spec}
\end{figure}

For the spectral analysis, we use HST UV spectra from the ULLYSES dataset and optical spectra from the XShootU collaboration \citep{Sana+2024} -- see, e.g., Fig.~\ref{fig:uv-op-spec}. The UV spectra were obtained with FUSE, from 950~$\mathrm{\AA}$ to 1190~$\mathrm{\AA}$, and COS plus STIS instruments for the rest of the UV range. FUSE operated using the LWRS grating offering a resolving power $R = \Delta\lambda/\lambda \sim 10000$. COS operated in medium resolution modes, yielding $R \sim 13000$. STIS E140M and E230M modes reach $R \gtrsim 45000$ and $R \gtrsim 30000$ respectively. The X-Shooter spectra were taken using the UVB and VIS arms, which cover respectively the range of 3100 to 5600\,$\mathrm{\AA}$ with $R \sim 8000$ and 5600\,$\mathrm{\AA}$ to 10240\,$\mathrm{\AA}$ with $R \sim 11000$. 
In Table~\ref{tab:data-source_1}, we list for each of the stars we analyzed the instruments that were used together with their respective gratings and the spectra acquisition dates for the UV and optical regions, which for most of BSGs were not obtained simultaneously. 

For targets that were not included in the XShootU data release (DR3), we retrieved the spectra from the ESO-Archive, also taken with the X-Shooter spectrograph, and normalized them by fitting the continuum manually. Some targets in our sample show peculiar line profiles in the optical and/or UV region. Such behavior may highlight potential binaries and/or atypical phenomena happening in the BSGs photospheres. To inspect eventual variability in spectral lines of such BSGs, we further used additional archival spectra retrieved from the ESO archive and MAST archive, which were subsequently normalized using the same technique. Information on these spectra can also be found in Table~\ref{tab:data-source_1}. We discuss the results of our brief variability investigation in Sect.~\ref{sec:stellar-props}.

\begin{table}
\caption{\label{tab:sample}Sample of BSGs analyzed in this paper.}
\label{tab:targets}
\centering
\begin{tabular}{lccr}
\hline\hline
Star  & Alt. name & Lit.\ sp.\ type  &  Updated sp.\ type\\  
% --    &  --  & -- & --  & --  \\ 
\hline
AzV\,235 & RMC~17  & B0\,Iaw & \textbf{B0.2\,Ia}  \\
AzV\,215 & Sk\,76   & BN0\,Ia & -- \\
AzV\,488 & Sk\,159  & B0.5\,Iaw & -- \\
AzV\,104 & -- & B0.5\,Ia & -- \\
AzV\,242 & RMC~18 & B0.7\,Iaw & -- \\
AzV\,266 & Sk\,95 & B1I & \textbf{B0.7\,Ia} \\
AzV\,264 & Sk\,94 & B1\,Ia & -- \\
AzV\,210 & Sk\,73 & B1.5\,Ia & -- \\
Sk\,191 & WDG~1 & B1.5\,Ia & -- \\
AzV\,78  & HD~5045 & B1.5\,Ia+ & \textbf{B1\,Ia+} \\
AzV\,18  & Sk\,13 & B2\,Ia & \textbf{B1.5\,Ia} \\
AzV\,472 & Sk\,150 & B2\,Ia & \textbf{B1.5\,Ia}  \\
AzV\,187 & Sk\,68 & B3\,Ia  & -- \\
AzV\,362 & RMC~36 & B3\,Ia & -- \\
AzV\,22  & Sk\,15 & B5\,Ia & \textbf{B3\,Ia} \\
AzV\,234 & Sk\,81 & B3\,Iab & \textbf{B2.5\,Ib} \\
Sk\,179 & -- & B6I & \textbf{B3II} \\
 AzV\,343 & Sk\,111 & B6\,Iab & \textbf{B8\,Iab} \\
\hline
\end{tabular}
\tablefoot{
Literature (``lit.'') spectral types are retrieved from \citet{Vink+2023}. Based on the new XShootU spectra, we provide updated spectral types for several of our targets.}
\end{table}

\subsection{Photometry} 
The \texttt{UBV} magnitudes were retrieved mainly from sources listed in \citep{Vink+2023}, namely: \citet{Massey+2002}, \citet{Ardeberg-Maurice1977}, \citet{Ardeberg1980}, and \citet{Azzopardi+1975}.
Additionally, we included \textit{Gaia} DR3 \citep{GaiaCollaboration+2023} magnitudes (\texttt{Gbp}, \texttt{G}, and \texttt{Grp}). The near-infrared JHK magnitudes from 2MASS, also retrieved from \citeauthor{Vink+2023}, are sourced from \cite{Cutri+2003} and \cite{Cioni+2011}. The only exception is AzV\,78, whose JHK magnitudes were collected from \cite{Bonanos+2010}. The far-infrared \textit{Spitzer} and WISE magnitudes were collected from \citeauthor{Bonanos+2010} and the AllWISE catalog \citep{Cutri+2013} respectively. Each magnitude value can be found in the appendix\,F, available at Zenodo\footnote{\url{https://zenodo.org/records/12700118}.}.

%%%%%%%%%%%%%%%%%%%%%%%%%%%%%%%%%%%%%%%%%%%%%%%%%%%%%%%%%%%%%%%%%%%%%%%%%%%%%%%%%%%%%%%%%%%%%%%%%%%%%%%%%%%%%%%%%%%%%

\section{Atmospheric modeling}
\label{sec:atmos_model}

For the detailed analysis of our sample stars, we use CMFGEN \citep{Hillier-Miller1998, Hillier+2003}, an open-source, non-LTE stellar atmosphere code that solves simultaneously and consistently the radiative transfer and the ionization/excitation population equilibrium equations. The program requires an input velocity structure and atomic data, which encodes the information about the transition levels and energies. Our atmosphere models included H, He, C, N, O, Ne, Al, Mg, Si, S, P, Ca, Cr, Ma, Fe and Ni, adding up to a total of more than $5 \times 10^5$ individual atomic transitions -- see \citet{Hillier+2003} for details on the atomic data sources. Our initial models were calculated with a metallicity of 0.2~Z$_\odot$ using \citet{Asplund+2009} for the elements listed in \citet{Vink+2023}. There is evidence for non-solar relative ratios of individual elements for the SMC, especially carbon (C), nitrogen (N), and oxygen (O). However, as BSGs in general present alterations in their CNO surface compositions \citep[e.g.,][]{Trundle+2005, Crowther+2006}, and we individually determine the abundances of each of the CNO elements, their initial baseline values were unimportant.

The procedure to obtain the stellar properties is similar to \citet{Bernini-Peron+2023}, albeit we explore more parameters in this work. Below, we summarize the procedure to obtain the stellar properties, highlighting the differences to \citeauthor{Bernini-Peron+2023}. Except for the stellar luminosity, which is determined by the overall spectral energy distribution (SED), all the properties were obtained via fitting specific diagnostic lines, whose strength and shape respond well to certain stellar properties.

{Given the high dimensionality, the non-linearity, the model computation times, and the intricate cross-dependence between the parameters in the atmosphere models, a fully automatic approach exploring a high-dimensional grid to obtain the stellar properties and their uncertainties is computationally unfeasible. Instead, we pick a suitable model from an existing set of models and iteratively adjust the stellar parameters to find the best compromise to reproduce the various diagnostics, as common in the analysis of hot stars with computation-intensive atmosphere codes such as CMFGEN and PoWR  \citep[e.g.,][]{Crowther+2006,Martins+2015,Ramachandran+2019}. Knowing the reactions of the lines to changes in the parameters from a sequence of model calculations, the quality of the spectral fitting is inferred from visually inspecting the different solutions (``by-eye fit''). This is a standard procedure in quantitative spectral analysis that has been successfully applied since the availability of complex non-LTE atmosphere models \citep[e.g.,][]{Hillier-Miller1999,Trundle+2005,Martins+2015,Ramachandran+2019}. As shown in the discussion of the different approaches and analysis methods in XShootU paper IV \citep[][, submitted]{Sander+2024}, this type of analysis yields in general coherent and similar results compared to more automatized approaches.}

To establish the uncertainties of fundamental stellar and wind parameters (e.g., $T_\mathrm{eff}$, $\log g$, abundances, $\dot{M}$, $\varv_\infty$), we test variations of the individual parameters deviating from the best fit we could achieve. Given the inherent systematic uncertainties within the atmosphere modelling and the scatter between different atmosphere codes \citep[cf.][]{Sander+2024}, we refrain from doing this individually for each star, but perform our error margin study for selected individual hot and cool BSGs, adopting the obtained error margins to the rest of our sample. 
For properties based on more approximate or ad-hoc descriptions (e.g., clumping, X-rays, $\beta$), we do not provide uncertainties in this study.
While it is formally possible to explore this parameter space, e.g., in the context of a Genetic Algorithm coupled with a fast atmosphere calculation code \citep[see, e.g.,][for applications with FASTWIND]{Brands+2022}, codes like CMFGEN, which compute the full spectrum in the comoving frame in much more detail are very time-consuming and computationally expensive. Therefore, in this analysis framework, uncertainties for each parameter for each star are much more challenging to be quantified. Usually one needs to rely on the analysis of a few objects/models \citep[e.g.,][]{Bernini-Peron+2023} and/or usually can only be better constrained in dedicated studies for specific properties \citep[e.g.,][]{Ruebke+2023}.

\subsection{Atomic data}
\label{sec:atomdata}

For the detailed NLTE modeling of the atmosphere, we followed the same approach as described in \citet{Bernini-Peron+2023}, who analyzed Galactic BSGs cooler than 20\,kK. However, as we also model hot BSGs in this work, we additionally consider higher ionization stages. The ion sets included in hotter and cooler models are slightly different, as certain ions are not significantly populated at some temperature ranges. We list our considered elements and ions in Table~\ref{tab:atomic}.

The atomic data we used are those present in the default, publicly available, installation of CMFGEN (v.05.05.2017). Most of the sources used by the code can be found in \citet{Hillier+2003}, though.

\subsection{Determination of the photospheric parameters}
\label{subsec:diagnostics_phot}

In the following, we describe the process to obtain the fundamental stellar (i.e., non-wind) properties, which are summarized and discussed later in Sect.\,\ref{sec:stellar-props}.

\paragraph{Luminosity and reddening} To obtain the stellar luminosity $L$ and the reddening $E(B-V)$, we fitted the SED from the UV to the infrared using the flux points derived from the different magnitudes (listed in Sect.\,\ref{sec:observ_data}) and flux-calibrated spectra from ULLYSES and the XShooter counterpart when available. We adopt the extinction law by \cite{Howarth+1983} with $R_V = 2.7$ and we consider a fixed galactic foreground extinction with $E(B-V) = 0.03$  and $R_V = 3.1$ using \cite{Cardelli+1989} law. The generally low extinction toward the SMC limits the diagnostic values of the absorption bump at $2175$~$\AA$. For inferring the $E(B-V)$, we thus focus on reproducing the overall spectral energy distribution given by the shape of the flux-calibrated UV and optical spectra as well as infrared photometry. We estimate a conservative uncertainty of $0.03$ for our derived $E(B-V)$ values. As the reddening also enters the determination of the luminosity $L$, we adopt an error of $15\%$ for our derived values of $L$.

\paragraph{Rotation and macroturbulence} The rotation $\varv \sin i$ and macroturbulence $\varv_\mathrm{mac}$ were obtained by using the IACOB Broad tool \citep{SimonDiaz-Herrero2014}. We used He and metal lines, namely \ion{He}{I}~$\lambda4713$ and \ion{Si}{III}~$\lambda4552$ for stars earlier than B2 and \ion{Mg}{II}~$\lambda4481$ for later types (see Fig.\,\ref{fig:AV18-IACOB_Me} and Fig.\,\ref{fig:AV18-IACOB_He} for  an example). The tool also provides the $\varv \sin i$ computed from the Fourier transform of the line profile (hereafter $\varv_\mathrm{fft}$), which we use as a sanity check and to compute uncertainties (see appendix\,\ref{sec:rotmac}).
As the metal lines are less affected by Stark and other additional broadening mechanisms \citep[see, e.g.,][]{SimonDiaz-Herrero2007,SimonDiaz-Herrero2014}, we give preference to their resulting $\varv \sin i$, $\varv_\mathrm{mac}$, and $\varv_\mathrm{fft}$ and report them in Table\,\ref{tab:stelprop}. To quantify the uncertainty, we also obtain the $\varv \sin i$, $\varv_\mathrm{mac}$, and $\varv_\mathrm{fft}$ via \ion{He}{I}~$\lambda4713$ and compute the standard deviation between the He and metal lines for each the quantities ($\Delta\varv \sin i$, $\Delta\varv_\mathrm{mac}$, and $\Delta\varv_\mathrm{fft}$). 
For all stars, we consider a minimum uncertainty of 25\% in $\varv \sin i$ and of 40\% for $\varv_\mathrm{mac}$ -- the average relative standard deviation of these quantities in the sample. In case of the values being smaller than $\Delta\varv_\mathrm{fft}$, we select the latter as the final uncertainty. This conservative approach avoids mathematically small errors and does not artificially underestimate the errors of targets whose uncertainties are larger.  
In the spectral analysis, the derived rotation is implemented by convolving an elliptical kernel, while for macro-turbulence a Gaussian kernel convolution is applied \citep[cf.][]{Gray2005}. We further evaluated the overall spectral fitting, and if necessary made adjustments to the initially obtained values of rotation and macroturbulence. This happened in particular for AzV\,187, where $\varv \sin i$ is close to $11$~km~s$^{-1}$,  which is near the limit imposed by the Nyquist frequency for the star's spectral resolution -- see \cite{SimonDiaz-Herrero2007}. Thus, the determinations were deemed as not reliable -- see discussion in appendix\,\ref{sec:rotmac}. 

\paragraph{Effective temperature and surface gravity} The effective temperature $T_\mathrm{eff}$ is obtained via evaluating the ionization balance of Si and He by fitting multiple ionization stages, when available. For the ``hot'' BSGs (B0 to B1.5) \ion{Si}{IV}/\ion{Si}{III} and \ion{He}{II}/\ion{He}{I} are used, namely \ion{Si}{IV}~$\lambda4089,4116$ and \ion{Si}{III}~$\lambda\lambda$4556,69,76 for silicon, and (ii) \ion{He}{I}~$\lambda4471$, \ion{He}{I}~$\lambda4387$ and \ion{He}{II}~$\lambda4552$ for helium. \ion{He}{II}~$\lambda4686$ was used as an auxiliary diagnostic as it can be affected by wind properties \citep{Martins+2015}. For the ``cool'' (B2 to B5) and ``cold'' BSGs (later than B5) we aim to fit \ion{Si}{III}/\ion{Si}{II} and the \ion{He}{I} lines (as the \ion{He}{II} lines are non-existent), namely \ion{Si}{II}~$\lambda\lambda$4128,32 and \ion{Si}{III}~$\lambda\lambda$4556,69,76 for silicon, and especially \ion{He}{I}~$\lambda4471$, \ion{He}{I}~$\lambda4387$ and \ion{He}{I}~$\lambda4713$. Additionally, other lines such as \ion{Mg}{II} $\lambda$4481 and the Balmer lines are taken into account. In some cases, $T_\mathrm{eff}$ went through further revisions during the determination of He abundance. As \citet{Vink+2023} discusses, there is evidence that the Si and Mg content in SMC could be individually lower than the 0.2~Z$_\odot$ scaling -- 0.14 and 0.17, respectively. However, this does not introduce major uncertainties in the temperature, as it affects Si lines more or less equally. Testing models with varying temperatures only, we find in general an accuracy of 1~kK. To account for influence of other parameters, we consider a standard error of $1.5$\,kK, which is within the range of similar studies. We specify otherwise in the case of targets which we encountered more difficulties finding a satisfactory spectral fitting.
The $\log g$ values in the photosphere are obtained by fitting the wings of the Balmer lines, given their high sensitivity to pressure in the stellar photosphere \citep[see, e.g.,][]{Gray2005}. Additionally, some lines such as \ion{Si}{III}~$\lambda$4550--65--79 and \ion{He}{I}~$\lambda$4471 are also affected and were used as a sanity check. Likewise for the temperature, we estimate an uncertainty of $0.1\,$dex for $\log g$.

\paragraph{Spectroscopic masses and radii} These quantities were indirectly determined from $\log g$, $T_{\mathrm{eff}}$ and $L$. The uncertainty for mass $M$ and $R$ are, respectively, 35\% and 15\%. These values were obtained considering our errors in the primary quantities to follow normal distributions \citep[see][]{Bernini-Peron+2023}. Because of the small relative errors in the adopted distance \citep[e.g., $\lesssim 0.01$ dex,][]{Graczyk+2020}, the luminosity is very well constrained in the case of SMC stars. Hence, $\log g$ is the dominant source of the mass error.

\paragraph{Microturbulence} The microturbulence near the photosphere $\xi_\mathrm{phot}$, reflecting small-scale perturbations \citep[see, e.g.,][]{Moens+2022, Debnath+2024}, is estimated mainly by analyzing the strength of \ion{Si}{III} lines as these are sensitive to the velocity field \citep[e.g.,][]{Catanzaro-Leone2008}. The intensity of other lines such as He, H (mostly in the core of the line), and other metals are also affected and thus used as sanity checks. For the Si lines, the most prominent in the optical and the UV are considered, namely, \ion{Si}{III}~$\lambda\lambda4553,69,76$ and lines around 1300 $\mathrm{\AA}$. $\xi_\mathrm{phot}$ affects temperature and surface gravity diagnostics, $T_\mathrm{eff}$ and $\log g$ were re-adjusted if necessary. Especially for the earlier-type stars, the central depth of the Balmer lines is also affected by the density of the inner wind, thus introducing a sensitivity to the mass-loss rate, clumping, and the velocity gradient -- see Sect.\,\ref{subsec:windparadeter}. The microturbulence is set to increase linearly with the velocity across the wind, following:
\begin{equation}
    \xi(r) = \xi_{\mathrm{phot}} + \frac{(\xi_\mathrm{max}-\xi_\mathrm{phot})\varv(r)}{\varv_\infty},
\end{equation}
where $\xi_\mathrm{phot}$ is the turbulence near the photosphere and $\xi_\mathrm{max}$ is the turbulence when the wind reaches $\varv_\infty$.
{ We obtain an accuracy of $\sim$3~km~s$^\mathrm{-1}$. However, in cases where the actual Si abundance is lower than the adopted baseline,
a higher microturbulence would be inferred to reproduce the same line strength. To include this effect in our uncertainty estimates, we assume a higher margin of about 5~km~s$^\mathrm{-1}$.}

\paragraph{He and CNO surface abundance} 
As presumably evolved objects, BSGs are expected to present surface alterations in H, He, and CNO. In this study, to obtain the helium mass-fraction $Y$, we adjusted the He abundance aiming for a better fit of the many \ion{He}{I} (and \ion{He}{II} for earlier types) lines along the optical range. From that procedure, we estimate an error of a factor of 2. { As discussed in \citet{Hillier+2003} or \citet{Crowther+2006}, a precise determination of the He abundance in supergiants is particularly challenging and we thus give this rather large margin. Enhanced He abundances should thus be rather seen as a qualitative sign of enrichment rather than a precise number yielding hard constraints.}
To determine the CNO abundances, we employ the following sets of lines: For carbon, we fit \ion{C}{II}~$\lambda4267$ and \ion{C}{II}~$\lambda\lambda6578,82$ in late BSGs and \ion{C}{III}~$\lambda4069$ and \ion{C}{III}~$\lambda\lambda4648-50$ in early BSGs. For nitrogen, the \ion{N}{II} series of lines at 4600$\AA$ (in late BSGs), \ion{N}{III}~$\lambda4097$, \ion{N}{II}~$\lambda4447$, \ion{N}{III}~$\lambda\lambda4510-20$ (early BSGs) were used. \ion{N}{III}~$\lambda4097$ is heavily affected by the temperature and, therefore was not considered as a main diagnostic. In the case of oxygen, the optical region between 4000 and 500 $\mathrm{\AA}$ is crowded with oxygen lines. From these, we give most emphasis on \ion{O}{II}~$\lambda4069-92$, \ion{O}{II}~$\lambda4590-96,4661$ and \ion{O}{III}~$\lambda4367$. The determinations of precise abundances for individual hot massive stars can be quite challenging, as different diagnostic lines may point to different abundances \citep[][XShootU Paper V]{Martins+2024} and fundamental parameters ($T_\mathrm{eff}$, $\log g$, and $\xi$) affect these lines significantly. Given that uncertainties in $T_\mathrm{eff}$, $\log g$, and $\xi$ affect CNO lines, we consider an error of 0.3~dex also for the CNO abundances. In Fig.~\ref{fig:diag_CNO} we show an example of how CNO diagnostic lines are affected by different abundances.

\subsection{Determination of wind parameters}
  \label{subsec:windparadeter}

After determining the photospheric parameters, we focused on the analysis of the UV and recombination lines, which encode most of the wind properties. The resulting properties are listed further in Table\,\ref{tab:wind-prop}.

\paragraph{Wind terminal velocity and microturbulence} The wind terminal velocity is obtained by fitting the width of the UV P Cygni wind lines, especially the absorption trough. For stars where the UV P Cygni profiles are very unusually shaped (e.g., AzV\, 104) or very weak (mostly for later type BSGs) the determination of the terminal velocity is not obvious. In those cases, we adopt the values that yield a better overall fit of the line. The adopted microturbulence at the terminal wind speed ($\xi_\mathrm{max}$), which is computed in the formal integral to producing the observer's frame spectrum, can influence the derived values of $\varv_\infty$. To keep uniformity with previous works that used CMFGEN, we took a value of $\sim 10\%$ of the respective $\varv_\infty$. For targets with unusual or very weak UV P Cygni profiles, we explored a wide range of values, reaching, in some cases, values of $\xi_\mathrm{max} \sim \varv_\infty$.

\paragraph{Mass-loss rates} To determine the mass-loss rates, we aim for a simultaneous fit of H$\alpha$ and the wind P Cygni profiles in the UV. For some targets, fitting the profile with CMFGEN is not possible, in particular, due to a peculiar H$\alpha$ shape. In these cases, we perform a different approach described in appendix\,\ref{sec:pecultargs}. Additionally, H$\alpha$ is severely affected by clumping parameters whereas the UV wind lines are in general very affected by the X-ray emission.
From exploratory model calculations, we infer a standard uncertainty of 0.15~dex. However, when considering the effects of clumping -- see below -- we take a cautious approach and estimate a combined error of 0.40\,dex.

\paragraph{Velocity profile} As CMFGEN does not compute the velocity stratification $\varv(r)$ consistently with the radiative acceleration, the velocity structure needs to be provided. This is described by a modified $\beta$-law, describing the wind, connected to a quasi-hydrostatic structure that describes the inner atmosphere (subsonic regime) which is updated based on the estimated line acceleration. 
Wind lines, in particular optically thick lines are affected by different values of the parameter $\beta$. \citep[e.g.,]{Lefever+2023}. In the BSG regime, H$\alpha$ is notoriously sensitive to the density profile. However, other Balmer lines and P Cygni profiles are also affected, though with much less sensitivity than H$\alpha$.
Thus we estimated $\beta$ by aiming to find a better fit to H$\alpha$ and used the UV features as auxiliary diagnostics. As clumping also affects significantly H$\alpha$, we also needed in many cases to revise our $\beta$ values.
In this study, we set the connection velocity $\varv_{\mathrm{con}}$, reflecting the transition between the photospheric/subsonic and wind/supersonic regime, to be 10\,$\mathrm{km\,s^{-1}}$ by default, which is around 70\% of the sound speed in the photosphere of BSGs. The value is lowered if necessary to ensure a monotonic velocity field (which is mandatory in co-moving frame atmosphere calculations).

\paragraph{Clumping parameters}

In CMFGEN clumping is pictured as optically thin overdensities, whose diameters are smaller than the photons' mean free path, and whose surroundings (interclump regions) are void. Mathematically, it is by standard implemented the following profile in radius stratification:
$$
f(r) = f_\infty + (1-f_\infty)e^{-\varv/\varv_\mathrm{cl}}  \, \text{,}
$$
describing a scenario where clumping reaches its maximum in the outer wind.
The two clumping parameters are $f_\infty$ and $\varv_\text{cl}$, which, respectively, represent the volume filling factor (i.e., how clumped is the wind at $r\rightarrow \infty$) and the characteristic onset velocity (describing where clumping starts to be relevant in the wind). These are obtained by aiming at a consistent fit of H$\alpha$ (and H$\beta$ in some cases) and the UV lines in general.
However, for some targets, due to their unusual H$\alpha$ profiles, we explored other approaches -- cf. appendix~\ref{sec:pecultargs}.
The Balmer lines as recombination lines are proportional to $\rho^2(r)$, while the UV P Cygni profiles as scattering lines are proportional to $\rho(r)$ only. Therefore, as essentially perturbations to the density, clumping will impact much more the latter. However, as e.g.\ \citet{Bouret+2012} has demonstrated, clumping also has some impact in the UV by altering the ionization stage of different ions via a radiative field which encounters a different distribution of matter. For the O and early-B targets, \ion{N}{IV} and \ion{P}{V} are particularly good diagnostics. As \citet{Bouret+2005} discusses, these ions increase their population in the wind due to the presence of clumps (i.e., enhanced density). Therefore we aimed for a reasonable fitting of these lines when obtaining clumping. For later-type stars, clear diagnostic lines specific to this parameter are weak or absent, so we can only aim for a good overall UV spectral line fitting. Moreover, in later BSGs profiles that are sensitive to wind stratification are also affected by X-rays \citep{Bernini-Peron+2023}.

\paragraph{X-ray parameters}

The X-ray emission in the wind is by default parametrized in CMFGEN as a \textit{Bremsstrahlung} component plus tables of emissivities generated by the APEC code \citep{Smith+2001}. Three free parameters define the emissivity profile, namely $T_\mathrm{X}$, which represents the typical temperature reached by the X-ray production mechanism; $\varv_\mathrm{X}$, the characteristic velocity of the production of X-rays that reflects the point in the wind where the X-ray emissivity profile would reach $\sim$30\% of its possible maximum; and $f_\mathrm{X}$, the X-ray volume filling factor. As this X-ray treatment is a simplification, $T_\mathrm{X}$ and $\varv_{\mathrm{X}}$ should not be interpreted as actual physical properties of the wind and their precise values do not bear larger physical significance. 
With the basic X-ray parameters, the code integrates the emissivity profile and gives the produced as well as the observable X-ray luminosity ($h\nu >$ 0.1 keV). The latter is usually given in terms of the stellar luminosity as $\log (L_\mathrm{X}/L)$.
The X-ray luminosity ($L_\mathrm{X}$) is constrained by fitting the UV lines. For stars earlier than B1.5, the main diagnostic used is the famous \ion{N}{V}~$\lambda$1238 line. This ion can only exist due to the presence of this additional ionization source. For cooler stars, \ion{C}{IV}~$\lambda$1550 becomes the main diagnostic, as analogously, \ion{C}{IV} would not be present in winds of stars cooler than $\sim$20\,kK without the presence of X-rays. Conversely, the \ion{C}{II}~$\lambda$1335 profile without the presence of X-rays tends to be overpredicted, as such lower temperatures are more fertile to the existence of this less ionized stage. 
The ``shock temperature parameter''  $T_\mathrm{X}$ is set to be 1.0 MK for models with $\varv_\infty \geq 900$\,km\,$\mathrm{s}^{-1}$ and 0.5\,MK otherwise, motivated by the results of \citet{Bernini-Peron+2023}. Finally, the characteristic velocity parameter $\varv_\mathrm{X}$ is initially set to $\sim80\%$ of $\varv_\infty$ and changed if necessary to improve the fitting. For some targets, this value had to be severely changed to improve the fitting. In Sect.~\ref{sec:xrays} we discuss these aspects in more detail.

%%%%%%%%%%%%%%%%%%%%%%%%%%%%%%%%%%%%%%%%%%%%%%%%%%%%%%%%%%%%%%%%%%%%%%%%%%%%%%%%%%%%%%%%%%%%%%%%%%%%%%%%%

\section{Stellar properties}
\label{sec:stellar-props}

\begin{figure*}
\centering
  \includegraphics[width=18cm]{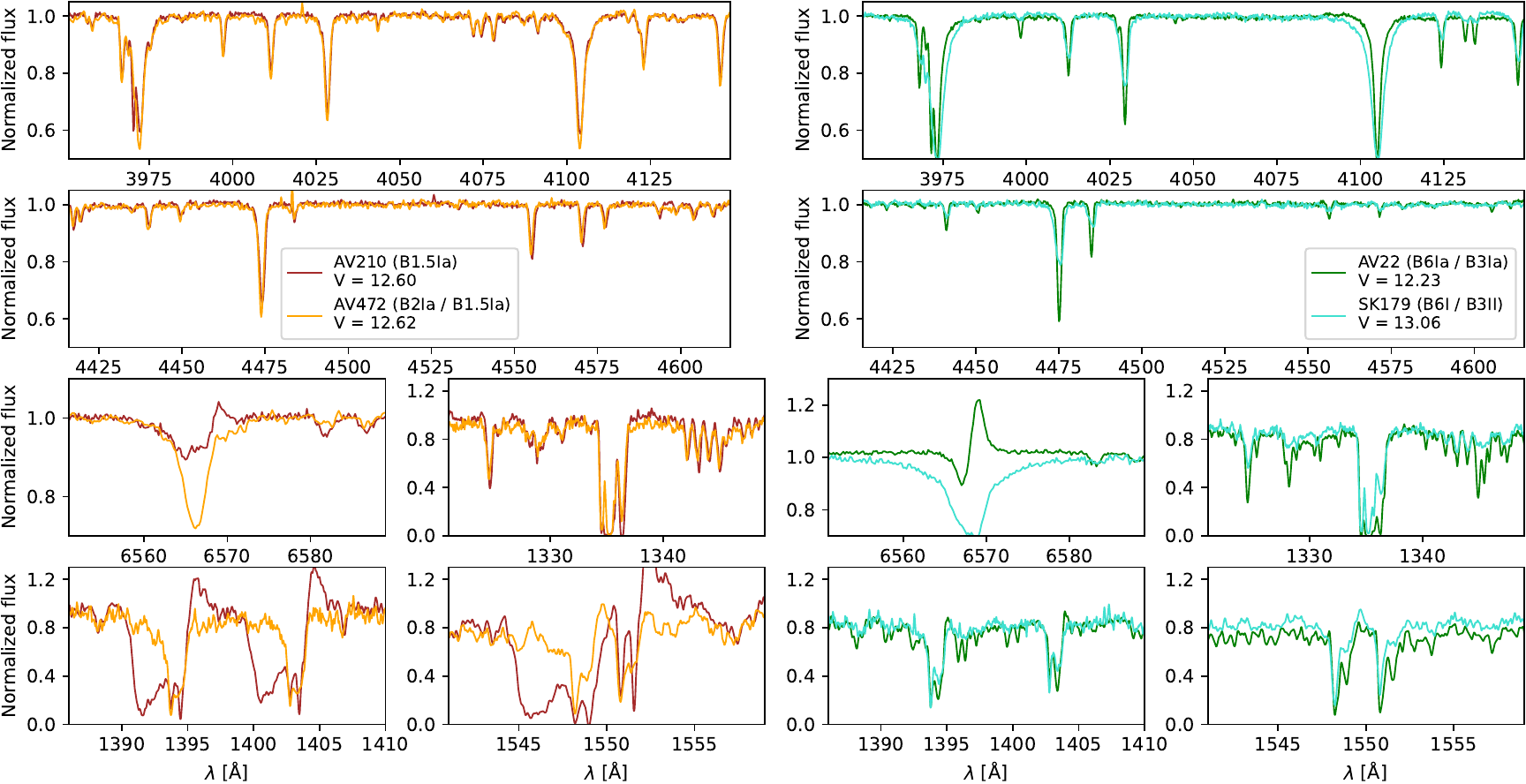}
  \caption{Comparison between observed spectra of AzV\,210 (B1.5\,Ia) vs. AzV\,472 (B1.5\,Ia), and AzV\,22 (B3\,Ia) vs. Sk\,179 (B3\,II). For the comparison between the B1.5 stars, their optical and  photospheric spectra are very similar to each other, but their wind lines at the UV and H$\alpha$ are considerably different, despite the  similar luminosity and temperature. In the juxtaposition between   B3 and B8, the profiles in their UV are similar, while their optical spectra differ, especially due to broadening and different surface gravities. In each comparison, we list the old and the updated spectral types (Crowther priv. comm.)}
  \label{fig:UV-same-spectra}
\end{figure*}

In the process of analyzing the spectra of BSGs, prior to the modeling phase, we noticed that some targets initially classified with similar spectral types had very similar optical spectra -- which foreshadow similar photospheric properties,  but looked very different on the UV. Conversely, some other targets present a similar UV spectrum between them, but very different optical spectra. This is illustrated in Fig.\,\ref{fig:UV-same-spectra}, where we can see that AzV\,210 and AzV\,472 have an almost identical optical spectrum (except for H$\alpha$) and similar luminosities, but very different UV lines. This highlights that their winds must be considerably different. 

Contrary to that, AzV\,22 and Sk\,179 display very similar UV features but quite different optical lines. The differences between the latter two could be related to their difference in luminosity (or proximity to the Eddington Limit), since AzV\,22 is much more luminous and has a lower $\log g$, noticeable by its much narrower Balmer lines. By covering all of these cases in our analysis sample, we aim to get a representative overview of the different kinds of BSGs in the SMC.

Some stars within our sample display peculiar signatures in their UV, optical, or IR spectra -- namely AzV\,235, AzV\,104, Sk\,191, and AzV\,362, which could not be properly modeled by CMFGEN. We examined the spectra from different epochs and found no clear signatures of multiplicity. Yet, we noticed variability in their Balmer profiles (except for AzV\,104), which is especially strong for AzV\,362. For Sk\,191, a mild variability can also be seen in \ion{He}{I}~$\lambda4471$, which we interpret as a wind feature. Specific comments on individual targets are further given in appendix\,\ref{sec:pecultargs}.

In Fig.~\ref{fig:diagnostics}, we provide an overview of the spectral fits in selected diagnostic ranges for the whole sample to illustrate the quality of the results and the trends across the BSG subtype range. The figure illustrates also some common issues in the quantitative spectroscopy of BSGs, such as the under-prediction of \ion{N}{III}~$\lambda4097$ in the earlier-type stars \citep[e.g.,][]{Crowther+2006,Searle+2008}. Full comparison plots between models and observations as well as the SED fits can be found in the Appendix\,F, available at the platform Zenodo\footnote{\url{https://zenodo.org/records/12700118}.}.  
The resulting photospheric properties we derived are listed in Table~\ref{tab:stelprop}.

\begin{figure*}
\centering
  \includegraphics[width=18cm]{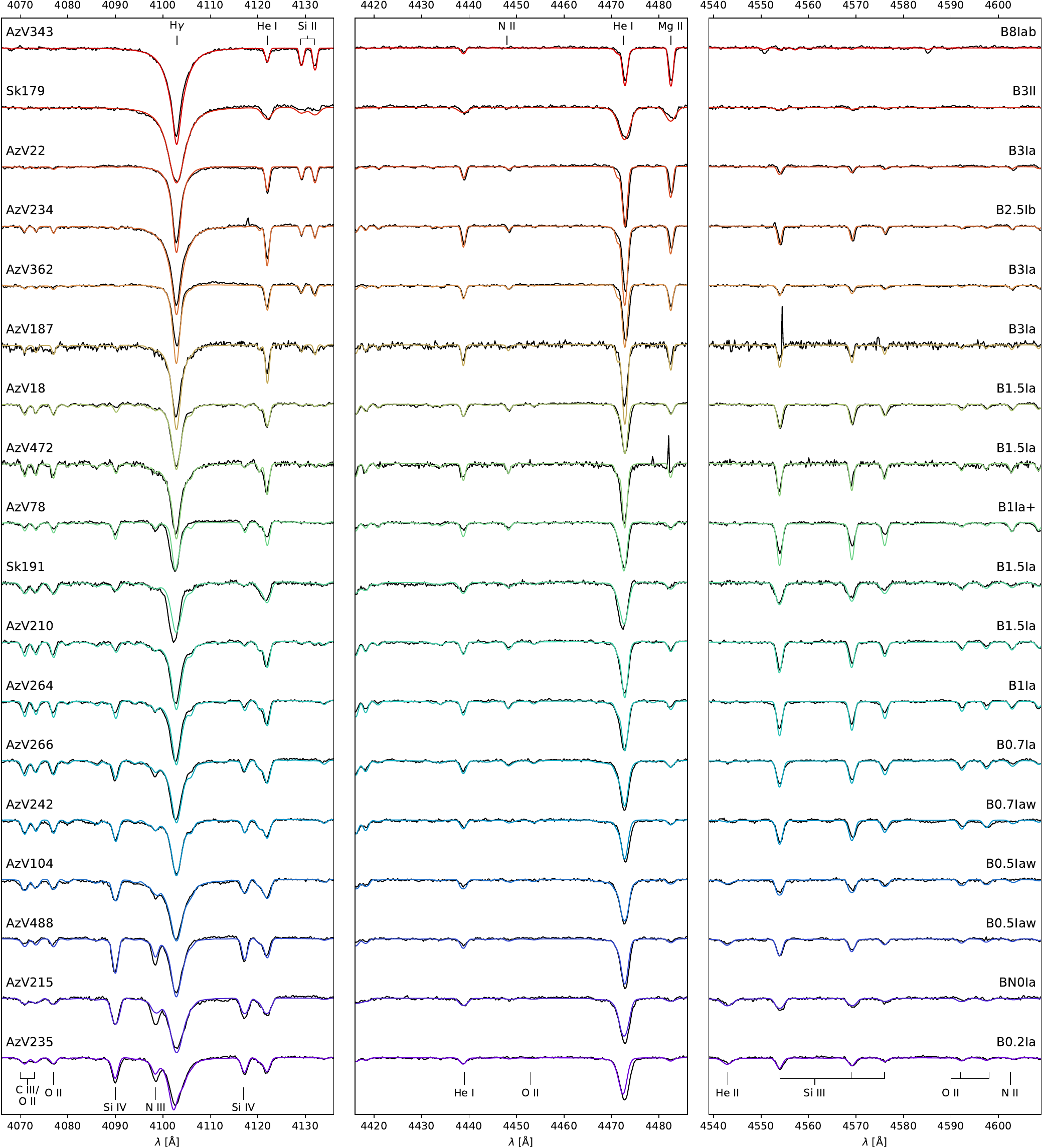}
  \caption{Spectral fits of exemplary subsets of diagnostic lines for the sample BSGs. The black curves are the X-Shooter data for each star, whereas the colored curves are their respective CMFGEN models.}
  \label{fig:diagnostics}
\end{figure*}

\subsection{Spectral-Type calibration}

\begin{figure}
\centering
  \includegraphics[width=\columnwidth]{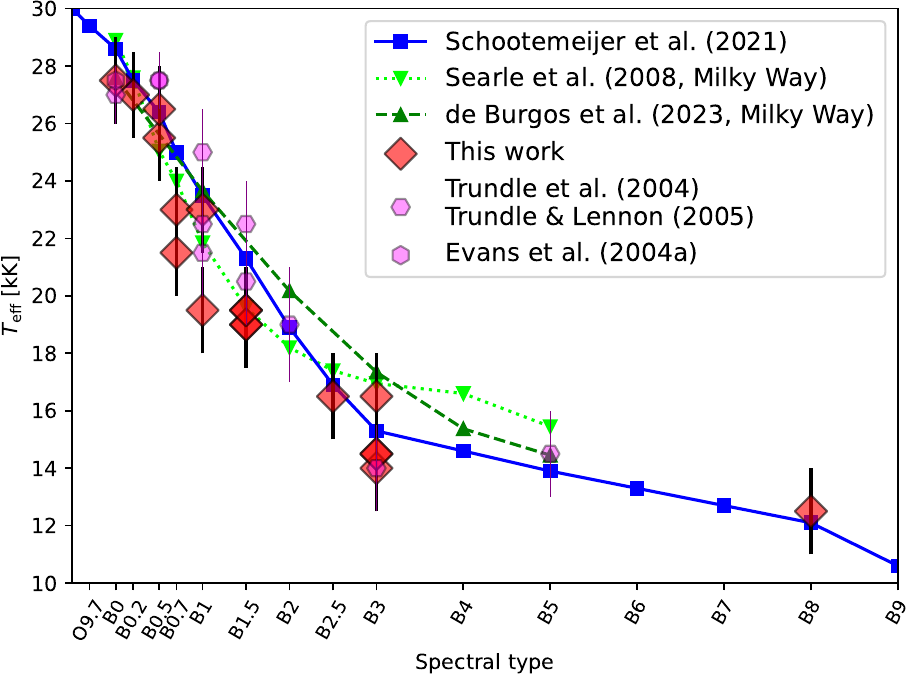}
  \caption{Comparison between effective temperatures and spectral types. The red diamond symbols represent our spectroscopic determinations, whereas the hexagons are literature values from \cite{Evans+2004-cmfgen}, \cite{Trundle+2004}, and \cite{Trundle+2005}. The plotted curves are the calibrations from \cite{Schootemeijer+2021} for the SMC stars (solid blue curve), \citet[][dotted green curve]{Searle+2008}, and \citet[][dashed dark green curve]{deBurgos+2023a} for Galactic BSGs.}
  \label{fig:teff-spt}
\end{figure}

As we analyzed a representative sample of SMC BSGs in this work, we can check the correspondence between the effective temperatures and spectral types. In Fig.~\ref{fig:teff-spt}, we show the comparison between $T_\mathrm{eff}$ and spectral type alongside the calibrations from \citet[][]{Schootemeijer+2021} for SMC supergiants, as well as with calibrations for Galactic BSGs from the studies by \citet{deBurgos+2023a} and \citet{Searle+2008}, who analyzed Galactic BSGs using FASTWIND and CMFGEN, respectively. We additionally plot results from \cite{Evans+2004-cmfgen}, \cite{Trundle+2004}, and \cite{Trundle+2005} for SMC BSGs as well.

In general, there is a good agreement between the spectral analysis results from our work and the literature when compared to the relation of \cite{Schootemeijer+2021}. Yet, up to a spectral type of B2, the $T_\mathrm{eff}$ of our sample BSGs fall below the calibration by \citet{Schootemeijer+2021}, which was based on previous literature results including those marked in fuchsia in Fig.\,\ref{fig:teff-spt}. The divergence is largest between spectral types B0.5 to B1.5. Within this range, we also tend to find lower temperatures than the Galactic calibration by \cite{deBurgos+2023a}. On the other hand, our results for this spectroscopic range agree better with the relation from \citet{Searle+2008}, even though falling slightly lower still. For spectral types later than B2, our temperatures appear to agree well with the \cite{Schootemeijer+2021} relation, but diverge from \cite{Searle+2008}. 

\subsection{Evolutionary status of the BSGs}
\label{sec:evol}

\begin{sidewaystable}
\caption{Main stellar parameters of our sample BSGs.
}
\centering
\begin{tabular}{lc|cccccccccccr}
\hline\hline
Star  & Sp.Type  & $T_\mathrm{eff}$     &  $\log g$  &  $\log L/\mathrm{L_\odot}$   & $M$ & $\Gamma_e$ & $Y$   & $X_\mathrm{C}$        & $X_\mathrm{N}$        & $X_\mathrm{O}$        & $\xi_\mathrm{phot}$ & $\varv \sin i$ & $E(B-V)$  \\

 -- & -- & [kK]  & [cgs] & -- & [$M_\odot$] & -- & -- & [$\times 10^{-4}$] & [$\times 10^{-4}$] & [$\times 10^{-4}$] & [km\,s$^{-1}$] & [km\,s$^{-1}$] & -- \\
\hline
AzV\,235 & B0.2\,Ia & 27.5 & 3.05 & 5.82 & 53 & 0.31 & 0.26 & 0.80 & 3.80 & 8.70 & 7      & 39 & 0.11\\
AzV\,215 & BN0\,Ia & 27.0 & 2.90 & 5.63 & 26 & 0.36 & 0.39 & 0.25 & 3.40 & 11.30 & 11    & 86 & 0.16\\
AzV\,488 & B0.5\,Iaw & 25.5 & 2.75 & 5.88 & 41 & 0.41 & 0.37 & 0.86 & 5.60 & 8.70 & 11   & 55 & 0.10 \\
AzV\,104 & B0.5\,Ia & 26.5 & 3.05 & 5.45 & 26 & 0.23 & 0.40 & 0.70 & 3.00 & 8.70 & 9     & 73 & 0.10 \\
AzV\,242 & B0.7\,Iaw & 23.0 & 2.60 & 5.62 & 24 & 0.40 & 0.33 & 0.63 & 3.20 & 10.0 & 8    & 40 & 0.10 \\
AzV\,266 & B0.7\,Ia & 23.0 & 2.65 & 5.44 & 17 & 0.36 & 0.49 & 0.43 & 3.60 & 7.50 & 8     & 44 & 0.11 \\
AzV\,264 & B1\,Ia & 21.5 & 2.50 & 5.43 & 16 & 0.31 & 0.45 & 0.37 & 4.15 & 6.90 & 7       & 44 & 0.07 \\
AzV\,210 & B1.5\,Ia & 19.5 & 2.45 & 5.39 & 19 & 0.31 & 0.26 & 0.63 & 4.30 & 11.30 & 15   & 25 & 0.16 \\
Sk\,191 & B1.5\,Ia & 19.0 & 2.25 & 5.67 & 26 & 0.38 & 0.42 & 0.90 & 3.20 & 11.30 & 14    & 57 & 0.15 \\
AzV\,78 & B1\,Ia+ & 19.5 & 2.15 & 5.96 & 37 & 0.49 & 0.45 & 0.30 & 5.80 & 4.00 & 18      & 26 & 0.14 \\
AzV\,18 & B1.5\,Ia & 19.0 & 2.35 & 5.45 & 20 & 0.35 & 0.26 & 0.60 & 3.50 & 9.00 & 10     & 41 & 0.25 \\
AzV\,472 & B1.5\,Ia & 19.5 & 2.50 & 5.29 & 17 & 0.27 & 0.26 & 0.52 & 3.90 & 8.20 & 12    & 32 & 0.09 \\
AzV\,187 & B3\,Ia & 16.5 & 2.20 & 5.33 & 19 & 0.28 & 0.26 & 0.11 & 3.00 & 11.30 & 15     & 40 & 0.04 \\
AzV\,362 & B3\,Ia & 14.5 & 1.80 & 5.58 & 22 & 0.40 & 0.32 & 0.85 & 4.20 & 11.30 & 17     & 14 & 0.13 \\
AzV\,22 & B3\,Ia & 14.0 & 1.90 & 5.27 & 16 & 0.29 & 0.26 & 0.70 & 4.05 & 11.30 & 18      & 42 & 0.13 \\
AzV\,234 & B2.5\,Ib & 16.5 & 2.40 & 4.97 & 13 & 0.18 & 0.26 & 0.11 & 3.00 & 11.30 & 17    & 14 & 0.08 \\
Sk\,179 & B3II & 14.5 & 2.30 & 4.81 & 12 & 0.13 & 0.26 & 0.83 & 3.26 & 11.30 & 15        & 83 & 0.05 \\
AzV\,343 & B8\,Iab & 12.5 & 2.05 & 4.72 & 10 & 0.13 & 0.26 & 0.21 & 3.26 & 11.30 & 30    & 45 & 0.09 \\
 \hline
Errors & & 1.5 & 0.1  & 15\% & 35\% & -- & $\times$2 &  $\times$2 & $\times$2 &  $\times$2 & 5 & $>25\%$ & 0.03 \\
\hline
\end{tabular}
\label{tab:stelprop}
\end{sidewaystable}

\begin{figure}
  \includegraphics[width=\columnwidth]{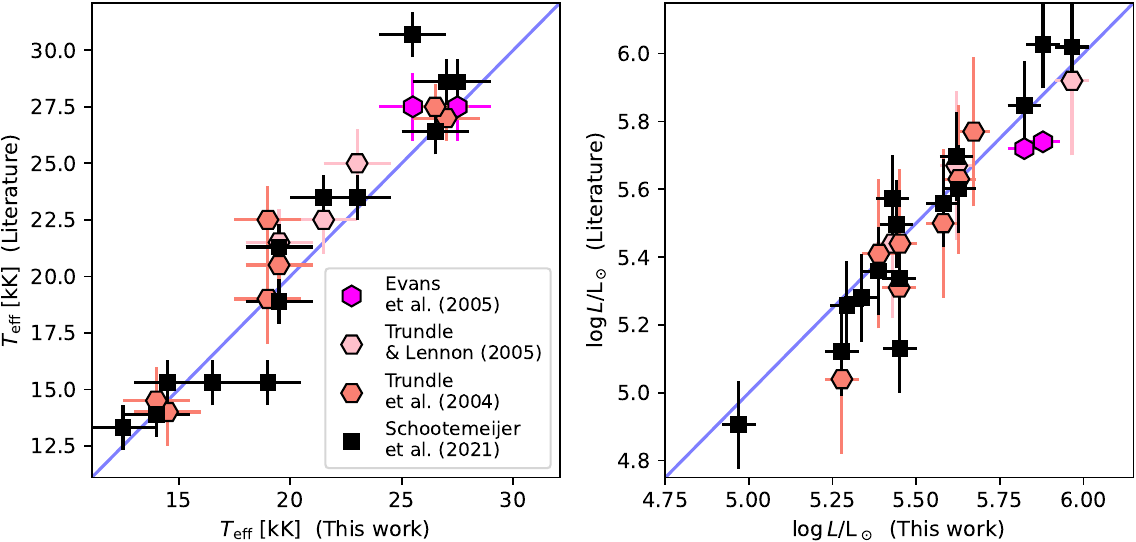}
  \caption{Comparison between the derived effective temperatures and luminosities of our sample SMC BSGs with results from the literature, namely: \citet[][fuchsia hexagons]{Evans+2004-cmfgen}, \citet[][pink hexagons]{Trundle+2004}, \citet[][orange hexagons]{Trundle+2005}, and \citet[][black squares]{Schootemeijer+2021}.}
  \label{fig:teff-l-lit}
\end{figure}

When comparing our derived luminosities and $T_\mathrm{eff}$ to previous spectroscopic analyses on SMC BSGs \citep[e.g.,][]{Evans+2004-cmfgen,Trundle+2004,Trundle+2005}, we notice no relevant systematic differences (see Fig.\,\ref{fig:teff-l-lit}). Likewise, we also do not find systematic disagreements between our investigation and the study of \cite{Schootemeijer+2021}, who used \textit{Gaia} distances to infer the stellar luminosities. Even the largest discrepancies in the parameters are within the expected frame of uncertainty when comparing different analysis methods \citep[see]{Sander+2024}.

\begin{figure}
  \includegraphics[width=\columnwidth]{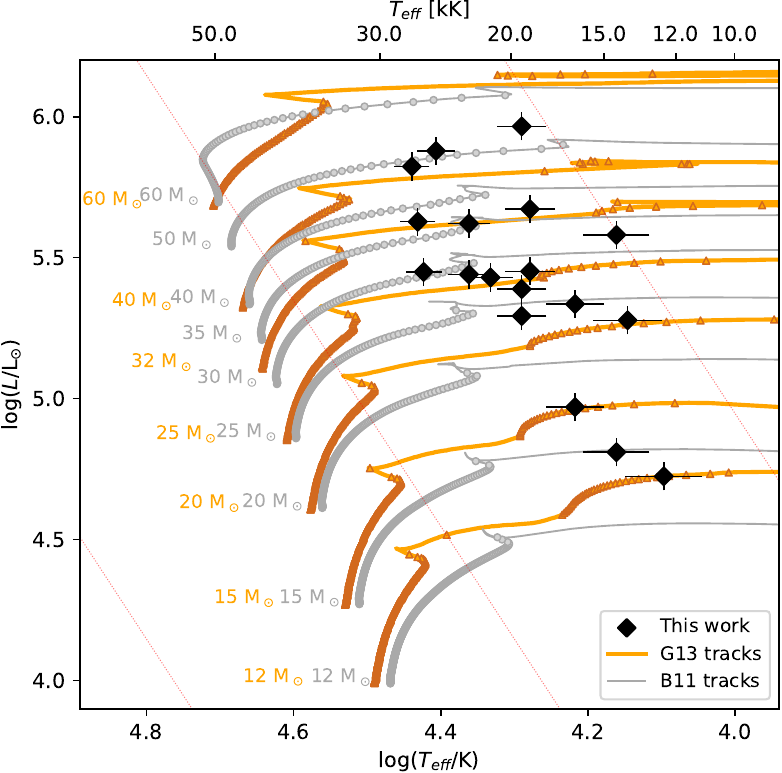}
  \caption{HR diagram of the sample stars (HRD). Tracks of \citet{Georgy+2013} (G13, thick orange lines) and \citet{Brott+2011} (B11, thin gray lines) with initial rotation of $\varv_{\mathrm{eq}}/\varv_\mathrm{crit} \sim 0.4$ (i.e., $\sim 300$ km/s). The small triangular and circular points on the tracks mark intervals of about 50 kyr for G13 and B11 respectively, and the red diagonal lines represent the iso-radii lines (1, 10, and 100, R$_\mathrm{\odot}$ from left to right). 
  }
  \label{fig:HRD}
\end{figure}

With the luminosities and temperatures of our 18 BSGs determined, we can locate the stars in the Hertzsprung Russell diagram (HRD) and discuss their evolutionary context. 
 In Fig~\ref{fig:HRD} we plot the HR diagram of the BSGs together with the evolutionary tracks' curves by \citet[hereafter B11]{Brott+2011} and \citet[hereafter G13]{Georgy+2013}. Both sets of evolution models have similar rotation, but their treatment of convection is considerably different. The B11 models determine convection using the Ledoux criterion \citep{Brott+2011}, while the G13 models use the Schwarzschild criterion \citep{Ekstrom2012, Georgy+2013}. We find our sample stars to be displaced relative to the zero age main sequence, which is consistent with previous results in the literature and the expectation that BSGs are already evolved objects \citep[e.g.,][]{Trundle+2004, Evans+2004}.
However, this does not automatically imply that BSGs cannot be core-hydrogen-burning. As \citet{HigginsVink2019} discuss in detail, evolutionary models with high convective mixing can produce tracks whose main sequence covers the BSG temperature and luminosity regime -- see also \cite{Martins-Palacios2013}.

Notably, the B11 tracks can reach the BSGs earlier than B1 in the HRD. As discussed in \cite{Brott+2011}, the overshooting adopted in B11 tracks is calibrated for stars with initial masses $M_\mathrm{init} \sim 16$\,M$_\odot$ to match the TAMS inferred from the VFTS sample \citep{Hunter+2008b}. This aligns with more recent investigations that argue more massive stars may have progressively higher mixing \citep[e.g.,][]{Castro+2014, Martinet+2021}. As a consequence, the main sequence might reach down to $T_\mathrm{eff} \sim 20$\,kK -- corresponding approximately to the B1 spectral type \citep{McEvoy+2015}.

Contrary to that, the G13 tracks interpret the HRD positions of the early BSGs as objects beyond core-H burning. The ``terminal age main sequence (TAMS) hooks'' of the G13 tracks happens at much hotter temperatures than those in the B11 tracks. This is rooted in the much lower convective mixing in the G13 high-mass tracks compared to B11. The G13 high-mass tracks are calibrated to match the TAMS width for intermediate-mass stars \citep{Georgy+2013}. According to the G13 models, more massive/luminous stars spend much less time as early-type BSGs (i.e., between $\sim$30 and $\sim$20~kK) than as later-type BSGs.

Below $T_\mathrm{eff} \lesssim 19$\,kK, the redward evolution predicted by the G13 models slows down and the stars spend a decent amount of time still in the hot star regime, where central He-burning eventually sets in. This is different for the B11 models, which evolve considerably fast in the Hertzprung gap, as indicated by the scatter symbol distribution along both track sets which mark intervals of $\sim$50\,kyr. This discrepancy is due to the different convection criteria between the two evolution model grids. The Schwarzschild criterion in the G13 models produces a larger intermediate convective zone (ICZ) at the onset of hydrogen shell burning \citep[see also][]{Sibony+2023,Josiek+2024}. The additional energy required to establish a larger ICZ is not available for the expansion of the star and thus the G13 models spend more time crossing the Hertzsprung gap than the B11 models.

In contrast to the Galactic BSGs discussed in \citet{Bernini-Peron+2023}, neither the B11 nor the G13 tracks predict a blueward evolution from the red supergiant (RSG) regime. Blue loops only happen for more massive G13 models and at temperatures more compatible with yellow supergiants/hypergiants and luminous blue variables (LBVs). For this regime, we do not have corresponding objects in our sample. Consequently, assuming the evolution framework of B11 and G13, none of the studied SMC BSGs would be post-RSG objects. However, certain mixing settings \citep[e.g.,][]{Schootemeijer+2019} or enhanced mass-loss in the RSG or LBV regime \citep[e.g.,][]{Schootemeijer+2024} could produce tracks evolving blueward from the RSG regime at SMC metallicity. Therefore, we cannot rule out that some of our targets are post-RSG objects. This would be supported by our finding of considerably lower masses compared to evolutionary predictions, which we further discuss in Sect.\,\ref{sec:mspec-mevol}.

\subsection{Surface chemical abundances}

\begin{figure}
  \includegraphics[width=\columnwidth]{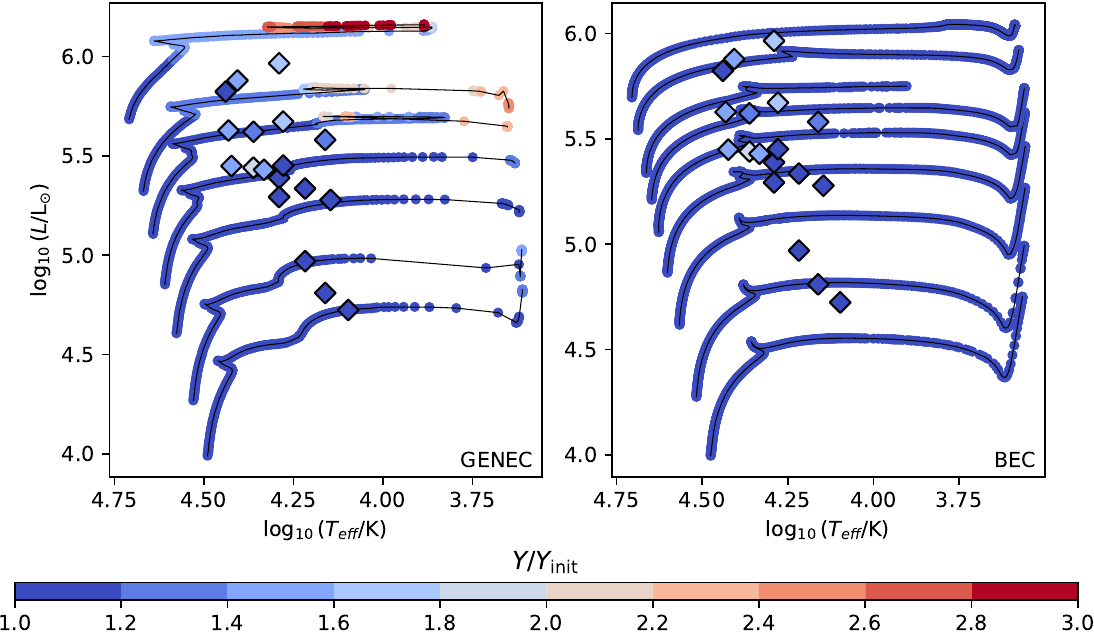}
  \caption{HR diagram showing helium enrichment ($Y/Y_\mathrm{init}$, where $Y_\mathrm{init}$ is the baseline value of 0.27) in the B11 and G13 SMC tracks compared to our sample BSGs (diamonds). The tracks are the same as in Fig.~\ref{fig:HRD}.}
  \label{fig:He-enrich}
\end{figure}

In Table\,\ref{tab:stelprop}, we also list the surface helium mass fraction $Y$. Several of our objects show signs of He enrichment, in particular the BSGs above 19\,kK, corresponding approximately to spectral types earlier than B1.5. This is qualitatively in line with the recent finding by \citet{deBurgos+2023b}, who also found hardly any He-enriched objects in the Galactic BSGs below this temperature.  Although, it is important to note that for $T_\mathrm{eff} \lesssim 20$~kK, He abundance determination rely solely on \ion{He}{I}. In our sample, the BSGs without He enrichment are in the clear minority in the hotter regime, which does not seem to be the case for the Galactic sample, albeit the work by \citet{deBurgos+2023b} mixes actual supergiants with objects from other luminosity classes. 
When considering the He abundances of both evolution models, we do not see the observed amount of He enrichment in the early stars (see Fig.~\ref{fig:He-enrich}), which hints at remaining issues.
One possibility is that the models do not assume sufficient mixing. \citet{Schootemeijer+2019} showed that for certain conditions of overshooting and semiconvection, it is possible to produce BSGs evolving blueward from the RSG regime, which would be richer in surface He. Still, these models cannot reach the $T_\mathrm{eff}$ of our earlier targets. Recent work from \cite{GormazMatamala+2024} showed that adopting a lower $\dot{M}$ (obtained via evolutionary models whose mass loss is computed consistently) in the main sequence may produce stellar models that reproduce the He enrichment for the early BSGs.
Another possibility, outside of the single-stellar evolution scenario, is that the hotter BSG population might have a more complex origin, like stripping or mergers \citep{Klencki+2022,Menon+2024} -- which we  discuss further below.

\begin{figure*}
  \includegraphics[width=18cm]{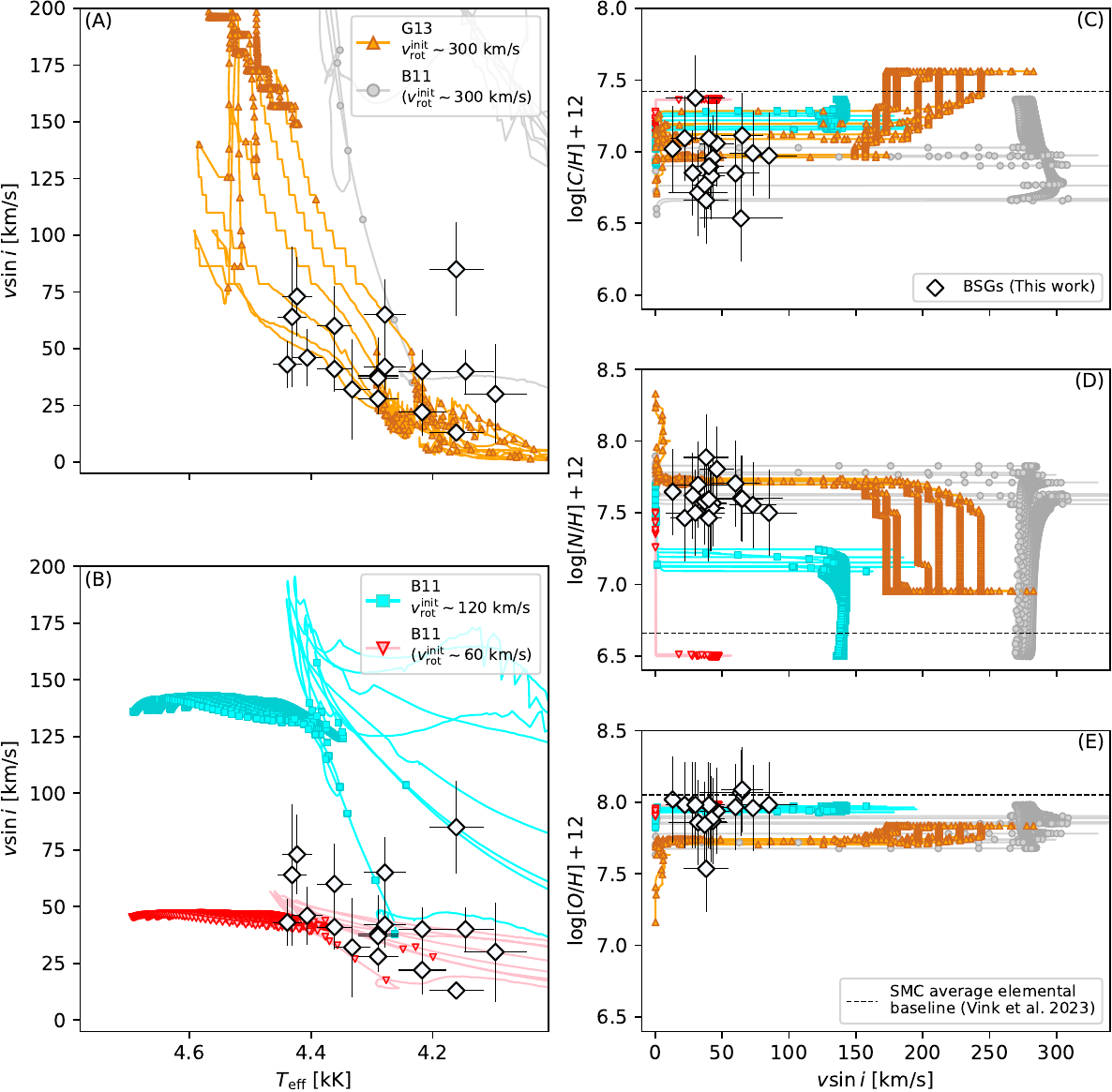}
  \caption{Comparison of effective temperature, rotation, and CNO enrichment--depletion between G13 \citep[orange tracks with triangle symbols,][]{Georgy+2013},  B11  \citep[gray, blue, and pink tracks,][]{Brott+2011}, and our spectroscopic determinations (white diamonds). In this figure, we include tracks with initial masses from 12 to 50\,M$_\odot$ (to 40\,M$_\odot$ for the G13 set). The tracks' equatorial rotation velocities are scaled by $\pi/4$ to account for the stars' inclinations \citep[e.g.,][]{Hunter+2009}. As in Fig~\ref{fig:HRD}, the points along the tracks mark intervals of about 50 kyr. Baseline CNO values of B11 models are respectively 7.37, 6.50, and 7.98, while for G13 those are 7.56, 6.95, and 7.83. The dashed horizontal lines mark the CNO average baselines listed in \citet[][Table\,2]{Vink+2023} determined as the average values from previous studies. Panel A and B show the evolution of the rotation against $T_\mathrm{eff}$ while panels C, D, and E show respectively the evolution of the surface C, N, and O vs. the rotation.
  }
  \label{fig:CNO-hunter}
\end{figure*}

When considering the evolution of the rotation and CNO abundances, we find a clear displacement from the SMC carbon and nitrogen baseline\footnote{The baseline considered is the average elemental baseline computed by \cite{Vink+2023} based on previous literature studies;  see their Table\,{2}.} (dashed lines in Fig. \ref{fig:CNO-hunter}). This confirms previous findings and underlines that our targets are evolved objects. For oxygen, we do not observe much variation and find values close to the baseline (see panel E in Fig. \ref{fig:CNO-hunter}), implying that the material in the outer layers has not reached full CNO equilibrium. The evolution models by B11 and G13 each have their own, slightly different, initial baseline abundances. Yet, the effect on the resulting tracks is likely minor compared to the larger differences in the physical treatments inherent to the two different evolution model codes. The G13 tracks and the slow-rotating  (i.e., with initial rotation velocity $\varv_\mathrm{rot}^\mathrm{init} \lesssim 100$\,km\,s$^{-1}$) B11 models reproduce the data better than the fast-rotating B11 models. Similar to our comparison in the HR diagram in Fig.\,\ref{fig:HRD}, it is evident that the G13 evolutionary models predict a fast evolution through the early-BSG stage (panel A in Fig\,\ref{fig:CNO-hunter}). 

Notably, our derived $\varv \sin i$ are lower than previous determinations \citep{Evans+2004-cmfgen,Trundle+2004,Trundle+2005}, whose values are higher than the rotation predicted from G13 for the late-O and BSG temperature range. This can be understood when considering that earlier work conflated $\varv \sin i$ and $\varv_{\mathrm{mac}}$ in their total line broadening, which can overestimate the rotation and thus explain their sistematically higher values of $\varv \sin i$.

Our obtained $\varv_\mathrm{mac}$ is generally high compared to the derived $\varv \sin i$, which is consistent with the results obtained by \cite{SimonDiaz+2017} for Galactic OB stars. Moreover, as e.g., discussed in \citet{Sundqvist+2013}, disentangling rotation and macroturbulence is considerably more challenging for stars with low rotation ($\varv \sin i \lesssim 50$~km\,s$^{-1}$). Additionally, as discussed in \cite{SimonDiaz-Herrero2014}, the photospheric microturbulence also hampers an accurate determination of $\varv \sin i$ below 40\,km\,s$^{-1}$. Hence, the derived $\varv_\mathrm{mac}$ could even be underestimated if $\varv \sin i$ is overestimated for our slowest rotating targets. A lower rotation for the late BSGs would increase the agreement with the predictions from G13, whereas the disagreement would be aggravated for the earlier ones.

We can only obtain a reasonable compromise between the stellar rotation, $T_\mathrm{eff}$, and partially the involved evolution timescales, when we consider B11 tracks with lower initial rotation, preferably with $\varv_\mathrm{rot}^\mathrm{init} = 60$ km\,$\mathrm{s}^{-1}$. However, for the chemical surface abundances, the B11 tracks with an initial rotation of 120 km\,$\mathrm{s}^{-1}$ would be preferred. This occurs especially for nitrogen, which also is very similar to the values predicted by G13. Only for oxygen, the B11 models of both 60 and 120 km\,$\mathrm{s}^{-1}$ give a better representation of our results than the G13 models. In general, none of our sample stars exceeds a $\varv \sin i$ of $100\,\mathrm{km}\,\mathrm{s}^{-1}$. Our object with the highest (projected) rotational velocity ($85\,\mathrm{km}\,\mathrm{s}^{-1}$) is Sk\,179, a B6 supergiant that defies the general trend seen in Fig.\,\ref{fig:CNO-hunter}. However, even this value would not classify the star as a fast rotator. We do not see any fast rotating star in our BSG sample, not even above 21\,kK. While fast-rotating BSGs are also not common at Galactic metallicity, there is an observed group of them, in particular above the presumed bi-stability jump \citep[e.g.,][]{deBurgos+2023c}.

\begin{figure}
  \includegraphics[width=\columnwidth]{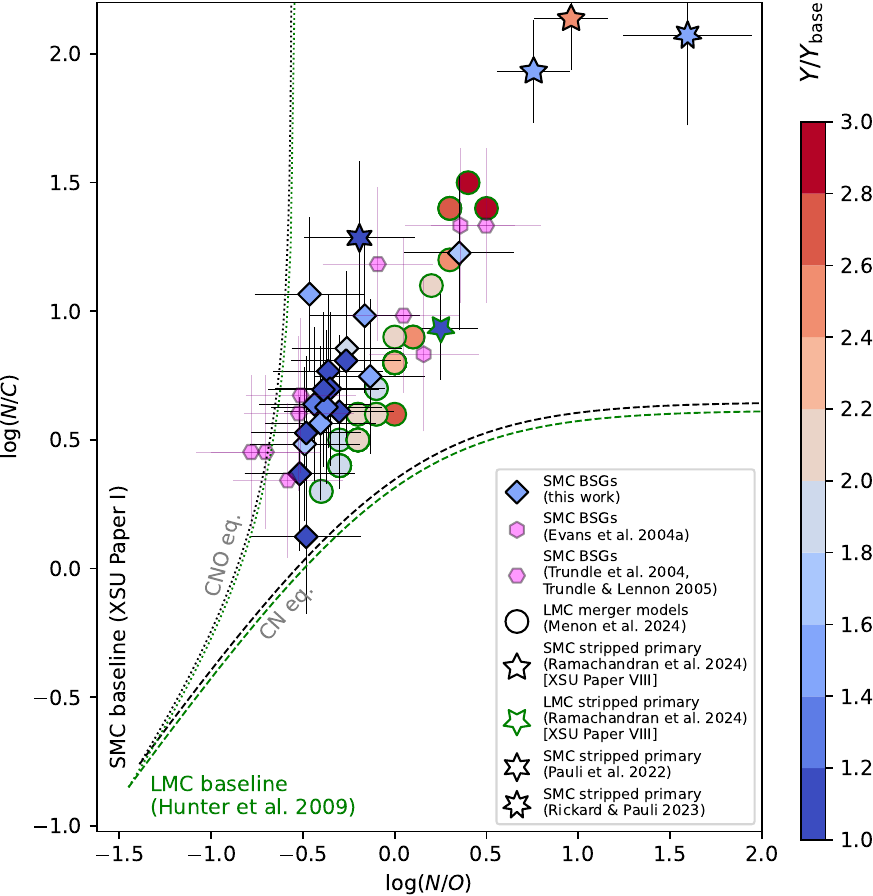}
  \caption{N/C vs. N/O diagram.  The dashed lines show the CN and CNO equilibrium curves for the SMC (in fuchsia) and LMC (in black) using the respective baseline abundances from XShootU Paper I \citep{Vink+2023} and from \cite{Hunter+2009}, which was the same used by the LMC merger models of \citet[][circles]{Menon+2024}. The star symbols with green contours represent the ratios determined for SMC stripped stars from \citet[][six-pointed star]{Pauli+2022}, \citet[][seven-pointed star]{RickardPauli2023}, and XShootU Paper VIII \citep[][five-pointed star]{Ramachandran+2024}. The diamonds show the ratios of our sample BSGs and the fuchsia hexagons represent ratios determined by \cite{Evans+2004-cmfgen, Trundle+2004, Trundle+2005}. Those do not provide He abundance determination, whose enrichment relative to the baseline is represented by the color scale applied to the symbols.}
  \label{fig:NCNO}
\end{figure}

In \citet{Bernini-Peron+2023}, we discussed stellar mergers as a possible solution for the presence of cooler BSGs (B2 to B5) in the Milky Way, given that evolutionary models predict a fast evolution in this regime. In the galactic context, mergers would produce core-He-burning objects, spending a considerable time of their post-main-sequence evolution in the observed parameter range. Given the high binary fraction among massive stars \citep[e.g.,][]{Mason+2009,Sana+2012,Sana+2013}, such a scenario would not be unlikely. Specifically, for apparently single stars, the merger incidence considering coalescence events at different moments of the evolution can amount up to 20\% \citep{deMink+2014}.
{ To get a basic idea of the enrichment of a merger scenario, we compare the N/C vs. N/O ratios derived for our SMC BSGs with those from Large Magellanic Cloud (LMC) BSG merger evolution models from \cite{Menon+2024} in Fig.~\ref{fig:NCNO}. The figure also shows data points for known LMC and SMC stars with (partially) stripped envelopes are plotted for comparison. Despite agreement within the large error bars, there seems to be a systematic shift between our BSGs and the merger models. The merger models further overpredict the He enrichment, notably in contrast to the underprediction by the G13 and B11 (single-star evolution) models. In terms of the predicted mass range, the LMC merger models predict similar values as obtained in our spectroscopic analysis. Yet, it is hard to draw any firm conclusions as none of these diagnostics can only be explained by a merger scenario. Moreover, tests for the SMC are necessary as the metallicity affects how the stars ought to interact \citep{Klencki+2022}, and thus at which stage they may merge or not.}

{ Considering the stripped OB stars and our sample BSGs, we find a good agreement in their He enrichment, except for 2dFS-163 (O8\,Ib). We further find that part of the stripped sample shows much higher N/C and N/O ratios. Yet, we see that two objects align well with the BSGs. One is an early O-giant primary and the other has a primary star with a BSG spectral type \citep[B1.5\,Ia,][]{Ramachandran+2024}, but is located in the LMC. The masses of our sample BSGs are higher than the stripped stars found by \citet{Ramachandran+2024}, but overall they agree with those from the more massive stars analyzed by \cite{Pauli+2022} and \cite{RickardPauli2023}, even though the latter are O giants.

}

\subsection{Spectroscopic and evolutionary masses}
\label{sec:mspec-mevol}

\begin{figure}
\centering
  \includegraphics[width=\columnwidth]{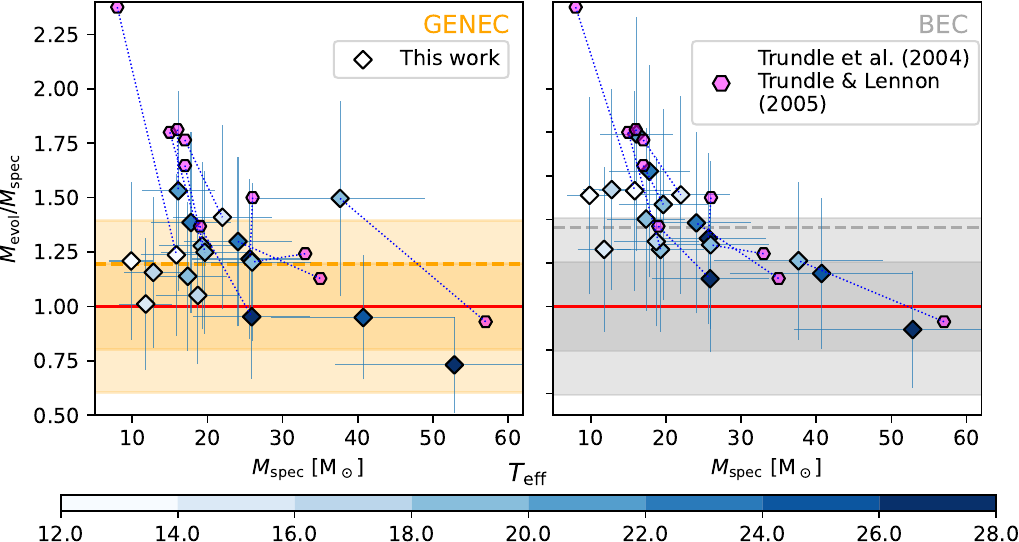}
  \caption{Comparison between evolutionary and spectroscopic masses. The color bar represents the effective temperature of the targets with lighter colors reflecting lower $T_\text{eff}$. 
  The fuchsia hexagons correspond to the data of \cite{Trundle+2004} and \cite{Trundle+2005} and the dotted blue lines connect the same stars. The darker horizontal orange and gray bands represent, respectively, the intervals of 1$\sigma$ obtained for G13 and B11. Likewise, the lighter colored bands are for 2$\sigma$.}
  \label{fig:Mspec-Mevol}
\end{figure}

In Fig\,\ref{fig:Mspec-Mevol}, we show a comparison between the spectroscopic and inferred evolutionary masses for each BSG in our studied sample. The evolutionary masses were determined by minimizing the $\chi^2$ between the spectroscopically derived stellar parameters and the available tracks of G13 and B11 each considering $T_\mathrm{eff}$, $\log g$, and $L$ \citep[Eq.\ 7 in][]{Schneider+2014}. This method, despite not considering interpolated tracks between the available set, is more systematic than what is commonly done by comparing visually in the HRD and/or $\log g$-$T_\mathrm{eff}$ diagram \citep[see, e.g.,][for O stars]{Bouret+2021}.

Globally, there is a clear trend of higher evolutionary masses relative to the spectroscopic mass, especially below $M_\text{spec} \approx 30\,M_\odot$ (G13) and $M_\text{spec} \approx 45\,M_\odot$ (B11), respectively. As shown in the figure, the B11 comparison trend aligns very well with the discrepancy obtained by \cite{Trundle+2004} and \cite{Trundle+2005}, especially the comparison with the B11 tracks. Juxtaposed to the  \citeauthor{Trundle+2004} results, the mass discrepancy is slightly lower for the comparison with G13 tracks. The more and less intense shaded regions represent deviations from unity of 1$\sigma$ and 2$\sigma$ respectively.

When considering the mean results, there is a decent agreement between the spectroscopic and evolutionary masses for both track sets with G13 yielding a ratio within 1$\sigma$, whereas the B11-derived mean mass ratio only agrees within 2$\sigma$. The amount of scatter is similar to what \citet{Bouret+2021} found for SMC O stars, but unlike them, our results show the aforementioned systematic shift. Similar to previous works which analyzed BSGs both in the Milky Way and in the SMC, we further find that for higher masses (or luminosities) the ratio between $M_\text{evol}$ and $M_\text{spec}$ tends to swap and the evolutionary models predict slightly lower masses \citep[see, e.g.,][]{Trundle+2005,Searle+2008}. This did not occur for the SMC O stars studies by \citet{Bouret+2021}, who attributed the discrepancy to a lack of photospheric turbulence in their CMFGEN O-dwarf models. This could lead to an underestimation of $\log g$ -- as also discussed in paper IV. Given that BSG atmospheres are more extended than OB-dwarfs (which implies higher scale heights), they are expected to have higher turbulent velocities, up to $\sim$20\,km\,$\mathrm{s}^{-1}$.

%%%%%%%%%%%%%%%%%%%%%%%%%%%%%%%%%%%%%%%%%%%%%%%%%%%%%%%%%%%%%%%%%%%%%%%%%%%%%%%%%%%%%%%%%%%%%%%%

\section{Limits of the X-ray luminosities}
\label{sec:xrays}

The UV spectra of BSGs reveal the presence of highly ionized ions (e.g., \ion{N}{V}~$\lambda1240$ and \ion{O}{VI}~$\lambda1038$), which are incompatible with their effective temperatures ($\lesssim 30$~kK) and the inferred wind stratification \citep[see, e.g.,][]{Puebla+2016}. This ``superionization'' can only be modelled when including a hot component in the form of an X-ray field into the atmosphere calculations, providing therefore an indirect evidence for the presence of X-rays in the wind of BSGs.

\begin{figure}
  \includegraphics[width=\columnwidth]{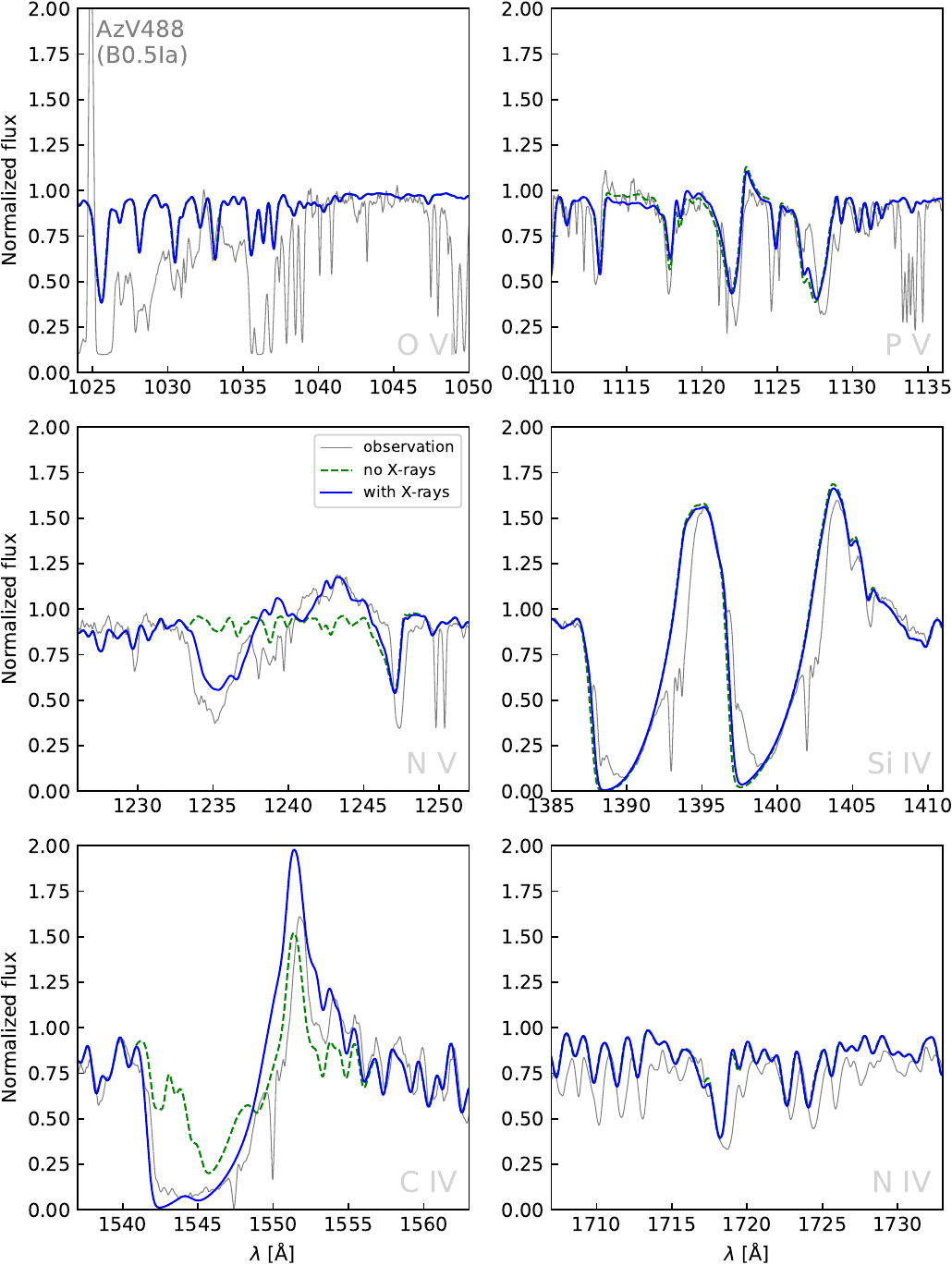}
  \caption{Comparison between models of AzV\,488 (B0.5\,Ia) with and without X-rays. The green dashed curves show the model without X-rays and the green full curves are the model with X-rays included. For this model, the $\log (L_\mathrm{X}/L)$ -7.4 with a $T_\mathrm{X}$ parameter of 1.0~MK. All the other properties are identical between each model.}
  \label{fig:xray-comparison-hot}
\end{figure}

In Fig.~\ref{fig:xray-comparison-hot}, we show a comparison between the models for AzV\,488 (B0.5\,Ia) with and without X-rays where we illustrate their impact on diagnostic wind lines. The biggest impact of the X-rays is seen in \ion{N}{V} (as discussed before) and \ion{C}{IV}. In the case of \ion{C}{IV}, the presence of X-rays is necessary to reach the observed profile saturation -- even though this comes at the cost of an over-prediction of the emission part. The Si lines are barely affected, as well as \ion{N}{IV}~$\lambda1720$, which shows that in this regime these lines are good mass-loss rate and clumping diagnostics. 
Regarding the \ion{O}{VI} and \ion{P}{V} lines, we noticed no significant changes after the inclusion of X-rays as these lines look essentially photospheric.

For Galactic BSGs later than B2, \citet{Bernini-Peron+2023} showed that the inclusion of X-rays is paramount to reproduce most of the superionization ubiquitously observed in the UV of such stars \citep[e.g.,][]{Cassinelli-Olson1979}. They also showed that the X-ray luminosity in general falls slightly below the ``standard relation'' of $\log L_\mathrm{X}/L = -7$ \citep{Sana+2006} observed in many early-B and O supergiants.
Unlike Galactic BSGs, the SMC targets show much less developed UV P Cygni profiles, an expected consequence of their low metallicity. However, one can still notice a clear superionization due to the presence of (i) \ion{C}{IV}~$\lambda$1548-50 and \ion{N}{V}~$\lambda$1238-42 as P Cygni profiles (the latter especially at the earlier type spectra), and reciprocally, (ii) \ion{C}{II}~$\lambda$1335 and \ion{Al}{III}~$\lambda$1855 as a weaker profile.
In comparison to the Galactic targets though, the $L_\mathrm{X}/L$ ratio of the SMC BSGs seems to be similar, suggesting that the X-ray luminosity in these objects could be independent of metallicity.

In \citet{Bernini-Peron+2023}, it was further shown that even the very low $\log (L_\mathrm{X}/L) \sim -12$ with temperatures of $\sim$ 0.1 MK produced successful simultaneous fits of \ion{C}{II} and \ion{C}{IV} lines of cool Galactic BSGs. However, the presence, even though weak, of \ion{N}{V} in the observations, which could not be produced by these models, showed that these values are likely not real. For the SMC stars, on the other hand, this ion is not observed in targets later than B2 ($T_\mathrm{eff} < 19$ kK), which prevents us from discarding those extremely low $L_\mathrm{X}/L$ ratios. Still, our modeling efforts apply the more likely value of $0.5$\,MK for the cooler BSGs in this work, which provides a more realistic upper limit for $L_\mathrm{X}/L$. This is also motivated by the fact that \ion{N}{v} is seen in the denser wind of the only B-hypergiant in our sample (AzV\,78, B1\,Ia+).

\begin{figure}
  \includegraphics[width=\columnwidth]{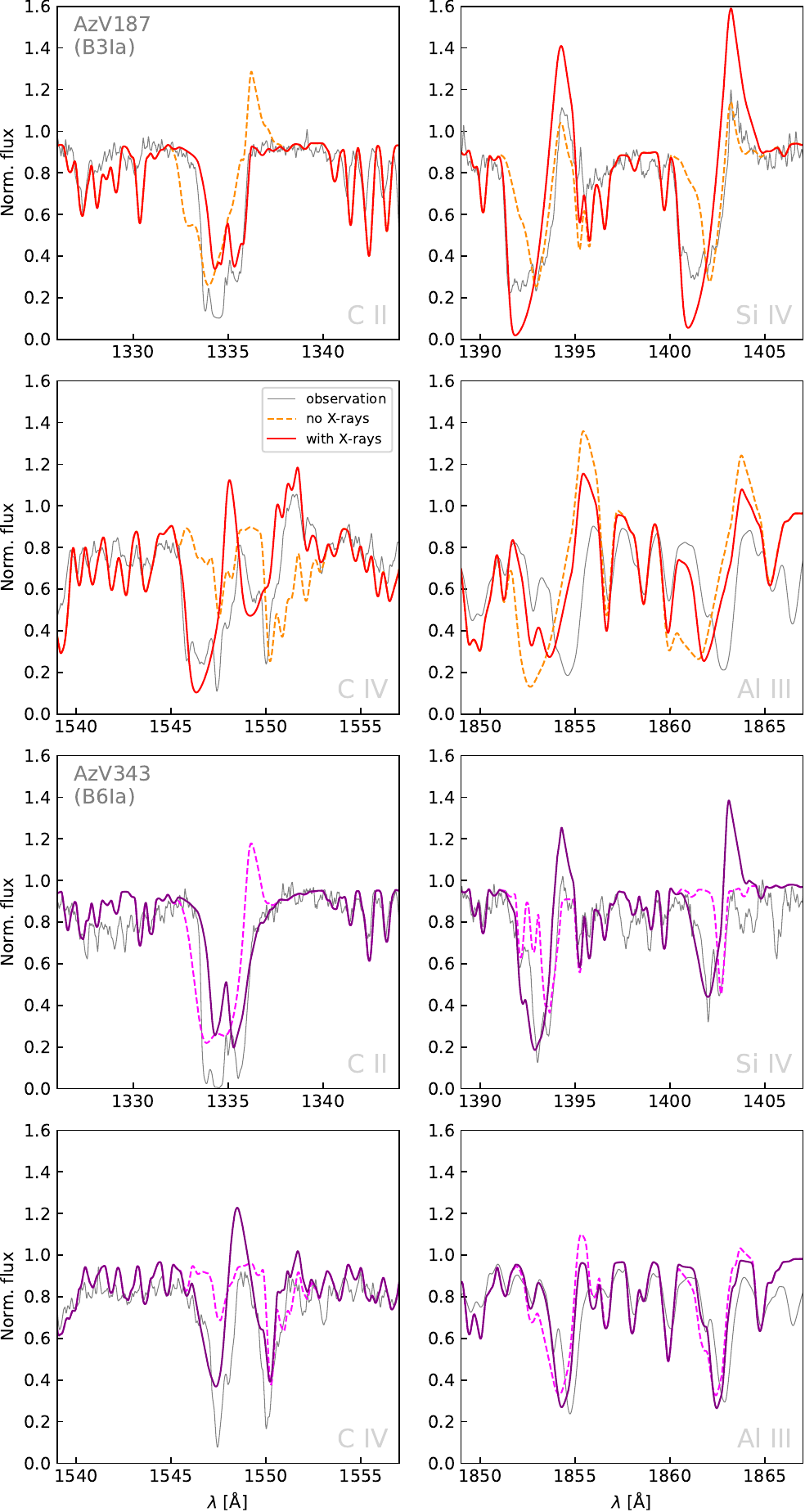}
  \caption{Comparison between models of AzV\,187 (B3\,Ia) and AzV\,343 (B8\,Iab) with and without X-rays. The dashed curves show the model without X-rays and the full curves are the model with X-rays included.}
  \label{fig:xray-comparison}
\end{figure}

In Fig.~\ref{fig:xray-comparison}, we compare models with and without X-rays for AzV\,187 (B3\,Ia), a star with $T_{\mathrm{eff}} = 16.5$ kK. It is evident that the inclusion of X-rays improves the fit of all the lines, with the possible exception of \ion{Si}{IV}~$\lambda1400$. The absence of this extra ionization source would yield a distinct population of \ion{C}{II} in the outer part of the wind, which is not observed in the \ion{C}{II}~$\lambda1335$ line, which appears essentially photospheric. Reciprocally, the models predict a photospheric profile for \ion{C}{IV}~$\lambda1550$, in clear disagreement with the clear P Cygni profile in the observations.

Similar to \ion{C}{II}, \ion{Al}{III} is predicted too strong in the model without X-rays. By including it, we managed to improve the fit, although the wind feature in the doublet is still too strong. An even higher ionization could improve the fit of this doublet, but would not align with the over-prediction by our model of \ion{Si}{IV}~$\lambda1400$, which, despite matching the width of the absorption trough better than without X-rays, predicts too much emission and a saturation.
This problem in simultaneously modeling \ion{Si}{IV} and \ion{Al}{III} is similar to what occurs for the Galactic BSGs in \citet{Bernini-Peron+2023}. As they discuss, optically thick clumps, which have strong evidence to be present in BSGs even at low metallicity \citep[see, e.g.,][]{Prinja-Massa2010, Parsons+2024}, may improve this slightly, but more in-depth studies to understand how different optically thick clumping formulations affect spectral lines in BSGs are necessary.

Even the coolest star in our sample, AzV\,343 (B8\,Iab, one of the latest BSGs in ULLYSES SMC sample), for which we determined an effective temperature of 12.5\,kK, shows spectroscopic features which can only be reproduced when including X-rays. This suggests that even much cooler BSGs with very low terminal velocities harbor somehow an extra source of ionization in their atmospheres. However, explaining these in the paradigm of shock-generated X-ray emission is challenging since winds as slow as $\varv_\infty \lesssim 300\,\mathrm{km}\,\mathrm{s}^{-1}$ should not produce the necessary high-velocity dispersions to generate X-rays. Specifically, from the relation $T_\mathrm{X} \sim (\Delta \varv/300\,\mathrm{km\,s^{-1}})^2 \cdot 10^6\,\mathrm{K},$ presented by \cite{Cohen+2014}, the velocity dispersion of the shocks $\Delta \varv$ would need to be of the same order of $\varv_\infty$ (or larger for some of the later-type BSGs) to be able to produce X-ray temperatures of $T_\mathrm{X} \sim 0.5$~kK.

\begin{figure}
  \includegraphics[width=\columnwidth]{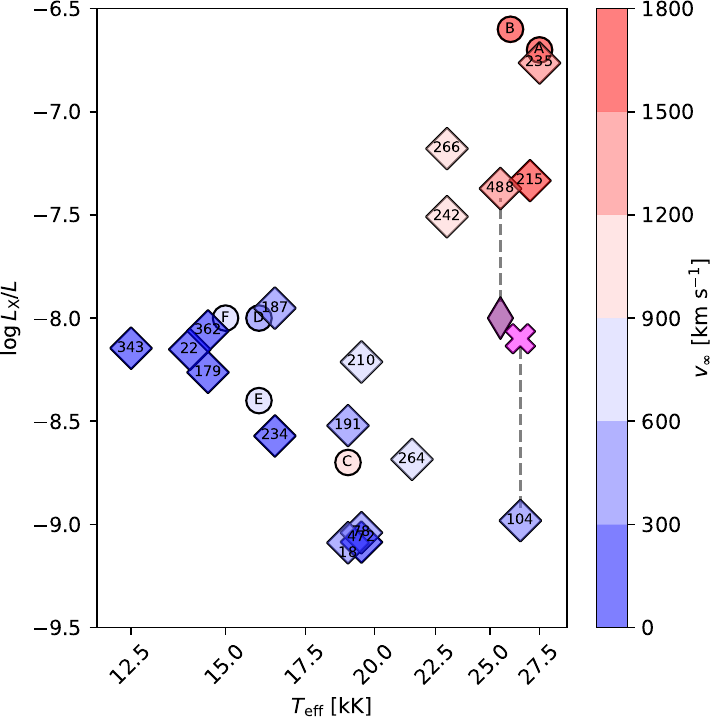}
  \caption{X-ray luminosity compared to effective temperature. The thick diamonds with numbers represent the SMC targets AzV (or Sk for 179 and 191), while the circles with letters, in alphabetical order, represent the Galactic targets $\epsilon$~Ori \citep[B0\,Ia,][]{Puebla+2016}, $\kappa$~Ori \citep[B0.5\,Ia,][]{Huenemoerder+2011,Haucke+2018}, 9~Cep (B2\,Ib), 55~Cyg (B2.5\,Ia), o$^2$~CMa (B3\,Ia), and 64~Oph \citep[B5\,Ib/II,][as for the latter three]{Bernini-Peron+2023}. The colors of the circle and broad diamond symbols represent the wind terminal velocity. The purple thin diamonds represent the alternative values considering a $T_\mathrm{X} = 0.5$~MK for the corresponding early target connected by gray dashed lines. Likewise, the fuchsia crosses represent the alternative value for AzV\,104 if considering a $T_\text{X}=1.0$~MK.}
  \label{fig:LX_Teff}
\end{figure}

Considering the full sample, we plot the relative X-ray luminosity as a function of the stellar temperature in Fig.~\ref{fig:LX_Teff}, color-coding also the terminal velocities. Our results indicate that BSGs with $\varv_\infty < 900 \,\mathrm{km}\,\mathrm{s}^{-1}$ have $\log (L_\mathrm{X}/L) \lesssim -8$. Except for AzV\,104, this regime also coincides with $T_\mathrm{eff} \lesssim 22$ kK. This aligns with the findings reported by \citet{Berghoefer+1997} \citep[but see][]{Naze+2009}, that in spectral types later than B1, BSGs fall in general below the X-ray detectability limit.

\begin{figure}
  \includegraphics[width=\columnwidth]{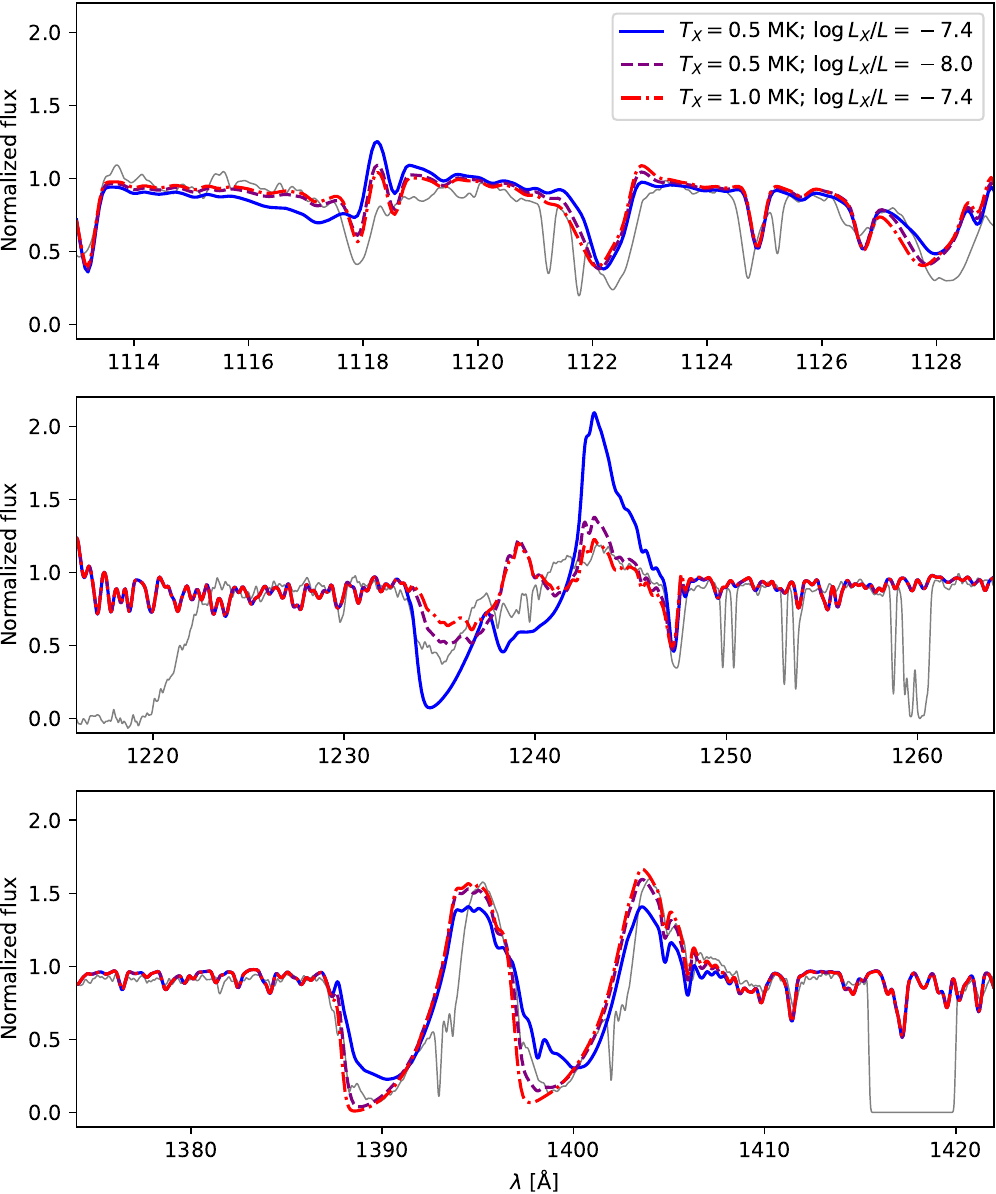}
  \caption{Comparison between models with different relative X-ray luminosity and shock temperature parameters. The thin gray line is the normalized UV spectrum and the colored lines are the CMFGEN models. The blue filled line is the model with $T_\mathrm{X} = 0.5$~MK and $\log (L_\mathrm{X}/L) = -7.4$. The purple dashed line has the same temperature, but with a lower $\log (L_\mathrm{X}/L) = -8.0$. The red dot-dashed line has the same $\log (L_\mathrm{X}/L) = -7.4$, but a higher $T_\mathrm{X} = 1.0$~MK. The different $L_\mathrm{X}$ are obtained by varying the X-ray emission filling factor. All the additional X-ray parameters are kept the same for all the models.}
  \label{fig:XR-TX-comparison-AV488}
\end{figure}

On the other hand, there is also the possibility that our result might be biased by our initial setup where models with $\varv_\infty > 900$ are assigned with $T_\mathrm{X} = 1.0$ MK \citep[after][]{Bernini-Peron+2023}. In Fig~\ref{fig:XR-TX-comparison-AV488}, we show a comparison between models for AzV\,488 with the same X-ray luminosity but employing different $T_\mathrm{X}$. It is noticeable that the model with $T_\mathrm{X} = 0.5$ MK produces a much stronger \ion{N}{V}~$\lambda1240$, the main diagnostic for $L_\mathrm{X}/L$ for these stars. Only when we reduce the X-ray luminosity (dashed line) by a factor of five -- thus becoming similar to what we find for cool BSGs -- we can obtain a compatible profile with the higher $T_\mathrm{X}$. This is also a manifestation of the parameter degeneracy of the current X-ray implementation. 
Under these conditions, our results point to limits of $ -8.0 \gtrsim L_\mathrm{X}/L \gtrsim -7.0$ for early BSGs ($<$ B1.5, not considering AzV\,104 lowest value) and $ -9.0 \gtrsim L_\mathrm{X}/L \gtrsim -7.5$ for later BSGs. This relation however might be rooted more in the wind terminal velocity \citep[which tends to be proportional to $T_\mathrm{eff}$, see][]{Hawcroft+2023,Parsons+2024} than in the temperature itself. Even AzV\,104, which is an early BSG but has a very low $\varv_\infty$, has a low $L_\mathrm{X}/L$ regardless of whether we employ a $T_\mathrm{X} = 0.5$ or 1.0~MK.

The derived X-ray luminosites of all stars in our sample are well below $\log \left(L_\mathrm{X}\,[\mathrm{erg}\,\mathrm{s}^{-1}]\right)< 32.5$  which correspond to X-ray fluxes $< 5\times 10^{-16}$\,erg\,s$^{-1}$\,cm$^{-1}$. This is far below current point source detection limits in the SMC. For instance, ``The X-ray point-source catalogue'' obtained during the \textit{XMM-Newton} survey of the SMC has a flux limit for point sources $10^{-14}$\,erg\,s$^{-1}$\,cm$^{-1}$ (0.2\,--\,4.5\,keV band) \citep{Sturm2013}. Similarly, even deep \textit{Chandra} observations of the SMC \citep{Laycock2010, Oskinova2013} were not sensitive enough to detect BSGs in the SMC. Therefore, it is not surprising that none of our sample stars are detected in X-rays so far. In the region of 30~Dor in the LMC, which is closer to us, X-ray properties of OB stars appear generally consistent with Milky Way counterparts \citep{Crowther+2022}. Therefore, it is possible that the X-ray emission in OB winds does not change significantly with metallicity.

We conclude this section by highlighting that since direct X-ray measurements for SMC BSGs are currently unfeasible, the only way to obtain information on the X-ray content of such stars is via UV spectroscopy. Within the current analysis framework, big systematic studies focused on specific targets \citep[analogous to e.g.,][]{Puebla+2016} could improve the constraints on $L_\text{X}/L$ by investigating the effects of each parameter independently. Additionally, more sophisticated formulations need to be tested \citep[e.g., two-component winds, multicomponent plasmas, different emissivity laws, see][]{Zsargo+2008}. However, as the theoretical understanding of how X-rays are generated in such stars is reaching its limit, detailed simulations of the wind launching in these stars are required to fundamentally constrain any such approximative treatments.

%%%%%%%%%%%%%%%%%%%%%%%%%%%%%%%%%%%%%%%%%%%%%%%%%%%%%%%%%%%%%%%%%%%%%%%%%%%%%%%%%%%%%%%%%%%%%%%%%%%%%%%%%%

\section{Wind properties}
\label{sec:wind-props}

The wind properties of our models are compiled in Table~\ref{tab:wind-prop}.

\begin{table*}
\caption{Wind properties of the sample stars.}
\centering
\begin{tabular}{lc|cccc|cc|ccc}
\hline\hline
Star  & SpType  & $\log \dot{M}$     & $\varv_\infty$    & $\beta$    & $\xi_\mathrm{max}$ & $f_\infty$  & $\varv_\mathrm{cl}$   & $\log (L/L_\mathrm{X})$        & $T_\mathrm{X}$        & $\varv_\mathrm{X}$   \\

 -- & -- & [$M_\odot \, \mathrm{yr}^{-1}$]  & [km\,s$^{-1}$] & -- & [km\,s$^{-1}$] & -- & [km\,s$^{-1}$] & -- & [MK] & [km\,s$^{-1}$]  \\
\hline
AzV\,235 & B0.2\,Ia   &   -6.17    &   1490   &   2.4   &  100  &  0.02    &      5    &     -6.8     &      1.0    &  700  \\
AzV\,215 & BN0\,Ia   &   -6.44    &  1600    & 2.8     & 100   &  0.20    &   30      &    -7.3      &     1.0     &  100    \\
AzV\,488 & B0.5\,Iaw &   -6.07    &   1200   &   1.7   &  100  &   0.20   &    30      &     -7.4     &      1.0    &  700  \\
AzV\,104 & B0.5\,Ia  &   -6.96    &  350    &    0.5  & 300   &    0.60   &    30     &     -8.9     &      0.5   &  400    \\
AzV\,242 & B0.7\,Iaw &    -6.39   &   900   &   1.8   & 200   &   0.20   &      30    &      -7.5    &      1.0     & 500     \\
AzV\,266 & B0.7\,Ia     &  -6.92     &   985   &   1.8   & 147   &    0.10  &      30    &       -7.2   &      1.0    &  500    \\
AzV\,264 & B1\,Ia    &  -6.82     &   820   &   1.8   & 120   &    0.20  &      30    &     -8.7     &       0.5   &  500    \\
AzV\,210 & B1.5\,Ia  &   -6.66    &    810  &   2.5   & 100   &   1.00   &       30   &    -8.2      & 0.5          &  730    \\
Sk\,191 & B1.5\,Ia  &   -6.02    &   490   &   2.8   & 60   &   0.80    &       400   &     -8.5     & 0.5          &  400    \\
AzV\,78  & B1\,Ia+ &    -5.82   &  520    &    3.0  &  90  &  0.75    &         30 &   -8.6       &     0.5     &   590   \\
AzV\,18  & B2\,Ia    &   -6.66    &   350   &    1.5  &   60 &  1.00    &     30     &   -9.1       &    0.5      &  270    \\
AzV\,472 & B2\,Ia    &    -7.52   &   190   &  1.3    &  150  &  0.60    &     30     &   -9.1       &   0.5       &  160    \\
AzV\,187 & B3\,Ia    &    -7.22  &  450    &    2.2  & 60   &   0.30   &       30   &      -8.0   &       0.5    &   380   \\
 AzV\,362 & B3\,Ia    &     -6.92  &  200    &  2.5    & 60   & 0.10     &       30   &      -8.1    &      0.5     &  170    \\
AzV\,22  & B3\,Ia    &    -7.15   &   197   &    3.7  &  25  & 0.10     &        30  &       -8.2   &  0.5         &  170    \\
AzV\,234 & B2.5\,Ib   &     -7.74  &  140    &  2.2    &  60   &  1.00   &       30   &    -8.6      &    0.5      &    120    \\
Sk\,179 & B3II     &   -8.00    &  240    & 1.1     &  60  &  0.35    &         30 &       -8.3   &     0.5      &   200   \\
 AzV\,343 & B8\,Iab   &   -8.00    &  260    &  2.0    &  150   & 1.00     &        30  &     -8.2     &     0.5      &    250  \\
\hline
Errors & -- & 0.40 & 10\% & -- & -- & -- & -- & -- & -- & --  \\
\hline
\end{tabular}
\tablefoot{For AzV\,235 we employed the clumping law introduced by \citet{Najarro+2011}, which uses two additional parameters to account for a decrease in the clumping factor in the outer wind. In our model, the two parameters are $C_3 = 300\,$km\,s$^{-1}$ and $C_4 = 1.0$.}
\label{tab:wind-prop}
\end{table*}

\subsection{Wind terminal and turbulence velocities}

\begin{figure}
  \includegraphics[width=\columnwidth]{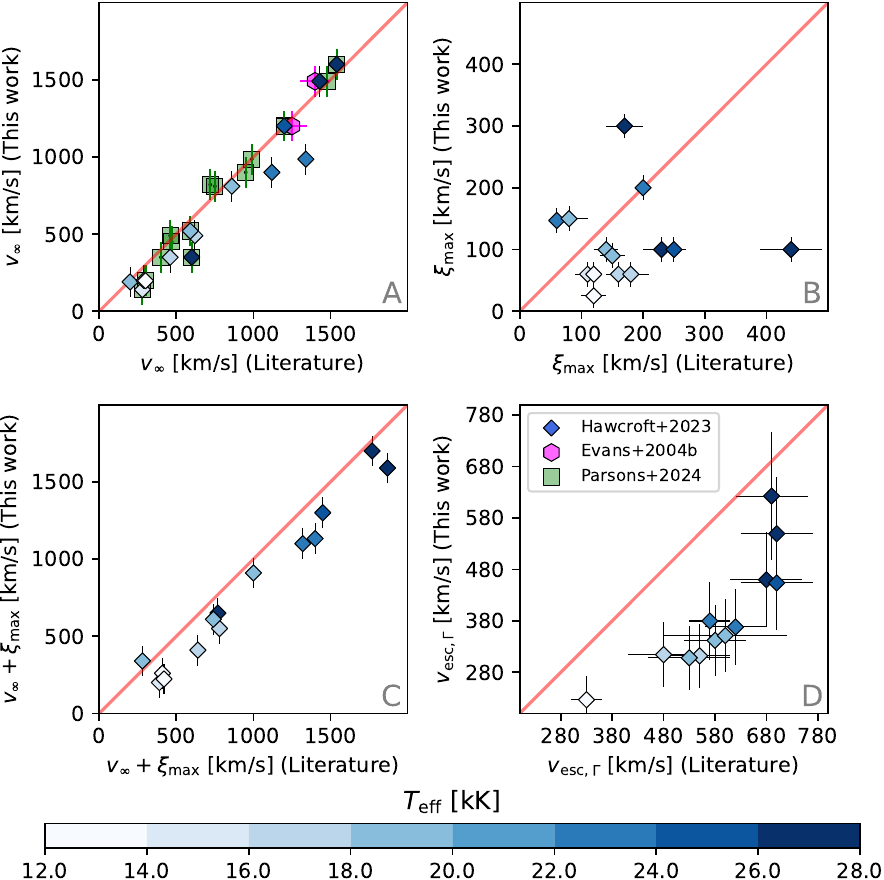}
  \caption{\textit{Panel A}: Comparison between derived terminal velocities ($\varv_\infty$) by this study and those from \citet[small blue diamonds]{Hawcroft+2023} and \citet[green squares]{Parsons+2024}. The fuchsia hexagons are data from \cite{Evans+2004-cmfgen}, which also obtained wind properties via optical and UV analysis with CMFGEN. \textit{Panel B}: Comparison of the wind microturbulent velocities ($\xi_\mathrm{max}$) between this work and \citeauthor{Hawcroft+2023}. \textit{Panel C}: Comparison between the sum of $\varv_\infty$ and $\xi_\mathrm{max}$, which would roughly translate the width of the absorption component of the UV P Cygni profiles. \textit{Panel D}: Comparison of the $\mathrm{\Gamma}$-corrected escape velocities ($\varv_\mathrm{esc,\Gamma}$) between this study and \citeauthor{Hawcroft+2023}
  }
  \label{fig:vinf-comparison-H23}
\end{figure}

In Fig.\,\ref{fig:vinf-comparison-H23}, we compare our inferred terminal velocities with previous literature results. For the small sample overlapping with \citet{Evans+2004-cmfgen}, who also derived their $\varv_\infty$-values by fitting CMFGEN models to the UV spectral lines, we find an excellent agreement.
\citet{Hawcroft+2023} employed the SEI method \citep{Hamann+1981,Lamers+1987} to obtain homogeneous values for $\varv_\infty$ and $\xi_\text{max}$ of most of the ULLYSES OB sample. In general, there is no striking discrepancy between their values and our analyses for the terminal velocities, although a fraction of our values is systematically lower, even when accounting for the error bars. 
We additionally compare our $\varv_\infty$ determinations with those from the recent study by \citet{Parsons+2024}, who analyzed BSG winds using the SEI method to constrain optically thick clumping. We find that except for AzV\,104, all of the $\varv_\infty$ values agree well with \citeauthor{Parsons+2024} within the error margins. Three key differences in their work compared to \cite{Hawcroft+2023} are that (i) \citeauthor{Parsons+2024} use different lines (\ion{Si}{iv} and \ion{Al}{iii} instead of \ion{C}{iv}), (ii) do not enforce $\beta = 1$, and (iii) consider the individual radial velocities of the stars. In general, \citet{Hawcroft+2023} assigned the lowest quality flag to many targets with $\varv_\infty < 1000\,\mathrm{km}\,\mathrm{s}^{-1}$, so that it is not surprising that more detailed follow-up studies that consider lines beyond \ion{C}{iv}\,1550\,\AA\ do find differences in the derived $\varv_\infty$ values. 
The comparison of $\xi_\text{max}$ instead, shows a large scatter between the studies accompanied by a lower trend of our values as well.

When we consider $\varv_\infty + \xi_\mathrm{max}$, which roughly translates to the width of the UV P Cygni absorption troughs, we notice that the comparison trend is tighter, also when looking at the temperature, represented by the color gradient of the scatter points in Fig.\,\ref{fig:vinf-comparison-H23}. However, our values are still systematically lower than their determinations.

\begin{figure}
  \includegraphics[width=\columnwidth]{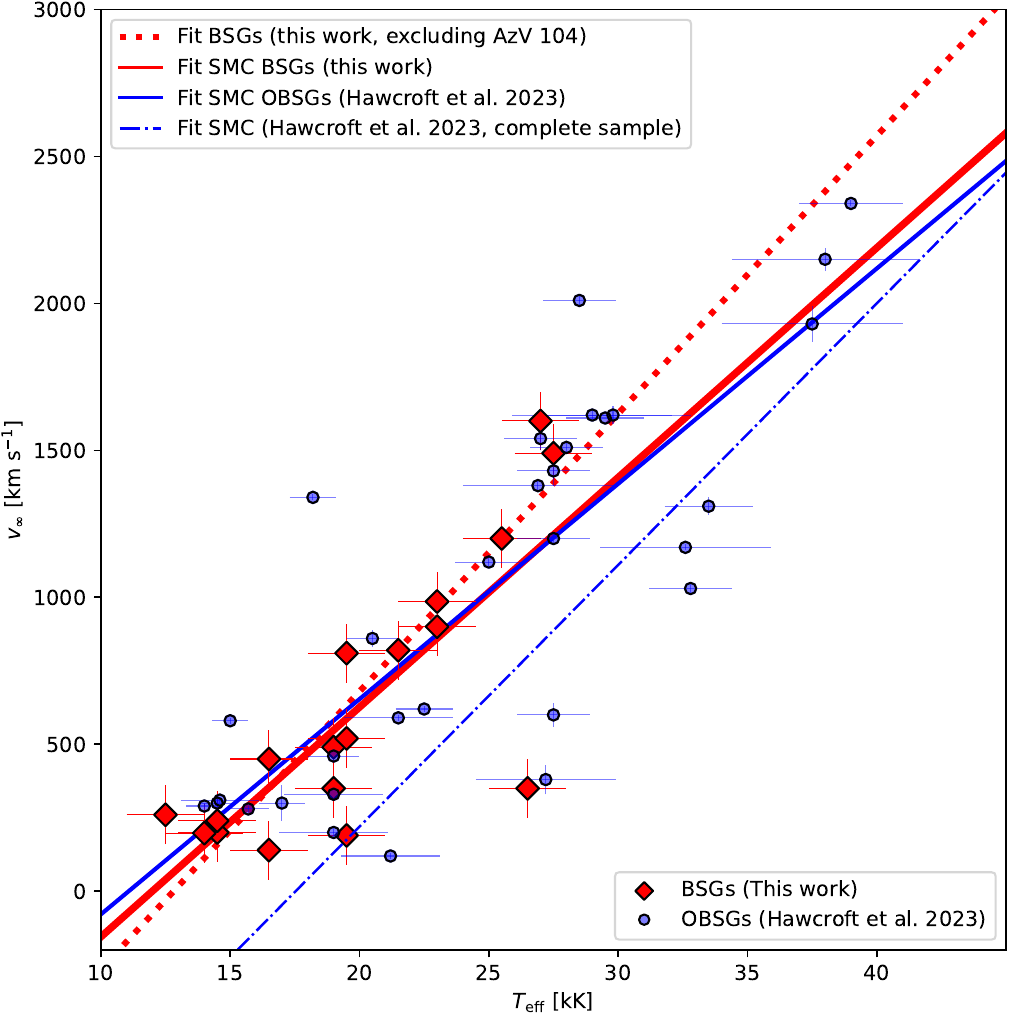}
  \caption{Terminal  velocities as a function of effective temperature: The results from this work (red diamonds) are compared to the values from \citet[blue circles]{Hawcroft+2023}. The thick solid red  and solid blue lines show the resulting relation when considering only supergiants (BSGs and OBSGs). The thick red dotted line shows the relation for only BSGs, but excludes the peculiar AzV\,104. The blue dot-dashed line shows the linear relation derived for all SMC OB stars (of all types and classes) in the \citeauthor{Hawcroft+2023} paper also considering  the  literature results. Conversely, the dark blue dashed line refers to the LMC.
  }
  \label{fig:vinf-Teff-trend}
\end{figure}

In the lower right (D) panel of Fig.~\ref{fig:vinf-comparison-H23}, we check the agreement of the effective escape velocity
\begin{equation}
   \varv_\mathrm{esc,\Gamma} = \sqrt{\frac{2GM}{R} \left(1 - \Gamma_\text{e}\right)} \text{,}
\end{equation}
where $\Gamma_\mathrm{e}$ is the Eddington parameter considering only the electron scattering opacity.
Again, we obtain lower values than those reported by \citet{Hawcroft+2023}. Interestingly, one can see a curved trend, where there is more agreement among the extreme early- and late-BSGs and less in the middle. In their study, $\varv_\mathrm{esc,\Gamma}$ was obtained via Bayesian analysis with B11 evolutionary tracks \citep[BONNSAI,][]{Schneider+2014}, where the masses are the corresponding evolutionary masses. As we consider spectroscopic masses for computing $\varv_\mathrm{esc,\Gamma}$, the systematic differences may be a consequence of the mass discrepancy discussed in Sect.\,\ref{sec:mspec-mevol}.

\paragraph{Temperature and wind terminal velocity relation}
With $T_\text{eff}$ and $\varv_\infty$ obtained from the spectral fits, we compare our results to the $\varv_\infty(T_\text{eff})$-relation derived by \citet{Hawcroft+2023} in Fig.\,\ref{fig:vinf-Teff-trend}. 
Considering our total BSG sample (red diamonds), we plot the derived $T_\text{eff}$ and $\varv_\infty$ values with the corresponding trend being reflected as a thick red line.
Interestingly, the fit hardly changes when using the values for SMC O and B supergiants from \citet[blue points and blue-thick line]{Hawcroft+2023}.
When comparing our BSG and O+BSG SMC relations to the ``total SMC'' relation from \citet{Hawcroft+2023}, which takes into account all ULLYSES SMC OB stars in their sample (thin dot-dashed-blue line), it is evident that the slope of the latter is notably steeper. 
However, for our BSG dataset, the difference in the slope between our BSGs' relation and \citeauthor{Hawcroft+2023} total SMC OB stars almost vanishes when excluding the peculiar slow-wind B0.5 supergiant AzV\,104 (cf.\ appendix~\ref{sec:pecultargs}), as shown by the thick-dotted red line. What remains is a systematic shift toward higher $\varv_\infty$ for the same $T_\text{eff}$. Additionally, the SMC O+BSG sample of \cite{Hawcroft+2023} also contains hot, ``slow-wind'' stars effectively reducing their obtained slope. If these are excluded, the obtained slope of the O+BSG relation (filled-blue line) would be much closer to the slope of their total SMC OB relation (dash-dotted-blue line).

In the regime of cooler BSGs below $\sim$19\,kK, it is hard to identify a clear trend between $\varv_\infty$ and $T_\text{eff}$. Still, it appears that a majority of the stars follow the general slope, but the amount of outliers, in particular below the curve, gets larger.
In principle, the weaker UV profiles make it more difficult to derive accurate values for $\varv_\infty$. Yet, the narrow profiles unequivocally show similar low velocities despite differences in $T_\mathrm{eff}$.

One could imagine that such stars, which have a comparably slow wind for their temperature, might have an increased mass-loss rate, but as we will see in Sect.\,\ref{subsec:mdotrecipies}, this is not the case and the mass-loss rates are usually even lower. Hence, the low terminal velocities of the ``outliers'' in the $\varv_\infty(T_\text{eff})$-diagram cannot be attributed to objects lying closer to the Eddington limit such as hypergiants or LBVs \citep[e.g.,][]{Sander+2014,Sander+2018,Vink2018}, but their winds seem to be suppressed instead for yet unknown reasons.

\begin{figure}
  \includegraphics[width=\columnwidth]{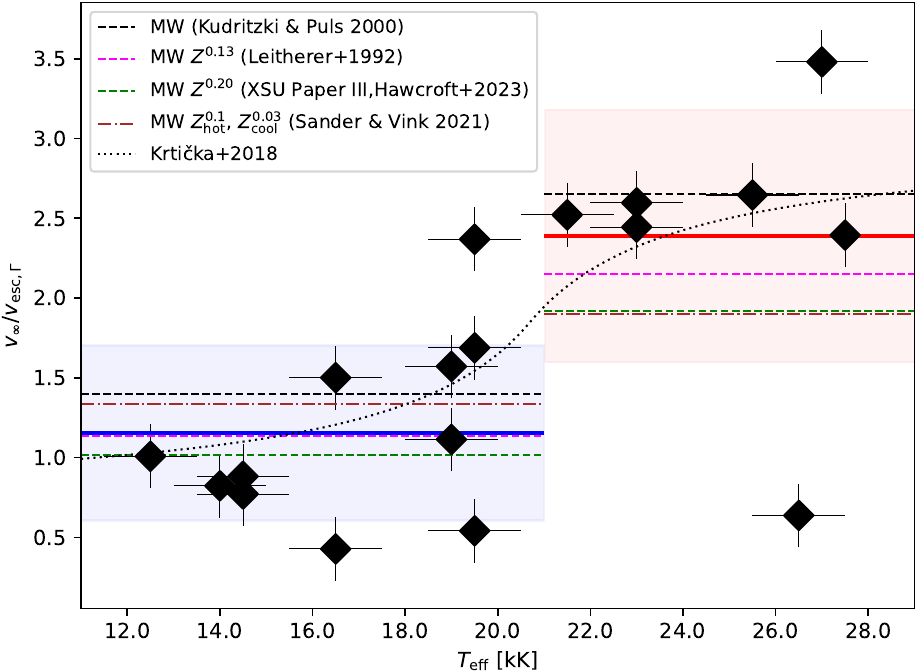}
  \caption{Escape-normalized terminal velocity vs.  effective temperature. The dashed black lines are the relation presented by \citet{Kudritzki-Puls2000} compiling previous data. The dot-dashed line is the relation presented by \citet{Krticka+2021}. The  red and blue thick solid lines are the average value of our determinations and their respective shaded regions are the standard deviation. The thin dashed fuchsia lines are the \citeauthor{Kudritzki-Puls2000} relation scaled by $(Z/\mathrm{Z_\odot})^{0.1}$ metallicity dependence from \citet{Leitherer+1992} and the thin dashed green line below it is the metallicity scaling found by \citet{Hawcroft+2023}. The brown horizontal lines indicate the velocity scaling proposed by \citet{Vink-Sander2021}.}
  \label{fig:BiSJ-velocity}
\end{figure}

\paragraph{Bi-stability jump in velocity}
The bi-stability jump has also been associated with a change in the ratio between the terminal velocity $\varv_\infty$ and the effective escape velocity $\varv_\mathrm{esc,\Gamma}$ along the temperature domain \citep{Lamers+1995}. To check whether our SMC BSG sample shows this, we plot the terminal velocity normalized by the escape velocity ($V^\infty_\mathrm{e} := \varv_\infty/\varv_\mathrm{esc,\Gamma}$) as a function of $T_\text{eff}$ in Fig.\,\ref{fig:BiSJ-velocity}. We find that the SMC BSGs seem to behave very similar to the Galactic BSGs.
The existence of a drop in $V^\infty_\mathrm{e}$ at $T_\mathrm{eff} \lesssim 25$\,kK is a well-established phenomenon, but whether this is a sharp ``jump'' or a more smooth drop is not clear so far \citep[cf.][]{Markova+2008,deBurgos+2023c}.
In Fig.~\ref{fig:BiSJ-velocity}, we show the relations from \citet{Kudritzki-Puls2000} compiling previous literature (black dashed lines) for Galactic stars in comparison to the results inferred from our sample. As evident, the average values (thick red and blue lines) at each side of the jump are very close to the Galactic ones. However, the individual data points suggest that the ``jump'' in $V^\infty_\mathrm{e}$ happens at a slightly lower $T_\text{eff} \sim 19\,$kK compared to $\sim$21\,kK in the Milky Way.

If we scale the Galactic relation from \citet{Kudritzki-Puls2000} for $V^\infty_\mathrm{e}$ by a presumed metallicity-dependence of $\varv_\infty(Z)$, we can test whether the resulting scaled relation align with our dataset for $V^\infty_\mathrm{e}$. Thus, we plot relations scaled by $Z/\mathrm{Z_\odot}^{0.1}$ \citep[][]{Leitherer+1992}, $Z/\mathrm{Z_\odot}^{0.2}$ \citep[][]{Hawcroft+2023}, as well as $Z/\mathrm{Z_\odot}^{0.19}$ (hot side) and $Z/\mathrm{Z_\odot}^{0.003}$ (cool side) from \citet[][]{Vink-Sander2021}. The resulting curves are all in agreement with our data within the 1$\sigma$ (shaded regions). On the hot side, however, the \citeauthor{Hawcroft+2023} and \citeauthor{Vink-Sander2021} scaling already start to deviate from most of the data. Notably, \citet{Hawcroft+2023} discuss that their $V^\infty_\mathrm{e}$ are mostly in agreement with the scaled Galactic values for $T_\mathrm{eff} > 21$ kK, but this is likely due to their larger values for $\varv_\mathrm{esc,\Gamma}$.

Overall, our empirical findings argue for a relatively weak scaling of $\varv_\infty(Z)$, even on the hot side of the bi-stability jump. This is further supported by our findings for $\varv_\infty(T_\text{eff})$ discussed above. Considering our  $\varv_\infty(T_\text{eff})$-relation without the ``outlier'' AV\,104 (cf.\ Fig.\,\ref{fig:vinf-Teff-trend}), we more or less obtain the slope of \citet{Hawcroft+2023}, but shifted upward by $\sim$400\,km\,s$^{-1}$. Notably, this shifted relation essentially aligns with the LMC relation found by \citet{Hawcroft+2023}, implying that there is not really a metallicity difference in the terminal velocities of ``normal'' OB stars. Instead, the lower metallicity environment of the SMC seems to bear more objects that defy the $\varv_\infty(T_\text{eff})$ and $V^\infty_\mathrm{e}(T_\text{eff})$ relations (``outliers'').

\subsection{Mass-loss rates and clumping}
  \label{subsec:mdot}

\begin{figure}
  \includegraphics[width=\columnwidth]{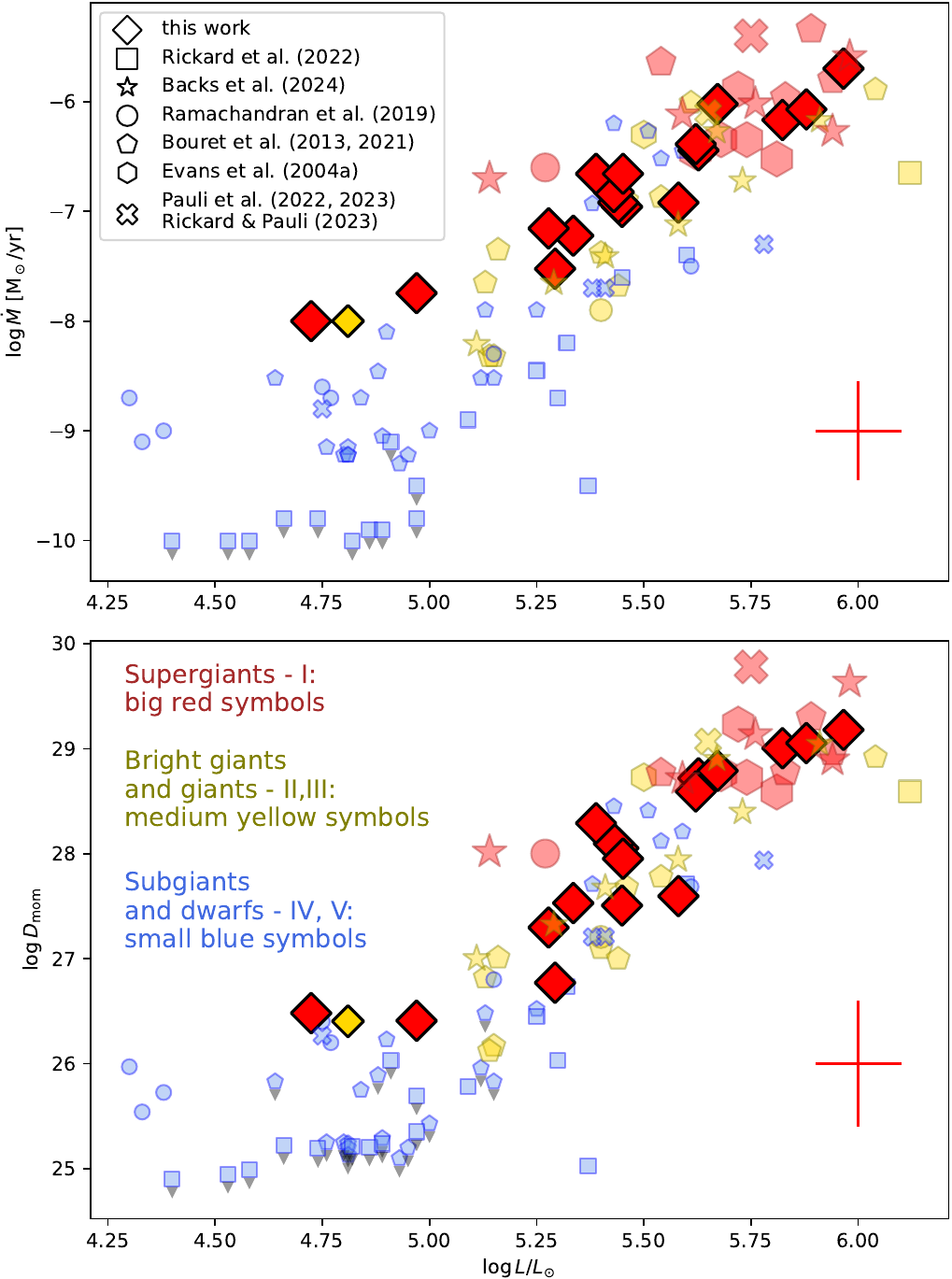}
  \caption{Mass-loss vs. luminosity (upper panel) and modified wind-momentum vs. luminosity (lower panel) diagrams of SMC OB stars of different sources and luminosity classes. The different symbols mark different studies and different colors/sizes mark different luminosity classes. { Error bars are not shown for visual clarity. The literature values are slightly translucent whereas the diamonds represent our sample BSGs. The error bars in the lower right corner indicate the typical associated errors for our sample}. The scatter points with downward black triangles denote upper limits.}
  \label{fig:Mdot-L-trend}
\end{figure}

The derived mass-loss rates of our SMC B supergiant sample are plotted against their luminosities in the upper panel of Fig.\,\ref{fig:Mdot-L-trend}, where we compare them to literature values for other OB stars in the SMC from \citet{Bouret+2013}, \citet{Ramachandran+2019}, \citet{Bouret+2021}, and \citet{Rickard+2022}. We further added the recent SMC binary analysis results by \citet{Pauli+2022,Pauli+2023} and \citet{RickardPauli2023}. It is evident that the SMC is an environment where objects with a wide range of mass-loss rates can be found. The results from \citet{Rickard+2022} of the O star population in NGC\,346 unveiled a sample of non-supergiant O stars with very weak mass-loss rates that fell below previous theoretical expectations, raising the question whether a star actually loses a relevant amount of mass via stellar winds during the main sequence in this low-metallicity environment. However, our BSG results as well as studies of O supergiants and other evolved stars underline that wind mass loss is definitely not negligible after the main sequence.

One has to be careful when assigning spectral types to mass-loss regimes as the implicit assumption about the evolutionary status may not be correct, as for example recently demonstrated by \citet{Pauli+2022} for the ULLYSES target AV\,476. Still, already the color-coding of the different luminosity classes in Fig.\,\ref{fig:Mdot-L-trend} highlights that there is no simple scaling of $\dot{M}$ with the stellar luminosity $L$ for all OB-type stars. Instead our B supergiants as well as the O supergiants analyzed by \citet{Bouret+2013,Bouret+2021} and \citet{Ramachandran+2019} are located above the bulk of the dwarfs in the $\dot{M}$-$L$-plane. Further detailed studies of more objects will have to see if the current overlap will remain or is largely an effect of missing wind information (due to the lack of UV data) or unresolved multiplicity.

In
addition to the direct scaling with the luminosity $L$, it can also be insightful to consider the ``modified wind momentum'' \citep[e.g.,][]{Kudritzki+1999} 
\begin{equation}
   D_\text{mom} = \dot{M} \varv_\infty \sqrt{R/\mathrm{R_\odot}}
,\end{equation}
which is expected to scale with stellar luminosity $L$ for radiation-driven winds following the (modified) CAK theory \citep{Castor+1975,Pauldrach+1986,Kudritzki+1989}. The lower panel of Fig.\,\ref{fig:Mdot-L-trend} showing the corresponding quantities reveals that even within the regime of our analyzed BSGs there is a considerable scatter, despite the overall trend toward larger $D_\text{mom}$ with higher $L$. For the highest luminosities, the scatter seems to get smaller, but there is also a smaller sample and a mixture of different types of objects. Notably, also the recently analyzed binary stars seem to follow the general trends. As the studies include both pre- and post-interaction binaries, this indicates that disentangled binary components do not severely differ in their wind properties from isolated stars with similar parameters.

Considering the clumping parameter $f_\infty$, we do not see a clear trend along the BSG domain (see Table\,\ref{tab:wind-prop}). For the cool Galactic BSGs, \citet{Bernini-Peron+2023} found a tendency toward smoother winds ($f_\infty \rightarrow 1$), in line with recent simulations \citep{Driessen+2019}. We still find some of the cool BSGs in the SMC preferring a smooth-wind solution, but also have objects that require higher amounts of clumping, in particular AzV\,362 and AzV\,22, where $f_\infty = 0.1$ is demanded by their H$\alpha$ profile -- the latter, in line with the significantly high clumping \cite{Parsons+2024} also found. However, \cite{Parsons+2024} also found evidence of clumping in AzV\,234, for which we required a basically smooth wind

The hot BSGs show generally a higher degree of clumping, but again with notable exceptions. For several targets, there is no strong preference for a particular value of $f_\infty$ \citep[see also the discussion in][]{Bernini-Peron+2023} and we took this into account by assigning a larger error bar to the corresponding mass-loss rates.

\begin{figure}
  \includegraphics[width=\columnwidth]{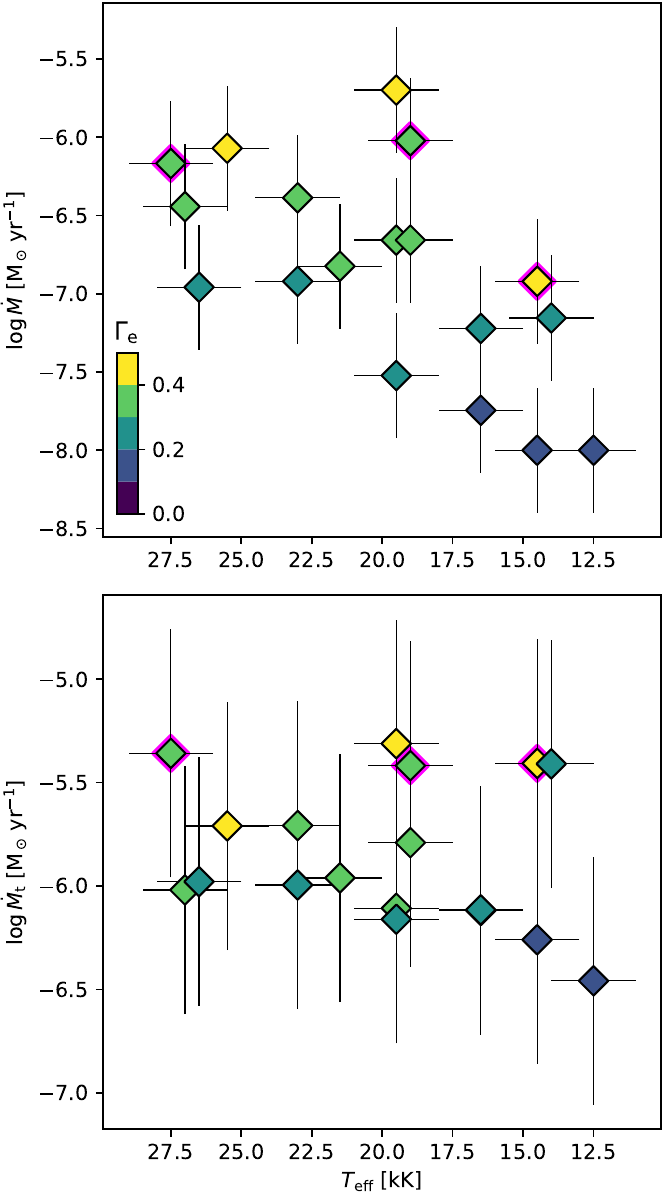}
  \caption{\textit{Upper panel}: Mass-loss rates vs. effective temperature. The scatter points are color-coded by their $\Gamma_\mathrm{e}$ values. The three points outlined in fuchsia in each plot are the BSGs whose Balmer line profiles were peculiar, and thus have less reliable empirical values for the  mass-loss rates.  \textit{Lower panel:} Transformed mass-loss rates \citep{Grafener-Vink2013} vs. effective temperature. The color-coding is the same as in the upper panel.} 
  \label{fig:MdotTrans}
\end{figure}

{
As the BSG regime is a stage which most of massive stars ought to pass -- and some stars may spend a considerable time in (cf.\ Sect.~\ref{sec:evol}) -- , it is important to benchmark theoretical predictions (which are applied to stellar evolution codes) with spectroscopic empirical values. Given that the temperatures of our BSGs span across the bi-stability jump region, we can use their properties to study the behavior of $\dot{M}$ versus $T_\mathrm{eff}$. In the first panel of Fig.~\ref{fig:MdotTrans} we compare both quantities for each star as well as with their respective $\Gamma_\mathrm{e}$. 
}

{
In the figure, we notice an overall downward trend, which is related to the progressively lower luminosities at lower $T_\mathrm{eff}$. There is also a correlation between $\Gamma_\mathrm{e}$ and $\dot{M}$, where stars farther from the Eddington limit are showing lower mass-loss rates, notably with a large scatter at $\sim$19~kK. This temperature coincides with the drop in $V_\mathrm{e}^\infty$ discussed above.

To better address whether or not there are ``jumps'' in $\dot{M}(T_\text{eff})$, it is worth correcting for the different intrinsic stellar luminosities in this comparison. In the lower panel of Fig.~\ref{fig:MdotTrans} we show the transformed mass-loss rates $\dot{M}_\mathrm{t}$, introduced by \citet[][]{Grafener-Vink2013}, where the stars are scaled to the same luminosity and terminal velocity $\varv_\infty$ -- essentially tracing the emission measure of the mass loss. 
Numerically, this scaling of the mass-loss rates is defined as $\dot{M}_\mathrm{t} = \dot{M}\cdot f_\infty^{-1/2}\cdot (10^3\, \mathrm{km\,s^{-1}}/\varv_\infty)\cdot (10^6\,\mathrm{L_\odot}/L)^{3/4}$.

From Fig.\,\ref{fig:MdotTrans}, it is evident that there is only a weak decline of $\dot{M}_\mathrm{t}(T_\text{eff})$. Moreover, the correlation between $\Gamma_\mathrm{e}$ and $\dot{M}$ is much more diluted. This result hints at a more constant behavior of the mass-loss rate with temperature.

}

\subsection{Comparison to mass-loss rate recipes}
  \label{subsec:mdotrecipies}

Our derived mass-loss rates can be compared with different theoretical predictions (``mass-loss recipes'') from the literature that apply to the SMC metallicity, namely: \citet[V2001]{Vink+2001}, \citet[K2024]{Krticka+2024}, and \citet[B2023]{Bjorklund+2023}. There are also recipes by \citet{Bestenlehner+2020} and \cite{GormazMatamala+2023}, but they only apply to the O-star regime. We discuss the three applicable to BSGs in the following paragraphs:

\paragraph{V2001} \citet{Vink+1999,Vink+2000,Vink+2001} derived mass-loss rates based on Monte-Carlo simulations for computing the radiative acceleration, ensuring a global hydrodynamical consistency of the wind (i.e., the consistency between radiative acceleration and mechanical acceleration can be violated locally, but the values integrated over the total wind are equal). This series of studies predicted that, due to the recombination of \ion{Fe}{IV} into \ion{Fe}{III}, the mass-loss rates ought to increase relatively sharply within a short effective temperature range between $27\,\mathrm{kK} \gtrsim T_\mathrm{eff} \gtrsim 20$\,kK. This was associated with the bi-stability jump region and established the paradigm that mass-loss rates are expected to increase when massive stars evolve toward cooler temperatures. Later on, Monte-Carlo models with local hydrodynamical consistency endorsed the presence of the jump \citep{Vink2018, Vink-Sander2021}. Considering metallicity effects, the derived mass-loss rates are expected to scale with $\dot{M} \propto Z^{0.64}$ for B stars and $\dot{M} \propto Z^{0.69}$ for O stars.

\paragraph{K2024} \citet{Krticka+2021,Krticka+2024} derived mass-loss rates using the METUJE stellar atmosphere code \citet{Krticka+2017} to obtain BSG properties. Their models solve the velocity structure consistently with the computed line acceleration, which is performed in the Sobolev approximation, but scaled to account for the difference to a comoving frame calculation. They found that the wind of BSGs on the hot side of the bi-stability jump should be driven mostly by C, Si, and S while on the cool side \ion{Fe}{III} should be dominant. For Galactic stars, they obtained an increase of the mass-loss rate on the cool side of the bi-stability jump. However, their increase is more gradual and less intense than the one in \citet{Vink+2000}, and happens at lower temperatures ($T_\mathrm{eff} \sim 19$ kK), which seems to align better with recently determined values for Galactic BSGs by \citep{Bernini-Peron+2023}. The \citeauthor{Krticka+2024} findings also align with the results by \citet{Petrov+2016}, who found a jump at $\sim$20 kK. When considering different metallicities, \citeauthor{Krticka+2024} find a scaling of $\dot{M} \propto Z^{0.6}$, which is in agreement with the empirical findings from \citet{Marcolino+2022}, who compiled literature CMFGEN results for O- and early-BSGs. Additionally, for lower metallicities, the increase in mass-loss rates is less pronounced, so such an increase manifests more as a mild interruption toward lower temperatures in an overall decreasing $\dot{M}(T_\text{eff})$-trend.

\paragraph{B2023} \citet{Bjorklund+2023} uses the comoving-frame version of FASTWIND to calculate a grid of hydrodynamically consistent models. The resulting mass-loss rates are then fitted assuming a multivariate power law to describe the dependence of $\dot{M}$ on the different stellar parameters. The physical and numerical basics of this method is similar to the approach in METUJE \citep{Krticka-Kubat2010,Krticka+2017} and PoWR \citep{Sander+2017,Sander+2020}, albeit the three codes differ severely in their detailed assumptions and implementations. However, the recipe derived by \citet{Bjorklund+2021,Bjorklund+2023} differs considerably from the other mass-loss descriptions as \citeauthor{Bjorklund+2021} obtain essentially a monotonic decrease in the relation between $T_{\mathrm{eff}}$ and $\dot{M}$ without any bi-stability jump.

\begin{figure}
  \includegraphics[width=\columnwidth]{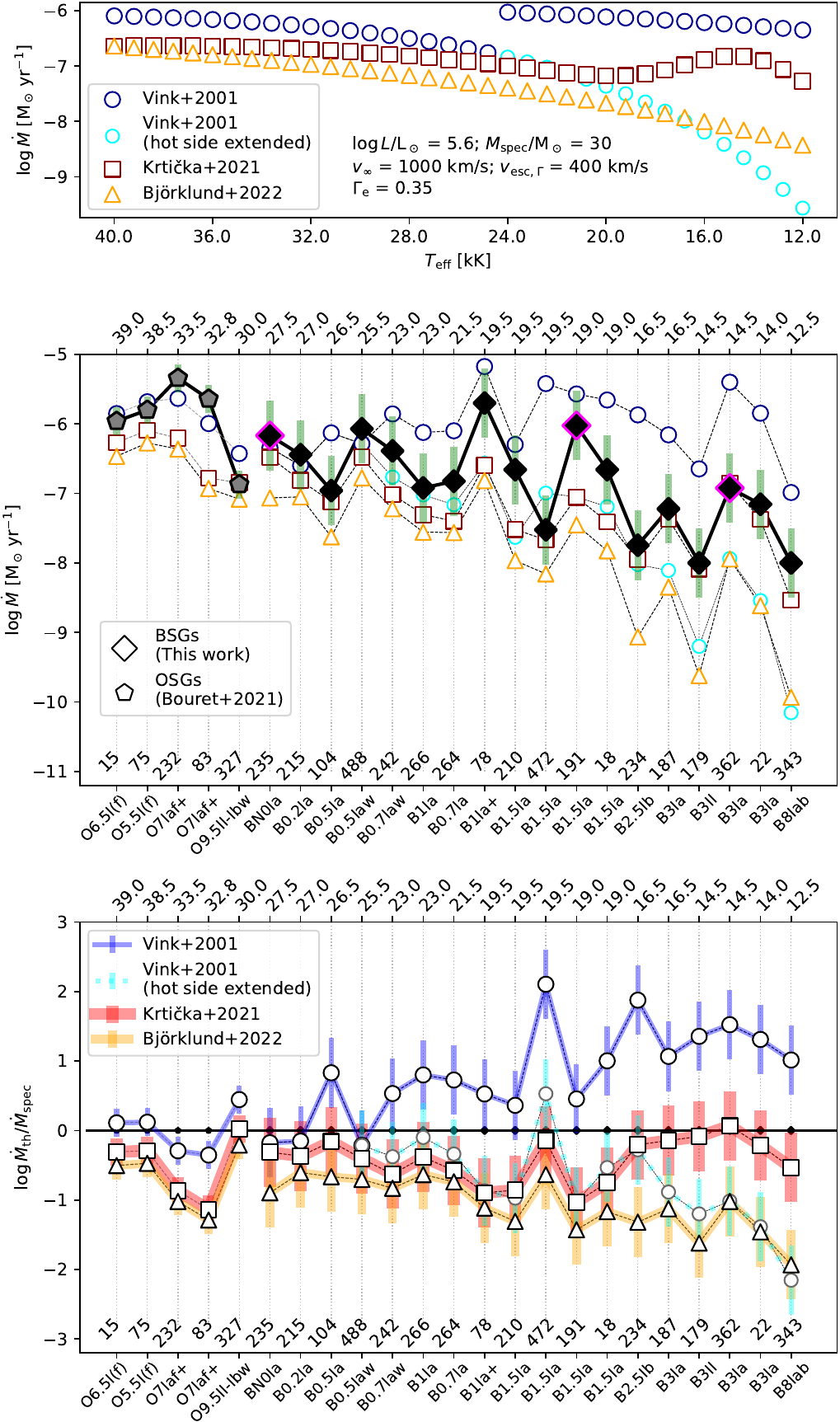}
  \caption{\textit{Upper panel}: Different mass-loss prescriptions applied to a generic star with the same properties but varying $T_\mathrm{eff}$ for illustrative purposes, to facilitate the visualization of the predicted $\dot{M}(T_\mathrm{eff})$ behaviors. \citet{Vink+2001} rates are represented by white circles, \citet{Krticka+2024} by white squares, and \citet{Bjorklund+2023} by white triangles. \textit{Middle panel}: Comparison between different mass-loss rate prescriptions and inferred values for the sample stars (black diamonds) and OSGs from \citet[][gray pentagons]{Bouret+2021}. The fuchsia-outlined diamonds are the targets where we did not find a satisfactory fitting of Balmer lines. The applied theoretical values have the same symbols as the upper panel. The stars are ordered according to their $T_\mathrm{eff}$ in the upper axis. The lower axes show the spectral types of each star. \textit{Lower panel:} Similar to the previous panel, but showing the ratio of the predicted according to each recipe ($\dot{M}_\mathrm{th}$) to our empirical determinations. The symbols follow the same encoding from the previous panels.}
  \label{fig:mdot-comp}
\end{figure}

\paragraph{Discussion} 
A comparison between our derived mass-loss rates and the predictions from the aforementioned recipes is shown in Fig~\ref{fig:mdot-comp}, assuming 0.2~$Z_\odot$ for the metallicity of the SMC. Previous studies of OB dwarfs in the SMC \citep[e.g.,][]{Ramachandran+2019,Rickard+2022} find mass-loss rates far below V2001. Yet, the picture is more complex for supergiants. Among the earliest targets, mostly with spectral types of B0 and B0.5, we find mass-loss rates more compatible with V2001 and K2024 predictions, which in general predict similar $\dot{M}$. 

These results align well with those from \citet{Evans+2004-cmfgen} and \citet{Trundle+2004}, who obtained $\dot{M}$ values slightly higher than \citeauthor{Vink+2001} predictions for 0.2~$Z_\odot$. Using the data from \citet{Bouret+2021}, whose analysis method is similar to ours, we can extend our discussion to also include the regime of O supergiants (OSGs). The trend of the early-type supergiants aligning with the V2001 predictions remains.

{ Other studies using different codes and methodologies also find a similar behavior for OSGs. In particular, \cite{Ramachandran+2019} found that their only OSG, unlike their O giants and dwarfs, agrees with V2001. 
However, only two of them should be considered seriously as they attribute a low confidence to the other three objects as they are suspected or known binaries. The early O3-supergiant component of SSN\,7 in NGC\,346 studied by \citet{RickardPauli2023} even has a mass-loss rate above V2001.
Unless all these early supergiants are subject to further multiplicity, we suggest there is a dichotomy in the mass loss between dwarfs and supergiants as partially indicated by Fig.\,\ref{fig:Mdot-L-trend}.
Interestingly, the agreements with V2001 for the SMC are contrary to the findings for Galactic OB stars (of all types/classes), where \citet{Vink+2000} overestimates $\dot{M}$ by a factor of $\sim$3 when considering the effects of clumping \citep[e.g.,][]{Najarro+2011,Krticka+2017,Hawcroft+2021}. 
This could imply that OSGs and early-BSGs might have a different $\dot{M}(Z)$-scaling compared to early OB-giants and -dwarfs.}

While the root of this remains so far unclear, recent theoretical studies of the winds of WR stars \citep{Sander-Vink2020,Sander+2023} revealed a more shallow metallicity scaling in the optically thick wind regime and a transition toward a steeper dependence when the winds get optically thin. Possibly, there could also be a different metallicity scaling between OB dwarfs, giants, and supergiants, assuming this coincides with different wind density regimes. Then, the scaling of the supergiant regime would likely be flatter than the scaling for the dwarf regime -- where we might see the ``weak wind phenomenon'' \citep{Martins+2005,Marcolino+2009} already at much higher luminosities in the SMC as recently discussed in \citet{Ramachandran+2019}. Remarkably, the proximity between empirical mass-loss determinations and the V2001 recipe for early BSGs appears to happen for LMC stars as well \citep{Verhamme+subm}, which strengthens this scenario. Yet, it is worth considering the potential origin of the supergiants in the SMC. In the lower metallicity of the SMC, evolution models such as \citet{Klencki+2022} predict a considerable number of (partially) stripped stars appearing as hot supergiants. Stripped stars have a higher $L/M$-ratio and thus show a stronger wind, which would naturally explain why the supergiants are not in line with the trend from dwarfs and giants. However, we do not find clear evidence that our sample BSGs are necessarily stripped.

Moving to cooler temperatures, most of the B1, B1.5, and B2 targets yield values between the predictions by V2001 (cool side) and K2024, with a slight preference for the former. Additionally, in this temperature regime, K2024 is very similar to the $\dot{M}$ values of V2001 (hot side) extended to the cool side (i.e., if we ignore the jump). 
For stars later than B2, the inferred mass-loss rates agree very well with the \citeauthor{Krticka+2024} SMC predictions, thus behaving very similarly to the cool BSGs in the Milky Way \citep{Bernini-Peron+2023}.

Considering the overall trend, we notice a general agreement between our findings and the theoretical $\dot{M}$ predictions by K2024 and V2001 for OSGs and early-BSGs. The predictions of K2024 further agree with our derived values for most early and later BSGs. Interestingly, K2024 seems to underpredict $\dot{M}$ in the temperature interval between $\sim$23 and $\sim$19~kK (or between $\sim$B1 and $\sim$B2). Conversely, the predicted increase in $\dot{M}$ by V2001 yields values higher than our empirical data, albeit within the considered error margin for some targets. For the vast majority of stars, B2023 underestimates the empirical $\dot{M}$.

From these results, we do not find clear evidence for a sharp increase in the mass-loss rate when passing the bi-stability jump region predicted by V2001 toward cooler temperatures. This is in line with a recent larger-sample optical analysis of BSGs in the Milky Way by \citet{deBurgos+2023c}, who also could not find an increase of the mass-loss rate.
For the SMC BSGs, we find the mass-loss rates on the cool side of the jump to be lower by more than one order of magnitude compared to V2001, despite the good agreement on the hot side. However, we notice that recipes ignoring the presence of a jump (B2023 and the extrapolated hot-side V2001) underpredict the amount of mass loss by about an order of magnitude. Hence, our results indicate that there is a change in the $\dot{M}(T_\mathrm{eff})$-behavior in the bi-stability region and not just a smooth downward trend.

%%%%%%%%%%%%%%%%%%%%%%%%%%%%%%%%%%%%%%%%%%%%%%%%%%%%%%%%%%%%%%%%%%%%%%%%%%%%

\section{Conclusions}
\label{sec:conclu}

In this work we presented a thorough analysis of a representative sample of BSGs in the SMC. Using CMFGEN model atmospheres as well as the ULLYSES/XShootU large and homogeneous dataset plus additional archival data, we could analyze the UV and optical regime of 18 BSGs with unprecedented quality, spanning a temperature range between 12.5 and 27.5\,kK and spectral types from B0 to B8. This comprises the widest multiwavelength study on B supergiants in the SMC where photospheric and wind properties were obtained. From the derived photospheric properties we could constrain their evolutionary context, from which we draw the following conclusions:

\begin{itemize}
    \item The  BSGs in general are evolved objects. From comparison with two evolutionary sets of single-star evolution models (G13 and B11), they are either main-sequence objects close to the  TAMS (for the early-type stars) or already in the H-shell burning stage for the late-type objects.
    \item While the G13 tracks reproduce the properties of later-type BSGs considering physical properties and appropriate timescales, they cannot explain the hotter BSGs. On the other hand, the B11 tracks, which have higher overshooting, can reproduce the early BSGs in the HRD, although not the observed  rotation and chemical abundances. The post-main-sequence evolution happens quickly in the B11 models, so they cannot explain the existence of late-type BSGs.
    \item The G13 models also yield a better overall agreement between the spectroscopic and the evolutionary masses compared to B11 and previous $M_\mathrm{spec}$-$M_\mathrm{evol}$ comparisons \citep{Trundle+2004, Trundle+2005}. Additionally, we confirm the previously reported trend of higher evolutionary masses for less massive and less luminous BSGs.
    \item While none of our objects have clear spectral indication for multiplicity, we cannot rule out multiplicity-based origins for our BSG sample stars. The obtained CNO abundance ratios are roughly in line with existing scenarios. However, the He enrichment is lower than currently predicted in merger scenarios, albeit with specific predictions for merger scenarios in the SMC still missing.
    \item When comparing the BSG sample with the abundances of (partially) stripped stars, we find partial agreement. However, several stripped stars show stronger enrichment in nitrogen than our sample stars.
\end{itemize}

Our study marks the first work analyzing  a big sample of UV spectra for BSGs in the SMC with a stellar atmosphere code. Moreover, this is the first time both clumping and X-rays have been included when studying UV wind diagnostic lines in this regime. Similarly to the Galactic study by \cite{Bernini-Peron+2023}, we can  put the following constraints on the X-ray emission of BSGs within the SMC:

\begin{itemize}
    \item We obtained a wide range of $L_\mathrm{X}/L$-ratios for the different stars in our sample. In general, the early-type BSGs have $-8.0 \lesssim \log (L_\mathrm{X}/L) \lesssim -7.0$, while late-type BSGs have $-9.0 \lesssim \log (L_\mathrm{X}/L) \lesssim -7.5$. 
    \item Our derived values are based on the parameterized {ad hoc} implementation within the wind-shock paradigm. We also find that there are degeneracies in the X-ray parameters with different values leading to a similar effect on the wind lines. However, some degeneracies can be lifted by making reasonable physical assumptions. For instance, it is not expected that cool BSGs, which in many cases have winds slower than 300~km~s$^{-1}$, can produce shocks that are energetic enough to generate $T_\mathrm{X} \sim 1.0$ MK. This in particular sets constraints for the later-type BSGs.
\end{itemize}

With the wind diagnostics given from the UV spectra and the sample covering a wide temperate range across the presumed bi-stability jump region, we gain the following insights into the wind velocities in this regime:

\begin{itemize}
    \item We find   good agreement between our obtained terminal velocities and those fro the  previous literature. However, we obtain systematically lower values relative to \citet[XShootU Paper III]{Hawcroft+2023} while the comparison with \cite{Parsons+2024},  both employing the SEI method,  yields a good agreement.
    \item We find a clear jump in the escape-normalized terminal velocity at $\sim$$19\,$kK, very similar to what was found for Galactic BSGs at about $21\,$kK \citep{Lamers+1995}.   
    \item For cooler BSGs, we obtained a large spread in the escape-normalized terminal velocities, meaning that none of the proposed $Z/$Z$_\odot$-scalings is clearly preferred. For early-type BSGs, a small, possibly even vanishing scaling term seems to reproduce the empirical findings better, while the scaling from \cite{Hawcroft+2023} and \cite{Vink-Sander2021} predict  velocities that are too low, albeit still within the 1$\sigma$ error.
\end{itemize}

Lastly, we can draw the following conclusions on the wind mass loss of BSGs by comparing our derived properties with literature results and theoretical predictions:

\begin{itemize}
    \item For the early-type BSGs, we find a good overall agreement between the different $\dot{M}$ recipes. However, we tend to find the best overlap with the predictions from \cite{Vink+2001},  similar to what \cite{Evans+2004-cmfgen} obtained,  with some objects even exceeding the predicted values. This result aligns very well with the literature studies for SMC OSGs using different methodologies.
    \item Given that literature studies for SMC OB giants and dwarfs yield mass-loss rates clearly below the predictions from \citet{Vink+2001}, we suggest that there may be a dichotomy in the scaling of mass loss with metallicity between supergiant and non-supergiant OB stars.
    \item In the bi-stability jump region (i.e., between $\sim$23 and $\sim$19~kK), the derived $\dot{M}$ is found to be between \cite{Vink+2001} and the other recipes that do not predict a jump. For stars cooler than this $T_\mathrm{eff}$ range, the \citet{Krticka+2024} predictions that include a small increase align best with the empirical values,  similarly to what was recently found for Galactic BSGs \citep{Bernini-Peron+2023,deBurgos+2023c}.
    \item Our findings do not support the smooth, monotonic decrease in $\dot{M}(T_\mathrm{eff})$ suggested by \cite{Bjorklund+2023}. Instead, our results outline a scenario where $\dot{M}(T_\mathrm{eff})$ does not have a clear bump or jump, but rather stays constant or increases very mildly between $\sim$25 and $\sim$16~kK.
    \item The escape-normalized terminal wind velocities for the SMC BSGs show a rather sharp drop at around $\sim$$19\,$kK. Notably, the temperature of this drop does not directly align with any change in the mass-loss rate behavior.
\end{itemize}

%%%%%%%%%%%%%%%%%%%%%%%%%%%%%%%%%%%%%%%%%%%%%%%%%%%%%%%%%%%%%%%%%%%%%%%%%%%%%%%%%%%%%%%%%%%%%%%%%%%%%%%%%%%%%%%%%%%%%
%%%%%%%%%%%%%%%%%%%%%%%%%%%%%%%%%%%%%%%%%%%%%%%%%%%%%%%%%%%%%%%%%%%%%%%%%%%%%%%%%%%%%%%%%%%%%%%%%%%%%%%%%%%%%%%%%%%%%

\begin{acknowledgements}
       MBP, AACS, and VR are supported by the German
      \emph{Deut\-sche For\-schungs\-ge\-mein\-schaft, DFG\/} in the form of an Emmy Noether Research Group -- Project-ID 445674056 (SA4064/1-1, PI Sander).  
       MBP, AACS, and VR are further supported by funding from the Federal Ministry of Education and Research (BMBF) and the Baden-W{\"u}rttemberg Ministry of Science as part of the Excellence Strategy of the German Federal and State Governments. JJ acknowledges funding from the Deutsche Forschungsgemeinschaft (DFG, German Research Foundation) Project-ID 496854903 (SA4064/2-1, PI Sander) and is a member of the International Max Planck Research School for Astronomy and Cosmic Physics at the University of Heidelberg (IMPRS-HD).
       DP acknowledges financial support by the Deutsches Zentrum f{\"u}r Luft und Raumfahrt (DLR) grant FKZ 50 OR 2005.
       JSV gratefully acknowledges support from STFC via grant ST/V000233/1. RK acknowledges financial support via the Heisenberg Research Grant funded by the Deutsche Forschungsgemeinschaft (DFG, German Research Foundation) under grant no.~KU 2849/9, project no.~445783058. We thank Danny Lennon and Philip Dufton for providing additional spectra.
\end{acknowledgements}

\bibliographystyle{aa}
\bibliography{biblio}

%%%%%%%%%%%%%%%%%%%%%%%%%%%%%%%%%%%%%%%%%%%%%%%%%%%%%%%%%%%%%%%%%%%%%%%%%%%%%%%%%%%%%%%%%%%%%%%%%%%%%%%%%%%%%%%%%%%%%%%
%%%%%%%%%%%%%%%%%%%%%%%%%%%%%%%%%%%%%%%%%%%%%%%%%%%%%%%%%%%%%%%%%%%%%%%%%%%%%%%%%%%%%%%%%%%%%%%%%%%%%%%%%%%%%%%%%%%%%%%

\begin{appendix}

\section{Additional observational data}

In Table~\ref{tab:data-source_1}, we list information on the observed spectra. We provide the observation dates for each BSG spectrum. For the UV, this information is retrieved via the ULLYSES search form and for the optical data this information is listed as output from search on ESO-archive portal. For the comments on individual stars provided in appendix \ref{sec:pecultargs}, we further used additional spectra for which analogous information is given in Table~\ref{tab:app_spec_tabs}.

\begin{table}[]
\caption{Additional spectra used in this study. The PI abbreviations read: \textit{Vi} for J. Vink, \textit{Ev} for C. Evans, \textit{Eh} for P. Ehrenfreund, \textit{Ro} for W.R.J. Rolleston, \textit{Co} for R. Cooke, and \textit{Ol} for C. Oliveira}
\centering
\label{tab:app_spec_tabs}
\begin{tabular}{lrrr}
\hline\hline
Instrument                                     & Obs. date                         & Resolution/Grating   & PI             \\
\hline
\multicolumn{3}{c}{AzV\,235}                                                                                            \\
\hline
X-Shooter (IR)                                  & 2020-10-28                        & 8000                      & \textit{Vi}      \\
UVES (blue)                                    & 2001-09-28                        & 42000                    & \textit{Ev}       \\
UVES (blue)                                    & 2001-09-28                        & 40000                    & \textit{Ev}       \\
\hline
\multicolumn{3}{c}{AzV\,362}                                                                                     \\
\hline
X-Shooter (IR)                                 & 2020-11-10                        & 8000                   &  \textit{Vi}        \\
UVES (blue)                                    & 2001-11-02                        & 30000                 &  \textit{Ev}          \\
UVES (red)                                     & 2001-11-02                        & 31000                 &  \textit{Ev}          \\
\hline
\multicolumn{3}{c}{Sk\,191}                                                                                            \\
\hline
UVES (blue)                                    & 2001-09-27                        & 53000                  & \textit{Eh}         \\
UVES (red)                                     & 2001-09-27                        & 56000                   & \textit{Eh}        \\
UVES (blue)                                    & 2001-11-01                        & 30000                    & \textit{Ro}       \\
UVES (red)                                     & 2001-11-01                        & 31000                     & \textit{Ro}      \\
UVES (blue)                                    & 2016-10-29                        & 75000                  & \textit{Co}         \\
UVES (red)                                     & 2016-10-29                        & 65000                   & \textit{Co}        \\
UVES (blue)                                    & 2016-11-01                        & 75000                   & \textit{Co}        \\
UVES (red)                                     & 2016-11-01                        & 65000                   & \textit{Co}        \\
UVES (blue)                                    & 2016-11-02                        & 75000                   & \textit{Co}        \\
UVES (red)                                     & 2016-11-02                        & 65000                   & \textit{Co}        \\
\hline
                                               & 2012-02-04                        & {G160M+G130M}  & \textit{Ol} \\
                                               & 2011-08-18                        & {G160M+G130M}  & \textit{Ol} \\
                                               & 2011-05-01                        & {G160M+G130M} & \textit{Ol} \\
                                               & 2011-01-07                        & {G160M+G130M} & \textit{Ol} \\
                                               & 2010-09-27                        & {G160M+G130M} & \textit{Ol} \\
HST-COS                                        & 2010-06-11                        & {G160M+G130M} &  \textit{Ol} \\
                                               & 2010-04-23                        & {G160M+G130M} & \textit{Ol} \\
                                               & 2010-02-11                        & {G160M+G130M} & \textit{Ol} \\
                                               & 2009-10-27                        & {G160M+G130M} & \textit{Ol} \\
                                               & 2009-09-16                        & {G160M+G130M} & \textit{Ol} \\
\hline
\end{tabular}
\end{table}
\FloatBarrier

\begin{sidewaystable}
    
\caption{Sources flagged with (*) indicate XShooter spectra obtained by the XShootU collaboration that we normalized. Targets flagged with (**) indicate archival X-Shooter spectra (whose PI was N. Przybilla) obtained via ESO-Archive. The acquisition date follows the \textit{yy-mm-dd} format. All FUSE observations were performed using the LWRS grating.}
\label{tab:data-source_1}
\centering
\begin{tabular}{lc|lll|rrr}
\hline\hline
Star  & SpType  & \multicolumn{3}{c|}{Ultraviolet (FUSE / HST)}                                                                     & \multicolumn{1}{l}{Optical (VLT)}             \\
\hline
      &         & Instrument & Grating               & Obs. date                                                  & \multicolumn{1}{l}{Obs. date}  & Instrument & Arms   \\
\hline
AV235 & B0.2\,Ia   & F/Hs       & LWRS / E140M, E230M               & {\tiny00-07-02 {\textbf/ }20-07-03, 20-07-03                        } & {\tiny 20-10-28 }                      & X-Shooter &  UVB/VIS                                     \\
AV215 & BN0\,Ia   & F/Hs       & LWRS / E140M, E230M               & {\tiny01-10-21 {\textbf/ }01-10-21, 03-07-11                        } & {\tiny 20-11-05 }                      & X-Shooter &  UVB/VIS                                      \\
AV488 & B0.5\,Iaw & F/Hs       & LWRS / E140M, E230M               & {\tiny00-10-11 {\textbf/ }20-07-17, 20-07-17                        } & {\tiny 20-10-29 }                      & X-Shooter &  UVB/VIS                                      \\
AV104 & B0.5\,Ia  & F/Hs       & LWRS / E140M, E230M               & {\tiny02-02-03  {\textbf/ }02-02-03, 03-07-11                       } & {\tiny 20-11-04 }                      & X-Shooter &  UVB/VIS                                      \\
AV242 & B0.7\,Iaw & F/Hs       & LWRS / E140M, E230M               & {\tiny00-10-04 {\textbf/ }15-12-21, 15-12                           } & {\tiny 20-12-26 }                      & X-Shooter &  UVB/VIS                                      \\
AV266 & B0.7\,Ia     & F/Hs       & LWRS / E140M, E230M               & {\tiny02-08-06 {\textbf/ }20-07-08, 20-07-08                        } & {\tiny 20-10-25 }                      & X-Shooter &  UVB/VIS                                      \\
AV264 & B1\,Ia    & F/Hs       & LWRS / E140M, E230M               & {\tiny00-10-06 {\textbf/ }22-03-15, 22-03-15                        } & {\tiny 20-11-04 }                      & X-Shooter &  UVB/VIS                                      \\
AV210 & B1.5\,Ia  & F/Hs       & LWRS / E140M, E230M               & {\tiny01-10-22 {\textbf/ }01-10-22, 03-07-18                        } & {\tiny 20-11-04 }                      & X-Shooter &  UVB/VIS                                      \\
SK191 & B1.5\,Ia  & F/Hs/Hc    & LWRS / E140M, E230M {\textbf/ }G130M, G160M& {\tiny01-10-20 {\textbf/ }01-10-21, 03-07-12 {\textbf/ }09-09-16, 09-09-16  } & {\tiny 20-11-06 }                      & X-Shooter &  UVB/VIS                                      \\
AV78  & B1\,Ia+ & Hs         & E140M, E230M               & {\tiny15-12-28                                           } & {\tiny *23-12-01 }                      & X-Shooter* &  UVB/VIS                                       \\
AV18  & B1.5\,Ia    & F/Hs       & LWRS / E140H, E140M, E230H, E230M & {\tiny00-05-29 {\textbf/ }01-10-17, 01-10-18 {\textbf/ }03-10-11, 03-10-13} & {\tiny 20-11-02 }                      & X-Shooter &  UVB/VIS                                      \\
AV472 & B1.5a    & F/Hc       & LWRS / G130M, G160M               & {\tiny02-07-22 {\textbf/ }21-08-03, 21-08-03                        } & {\tiny **13-11-11} & X-Shooter** &  UVB/VIS                  \\
AV187 & B3\,Ia    & F/Hs/Hc    & LWRS / E230M {\textbf/ }G130M, G160M       & {\tiny02-07-28 {\textbf/ }20-07-20 {\textbf/ }20-07-22, 2020-07-22           } & {\tiny **13-11-11} & X-Shooter** &  UVB/VIS                  \\
AV362 & B3\,Ia    & F/Hs       & LWRS / E140M, E230M               & {\tiny01-10-23 {\textbf/ }01-10-23, 02-08-02                        } & {\tiny 20-11-10 }                      & X-Shooter &  UVB/VIS                                      \\
AV22  & B3\,Ia    & Hs       & E140M, E230M               & {\tiny00-10-04 {\textbf/ }15-12-21, 15-12-22                        } & {\tiny 20-11-02 }                      & X-Shooter &  UVB/VIS                                      \\
AV234 & B2.5\,Ib    & Hs/Hc      & E230M {\textbf/ }G130M, G160M       & {\tiny20-07-24 {\textbf/ }20-08-19                                 } & {\tiny 20-10-27 }                      & X-Shooter &  UVB/VIS                                      \\
SK179 & B3II     & Hc/Hs      & E230M {\textbf/ }G130M, G160M, G185M& {\tiny21-08-28 {\textbf/ }21-08-28, 21-08-30, 21-10-22              } & {\tiny 20-11-14 }                      & X-Shooter &  UVB/VIS                                      \\
AV343 & B8\,Iab   & Hc/Hs      & E230M {\textbf/ }G130M, G160M, G185M& {\tiny 22-01-29 {\textbf/ }22-01-29, 22-01-29, 22-03-24             } & {\tiny 20-11-21 }  & X-Shooter &  UVB/VIS                  \\                 
\hline
\end{tabular}
\end{sidewaystable}
\FloatBarrier

%%%%%%%%%%%%%%%%%%%%%%%%%%%%%%%%%%%%%%%%%%%%%%%%%%%%%%%%%%%%%%%%%%%%%%%%%%%%%%%%%%%%%%%%%%%%%%%%%%%%%%%%
\section{Atomic data}
\label{sec:atomicdata}

In Table\,\ref{tab:atomic} we list the ions as well as the number of levels, superlevels, and transitions used in our models for the earlier (E) or later (L) spectral types of the BSG sample. The split between these ``groups'' is set at 21\,kK, roughly corresponding to the B1 spectral type.

\begin{table}
\caption{Ions and number of levels, superlevels, and transitions considered per ion used in our B supergiant models. The group letters indicate whether the ion is used in the early set (E), representing models hotter than 21 kK, or in the late set (L).}
\centering
\begin{tabular}{lcccc}
\hline
\hline
Ion  & group & levels & superlevels & transitions \\
\hline
H I    & EL & 30                         & 30                              & 435                          \\ \hline
He I   & EL & 69                         & 69                              & 905                          \\
He II  & EL & 30                         & 30                              & 435                          \\ \hline
C II   & EL & 39                         & 21                              & 202                          \\
C III  & EL & 243                        & 99                              & 5528                         \\
C IV   & EL & 64                         & 64                              & 1446                         \\ \hline
N I    & L & 104                        & 44                              & 855                         \\
N II   & EL & 105                        & 59                              & 898                       \\
N III  & EL & 287                        & 57                              & 6223                         \\
N IV   & EL & 70                         & 44                              & 440                          \\
N V    & EL & 49                         & 41                              & 519                      \\ \hline
O I    & L & 199                        & 58                              & 4193                         \\
O II   & EL & 340                        & 137                             & 8937                         \\
O III  & EL & 104                        & 36                              & 761                          \\
O IV   & EL & 64                         & 30                              & 359                          \\
O V    & EL & 56                         & 32                              & 314         \\
O VI   & E  & 65                         & 65                              & 1569 \\ \hline
Ne II  & EL & 48                         & 14                              & 328                          \\
Ne III & EL & 71                         & 23                              & 460                      \\
Ne IV  & EL & 52                         & 17                              & 315                           \\
Ne V   & EL & 166                        & 37                              & 1813                           \\ \hline
Mg II  & EL & 45                         & 18                              & 362                          \\
Mg III & EL & 201                        & 29                              & 3052                         \\ \hline
Al II  & EL & 58                         & 38                              & 270                          \\
Al III & EL & 65                         & 21                              & 1452                         \\ \hline
Si II  & EL & 80                         & 52                              & 628       \\
Si III & EL & 147                        & 99                              & 1639      \\ 
Si IV  & EL & 66                         & 66                              & 1090      \\ \hline
P IV   & EL & 90                         & 30                              & 656       \\
P V    & EL & 62                         & 16                              & 561       \\ \hline
S III  & EL & 44                         & 24                              & 193       \\
S IV   & EL & 142                        & 51                              & 1503      \\
S V    & EL & 101                        & 40                              & 831        \\   \hline
Ca III & EL & 110                        & 33                              & 868       \\
Ca IV  & E & 378                       & 43                                & 8532           \\
Ca V   & E & 613                       & 73                                & 18272          \\ \hline
Cr II  & L & 1000                       & 84                              & 66400     \\
Cr III & EL & 1000                       & 68                              & 73962     \\
Cr IV  & EL & 234                        & 29                              & 6354      \\
Cr V   & E & 223                       & 30                                & 4124    \\
Cr VI  & E & 215                       & 30                                & 4406           \\ \hline
Mn II  & L & 1000                       & 58                              & 49066     \\
Mn III & EL & 1000                       & 47                              & 70218     \\
Mn IV  & EL & 464                        & 39                              & 19176     \\
Mn V   & E & 80                        & 16                                &  867     \\
Mn VI  & E & 181                       & 23                                &  2005     \\ \hline
Fe II  & L & 827                       & 62                              & 13182     \\
Fe III & EL & 607                        & 65                              & 6670      \\
Fe IV  & EL & 1000                       & 100                             & 37899     \\
Fe V   & EL & 1000                       & 139                             & 37737     \\
Fe VI  & E &  1000                     & 59                                & 6670      \\ \hline
Ni II  & L & 1000                       & 59                              & 33555     \\
Ni III & EL & 150                        & 24                              & 1345      \\
Ni IV  & EL & 200                        & 36                             & 2337   \\
Ni V   & EL & 183                        & 46                              & 1524    \\ 
\hline
\end{tabular}
\label{tab:atomic}
\end{table}
\FloatBarrier

%%%%%%%%%%%%%%%%%%%%%%%%%%%%%%%%%%%%%%%%%%%%%%%%%%%%%%%%%%%%%%%%%%%%%%%%%%%%%%%%%%%%%%%%%%%%%%%%%%%%%%%%%%

\section{Rotation and macroturbulence}
\label{sec:rotmac}

In Table~\ref{tab:rotmac}, we list the complete set of broadening velocities for our sample stars obtained with the IACOB Broad tool for the \ion{He}{I}~$\lambda$4713 and \ion{Si}{III}~$\lambda$4552 (or \ion{Mg}{II}~$\lambda$4481 for later BSGs) lines.

For AzV\,187, the minimum value derived by the tool is identical to the lower limit that can be obtained. Therefore, we instead use a more realistic value obtained by convolving the observed spectrum. We indicate this in Table~\ref{tab:rotmac} by providing the automatic and the finally adopted value separated by a double-lined arrow (``$\Rightarrow$''). Likewise, the uncertainties $\Delta \varv \sin i$ and $\Delta \varv_\mathrm{mac}$ can be severely underestimated if both diagnostics yield a very similar value. In such cases, we increase the adopted uncertainty to the average of the standard deviation of the overall sample. This is indicated in Table\,\ref{tab:rotmac} with a single-lined arrow ($\rightarrow$) where the values on the left are the derived standard deviations between the He and metal lines, and the new uncertainty is given on the right. However, in some cases, we notice that $\Delta\varv_\mathrm{fft}$ is larger than the updated uncertainties in $\varv \sin i$. Hence, we further update $\Delta\varv \sin i$ in those cases yielding the final values given to the right of the half-arrows (``$\rightharpoonup$''). The final values for the broadening values and their respective uncertainties employed in our analysis are written in boldface.

\begin{table*}[]
\centering
\caption{Rotation and macroturbulence values of the sample BSGs obtained by applying the IACOB Broad tool to \ion{He}{I}~$\lambda$4713 and a metal line (\ion{Si}{III}~$\lambda$4552 for BSGs earlier than B3 and \ion{Mg}{II}~$\lambda$4481 for later BSGs). Values to the right of the arrows ($\Rightarrow$, $\rightarrow$, and $\rightharpoonup$) and/or in boldface indicate the values officially adopted (see text).}
\label{tab:rotmac}
\begin{tabular}{lc|ccc|ccc|rrr}
\hline\hline
Star  & Sp. type                      & $\varv \sin i ^{\mathrm{He}}$ & $\varv_\mathrm{mac}^\mathrm{He}$ & $\varv_\mathrm{fft}^\mathrm{He}$ & $\varv \sin i ^{\mathrm{Metal}}$ & $\varv_\mathrm{mac}^\mathrm{Metal}$ & $\varv_\mathrm{fft}^\mathrm{Metal}$ &  $\Delta \varv \sin i$        & $\Delta \varv_\mathrm{mac}$    & $\Delta \varv_\mathrm{fft}$      \\
\hline
AV235 & B0.2\,Ia   & 46        & 114      & 81       & \textbf{39}        & \textbf{88}       & 64        &  5 $\rightarrow$ 10 $\rightharpoonup$ \textbf{12}        & 18 $\rightarrow$ \textbf{39}          & 12         \\
AV215 & BN0\,Ia   & 42        & 119      & 81       & \textbf{86}        & \textbf{63}       & 87        & \textbf{31}          &\textbf{ 40}         & 4          \\
AV488 & B0.5\,Iaw & 37        & 108      & 75       & \textbf{55}        & \textbf{60}       & 62        & \textbf{13}          & \textbf{34}         & 9          \\
AV104 & B0.5\,Ia  & 73        & 55       & 78       & \textbf{73}        & \textbf{33}       & 73        & 0 $\rightarrow$ \textbf{18}          & \textbf{17}         & 4          \\
AV242 & B0.7\,Iaw & 42        & 75       & 58       & \textbf{40}        & \textbf{78}       & 74        & 1 $\rightarrow$ 10 $\rightharpoonup$ \textbf{12}          & 2 $\rightarrow$ \textbf{30}          & 11         \\
AV266 & B0.7\,Ia     & 75        & 10       & 70       & \textbf{44}        & \textbf{75}       & 61        & \textbf{22}                          & \textbf{46}         & 6          \\
AV264 & B1\,Ia    & 19        & 107      & 59       & \textbf{44}        & \textbf{52}       & 51        & \textbf{18}                          & \textbf{39}         & 6          \\
AV210 & B1.5\,Ia  & 31        & 87       & 56       & \textbf{25}        & \textbf{73}       & 49        & 4 $\rightarrow$ \textbf{10}          & 10 $\rightarrow$ \textbf{30}         & 5          \\
SK191 & B1.5\,Ia  & 73        & 65       & 78       & \textbf{57}        & \textbf{87}       & 81        & 11 $\rightarrow$ \textbf{16}          &  12 $\rightarrow$ \textbf{30}         & 2          \\
AV78  & B1\,Ia+ & 50        & 45       & 53       & \textbf{26}        & \textbf{63}       & 48        & \textbf{17}                          &  13 $\rightarrow$ \textbf{21}         & 4          \\
AV18  & B1.5\,Ia    & 43        & 68       & 54       & \textbf{41}        & \textbf{52}       & 52        & 1 $\rightarrow$ \textbf{10}          & 11 $\rightarrow$ \textbf{23}         & 1          \\
AV472 & B1.5\,Ia    & 42        & 59       & 57       & \textbf{32}        & \textbf{41}       & 48        & 7 $\rightarrow$ \textbf{10}         & 13 $\rightarrow$ \textbf{19}        & 6          \\
AV187 & B3\,Ia    & 11        & 57       & 50       & 11 $\Rightarrow$ \textbf{40}  & \textbf{39}        & 48          & 0 $\rightarrow$ \textbf{10}           & 13 $\rightarrow$ \textbf{18}         & 2          \\
AV362 & B3\,Ia    & 11        & 76       & 47       & \textbf{14}        & \textbf{55}       & 42        & \textbf{3}           & 15 $\rightarrow$ \textbf{25}         & 4          \\
AV22  & B3\,Ia    & 37        & 47       & 48       & \textbf{42}        & \textbf{11}       & 44        & \textbf{10}          & \textbf{25}         & 3          \\
AV234 & B2.5\,Ib   & 29        & 45       & 38       & \textbf{14}        & \textbf{42}       & 34        &  \textbf{11}                         & 2 $\rightarrow$ \textbf{17}         & 3          \\
SK179 & B3\,II     & 86        & 27       & 88       & \textbf{83}        & \textbf{115}      & 117       & 2 $\rightarrow$ 20 $\rightharpoonup$ \textbf{21}          &\textbf{62}         & 21         \\
AV343 & B8\,Iab   & 14        & 63       & 45       & \textbf{45}        & \textbf{23}       & 49        & \textbf{22}          & \textbf{28}        & 3     \\     
\hline
\end{tabular}
\end{table*}
\FloatBarrier

\begin{figure}
  \includegraphics[width=\columnwidth]{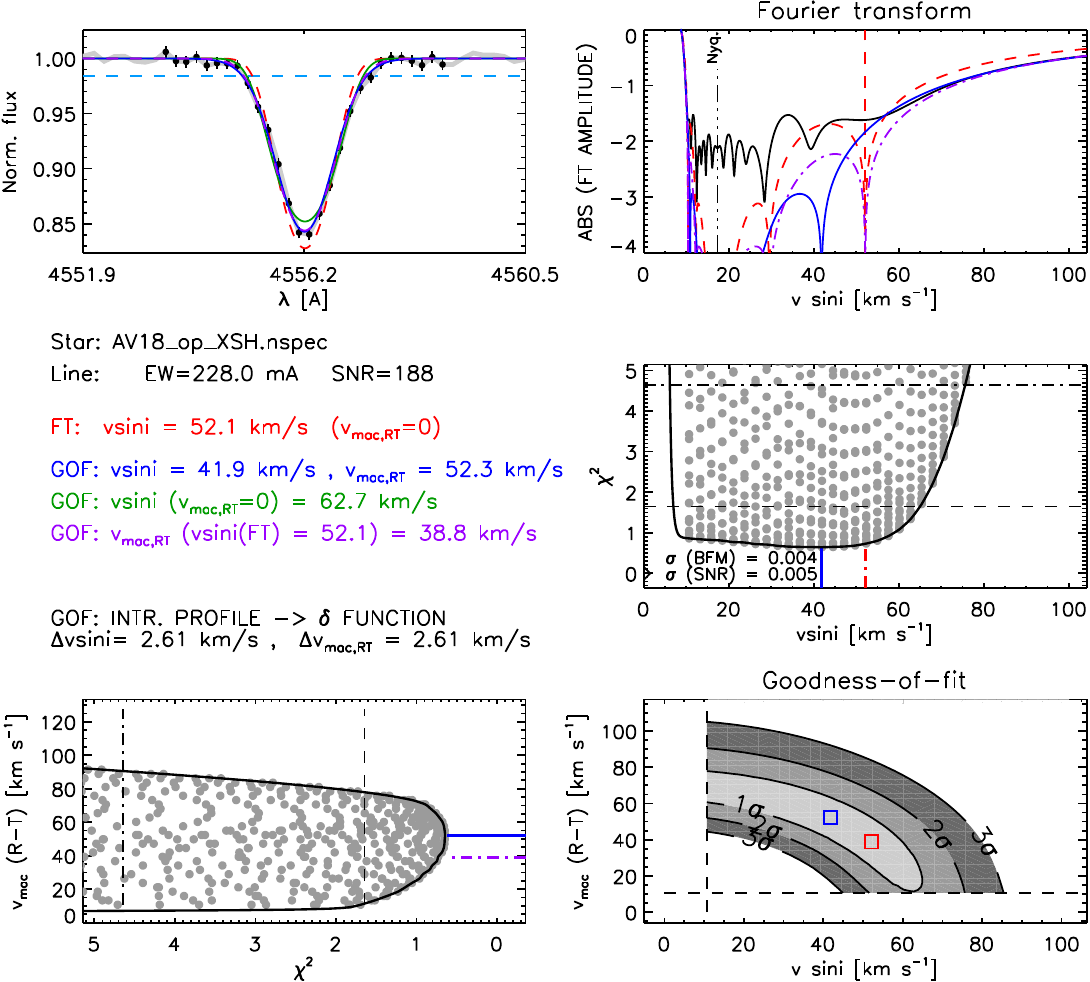}
  \caption{Output of IACOB Broad tool for AzV\,18 applied to \ion{Si}{III}~$\lambda$4552.}
  \label{fig:AV18-IACOB_Me}
\end{figure}

\begin{figure}
  \includegraphics[width=\columnwidth]{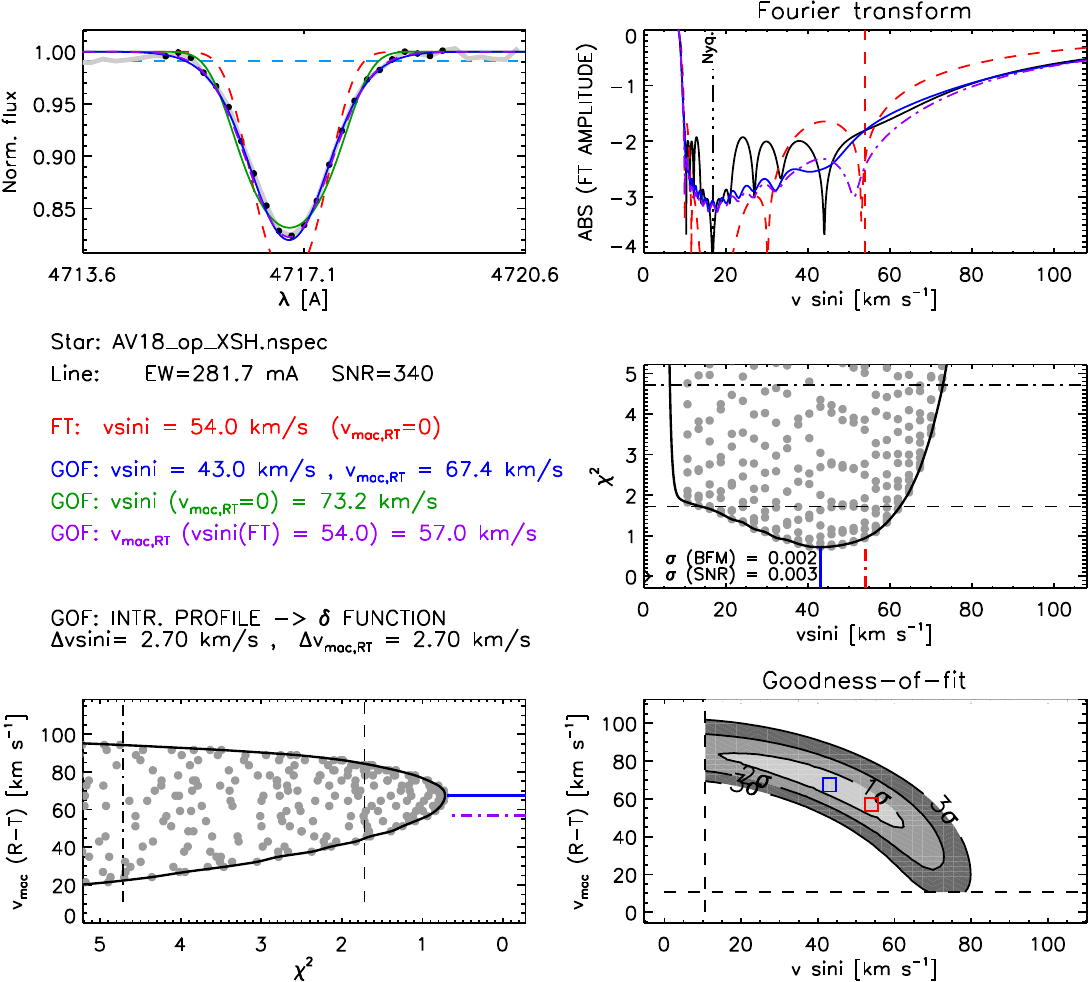}
  \caption{Output of IACOB Broad tool for AzV\,18 applied to \ion{He}{I}~$\lambda$4713.}
  \label{fig:AV18-IACOB_He}
\end{figure}

\begin{figure}
  \includegraphics[width=\columnwidth]{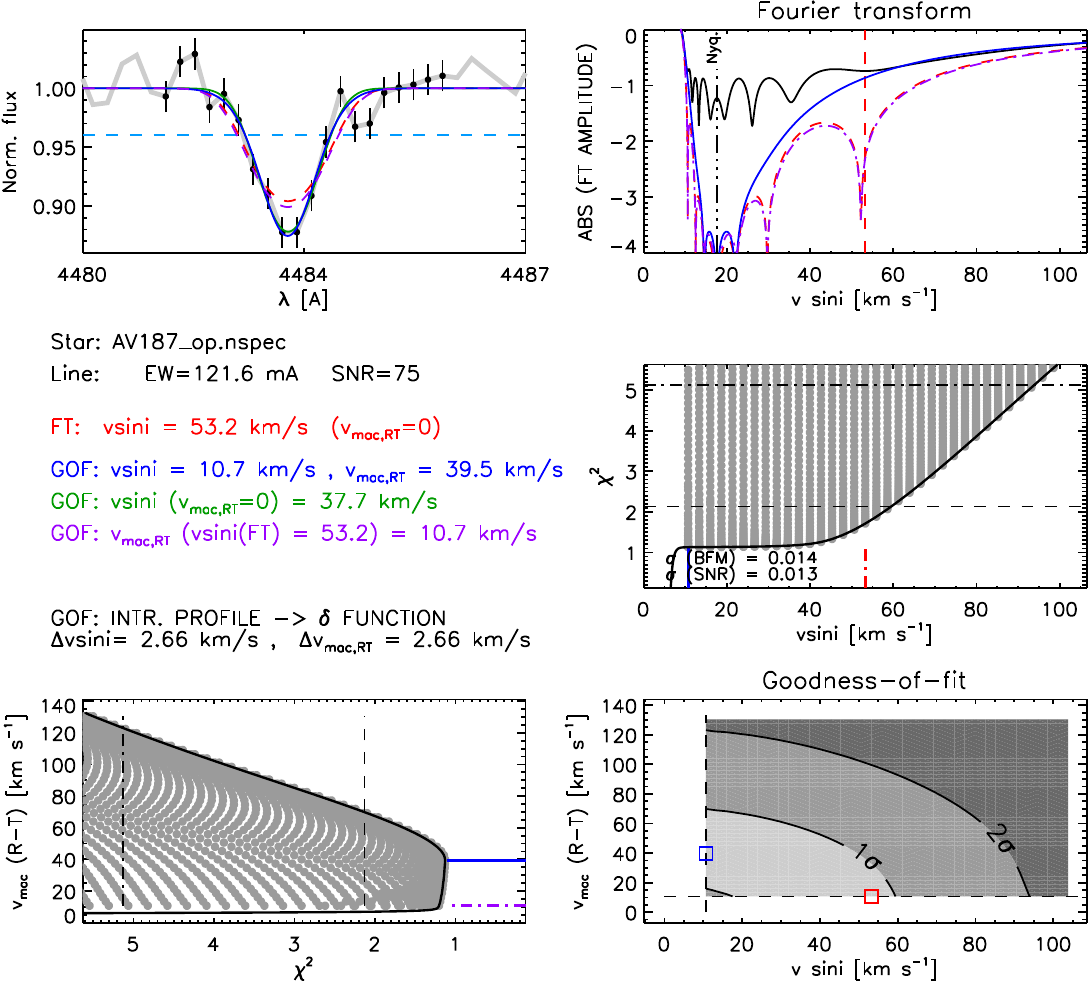}
  \caption{Output of IACOB Broad tool for AzV\,187 applied to \ion{Mg}{II}~$\lambda$4481.}
  \label{fig:AV187-IACOB_Me}
\end{figure}
\FloatBarrier

\begin{figure}
  \includegraphics[width=\columnwidth]{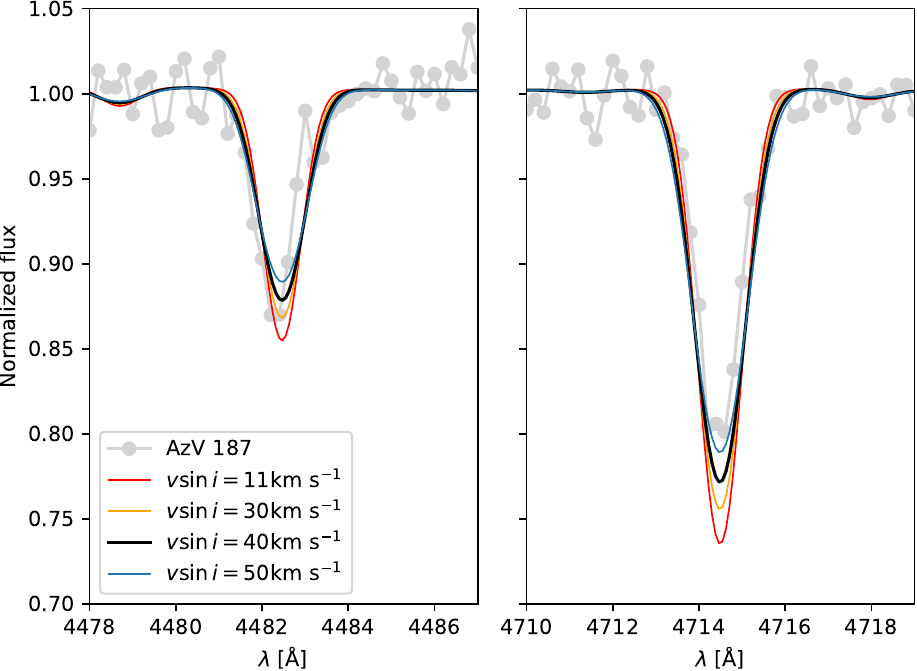}
  \caption{Model spectra of AzV\,187 with different rotational broadening illustrating the fitting process after the unsuccessful application of IACOB Broad tool for this target in particular. The macroturbulence is kept fixed ($\varv_\mathrm{mac} = 48$~km\,s$^{-1}$) in all the spectra.}
  \label{fig:AV187-rots}
\end{figure}

In Fig.~\ref{fig:AV18-IACOB_He} and Fig.~\ref{fig:AV18-IACOB_Me}, we show the output of the IACOB Broad tool for obtaining the $\varv \sin i$ and $\varv_\mathrm{mac}$ of AzV\,18. The first figure depicts the application to \ion{Si}{III}~$\lambda$4552 while the second figure shows the application to \ion{He}{I}~$\lambda$4713. In this case, the values from both lines are in good agreement. 
However, this is not the case for all targets, as for example shown in Fig.~\ref{fig:AV187-IACOB_Me} where we depict the results for the \ion{Mg}{II}~$\lambda$4481 line in AzV\,187. For this object, the best-fit $\varv \sin i$ is the minimum value (10.7\,km\,s$^{-1}$) and the $1\sigma$ and $2\sigma$ intervals (i.e., the uncertainty)  in the goodness-of-fit plot are very large. In such situations, we disregard the derived $\varv \sin i$ value and instead test different values by convolving the synthetic spectrum. Still, we keep the $\varv_\mathrm{mac}$ provided by IACOB Broad tool for the metal line $\varv_\mathrm{mac}^\mathrm{Metal}$. 
Such procedure is illustrated in Fig.\ref{fig:AV187-rots}.

\FloatBarrier

%%%%%%%%%%%%%%%%%%%%%%%%%%%%%%%%%%%%%%%%%%%%%%%%%%%%%%%%%%%%%%%%%%%%%%%%%%%%%%%%%%%%%%%%%%%%%%%%%%%%%%%%%%%%%

\section{Comments on Peculiar Targets}
  \label{sec:pecultargs}

\paragraph{AzV\,235 (B0.2\,Ia)}

\begin{figure}
  \includegraphics[width=\columnwidth]{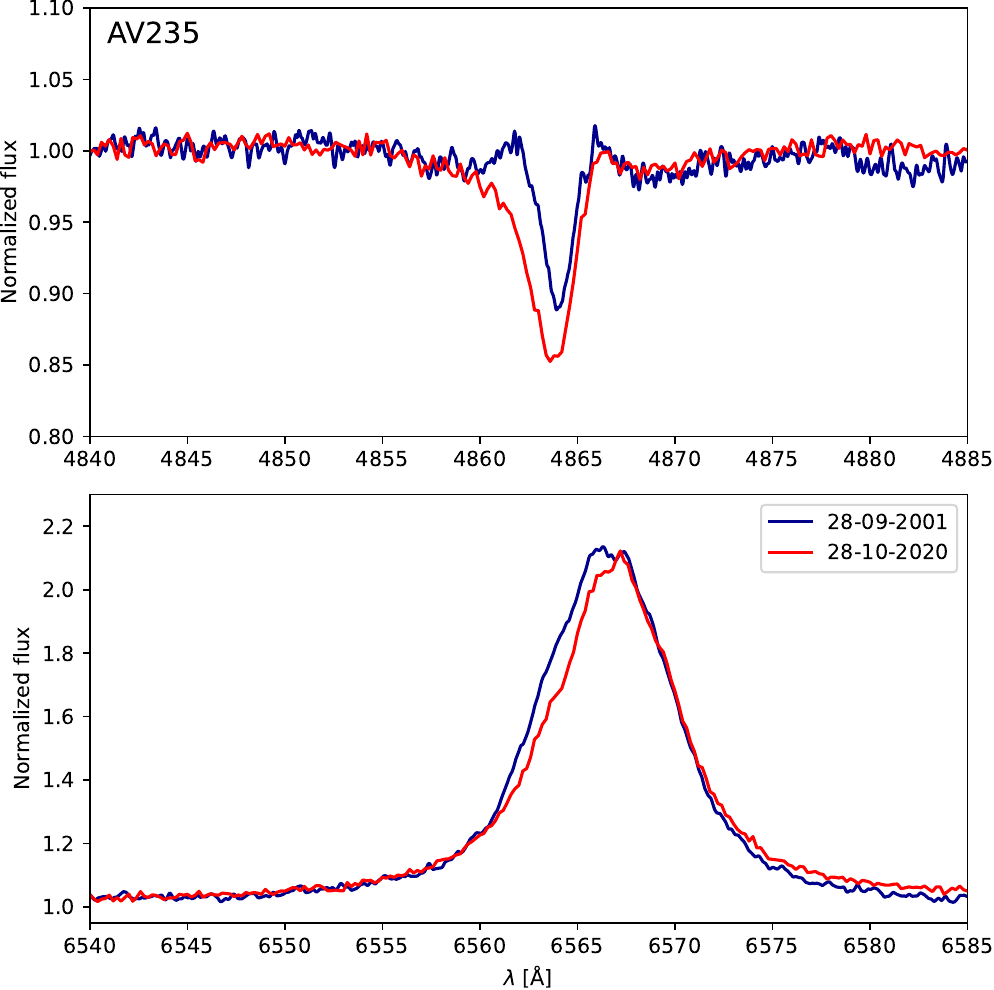}
  \caption{Time variation between 2001 (dark blue line) and 2020 (red line) of the spectrum of AzV\,235 in the regions of H$\alpha$ and H$\beta$.}
  \label{fig:AV235-Hab}
\end{figure}

The peculiarity of this target is the presence of an unusually strong H$\alpha$, which was reported by \citet{Evans+2004-cmfgen}. Compared with the old archival UVES spectrum, we see that the strong H$\alpha$ is still there. However, the data for 2001 seems to have an extra increase on the blue part of the line, which is mirrored by the H$\beta$ profile. See Fig.~\ref{fig:AV235-Hab}.

\begin{figure}
  \includegraphics[width=\columnwidth]{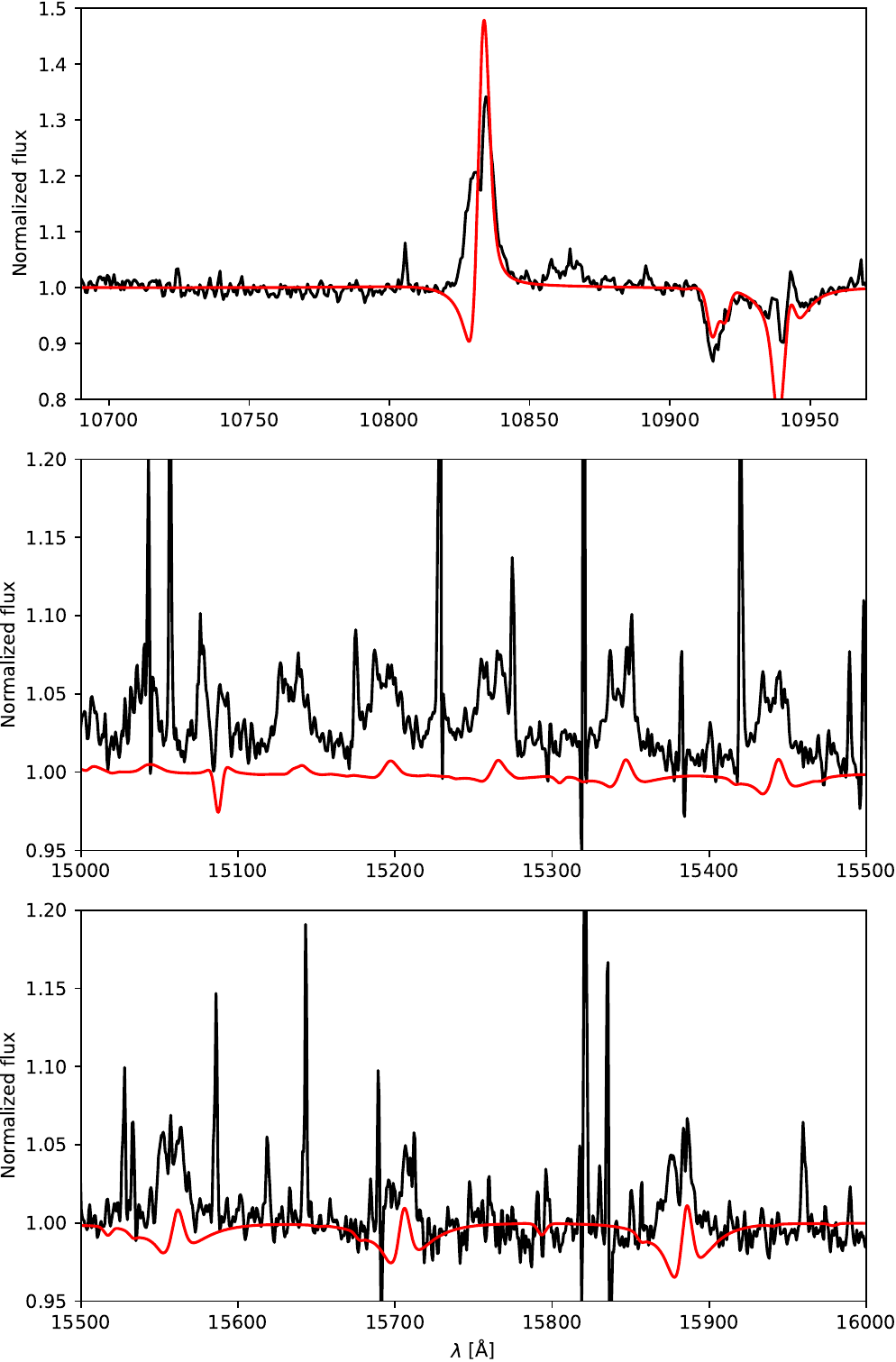}
  \caption{Observed (black line and dots) and model (red line) spectrum of AzV\,235 around   \ion{He}{I}~$\lambda10830$ (upper panel) and Part of the Bracket series (remaining panels).}
  \label{fig:AV235-ir}
\end{figure}

In our modeling we were not able to have the intensity of H$\alpha$ and H$\beta$ compatible with the observed spectrum even by applying very different clumping settings (i.e., very low $f_\infty = 0.02$), enforcing clumping to start deeper in the wind and using a clumping-profile law that produces a smooth wind at terminal speed \citep{Najarro+2011}. The exploration of other clumping settings was also motivated by the presence of an unusual behavior at the IR spectrum. For example, we noticed that the Brackett series is also in strong emission with possible double-peaked structure (see Fig.~\ref{fig:AV235-ir}). This double-peaked feature appears to happen with \ion{He}{I}~$\lambda10830$ as well.

In general, double peak profiles are sign of disk emission, which could also explain the presence of the super strong H$\alpha$ emission. Moreover, we also notice a slight IR excess in this star (see first panel of Fig.\,F.1 in the appendix\,F available at Zenodo), which can also be due simply to higher wind-density via free-free emission. Another possible explanation for the H$\alpha$ is the presence of a circumstellar nebula. However, there is to date no clear signs of nebula region around this star.

\FloatBarrier

\paragraph{AzV\,104 (B0.5\,Ia)}
AzV\,104 is found to be a peculiar target because its UV P Cygni profiles are considerably narrow, indicating a very low wind speed. Additionally, the profiles of \ion{Si}{IV} and \ion{C}{IV}, which usually are well developed in stars as early as B0.5, do not have emission components. We could only partially reproduce the lack of emission component by using a very high wind microturbulence velocity. \citet{Evans+2004-vinf} and \citet{Parsons+2024} also had problems in fitting the profiles of \ion{Si}{IV} using SEI.

\begin{figure}
  \includegraphics[width=\columnwidth]{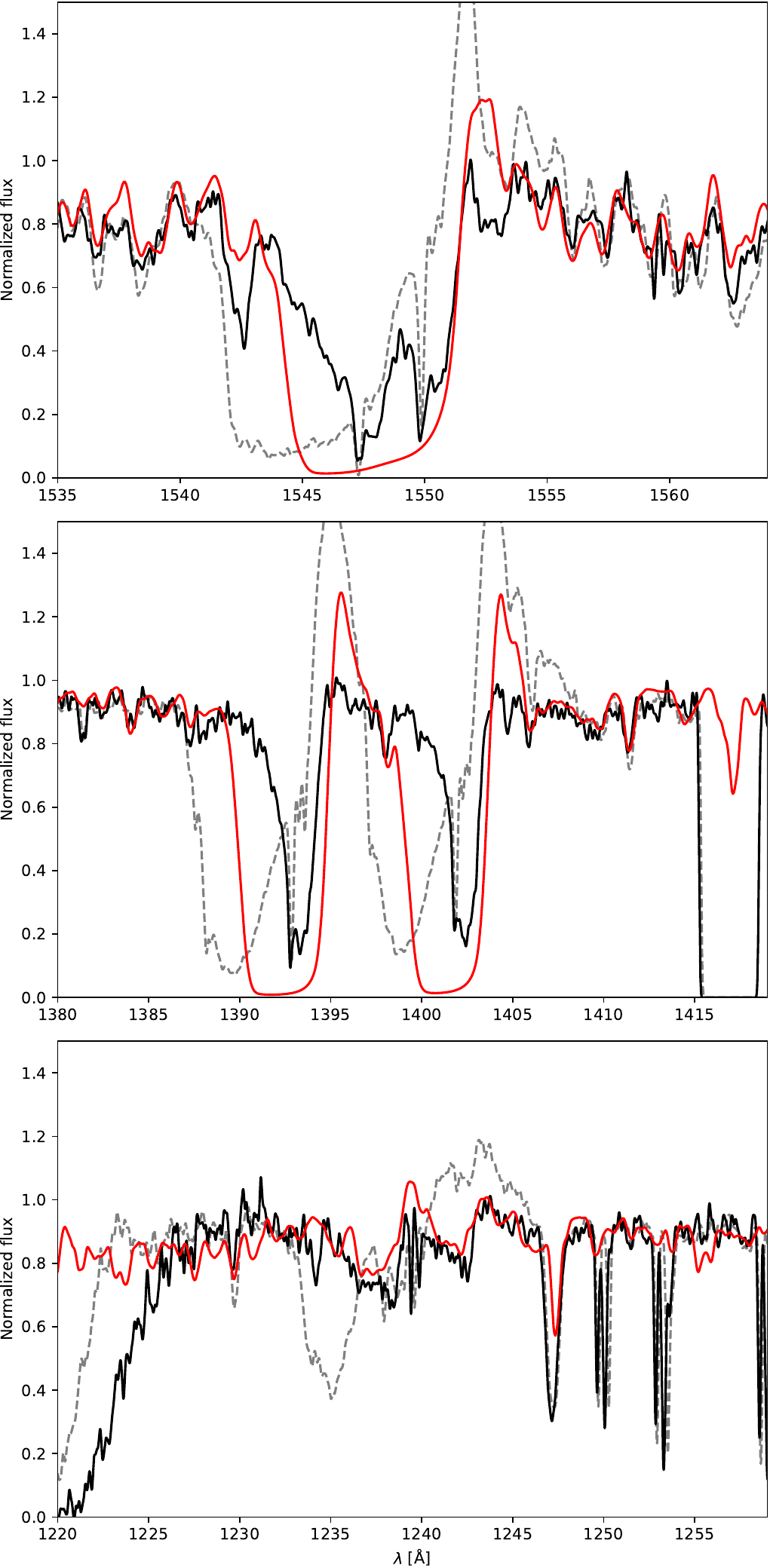}
  \caption{Observed (black line) and model (red line) spectrum of AzV\,104 in the regions of \ion{N}{V}~$\lambda1240$, \ion{Si}{IV}~$\lambda1400$, and \ion{C}{IV}~$\lambda1550$. The UV spectrum of AzV\,488 is shown as a gray dashed line for comparison.}
  \label{fig:AV104-UV}
\end{figure}

In Fig~\ref{fig:AV104-UV} we show the corresponding profiles and \ion{N}{V}~$\lambda1240$  for AzV\,104 and AzV\,488 as a comparison with a ``typical'' BSG of the same spectral type. The \ion{N}{V}~$\lambda1240$ which also seems pretty weak in AzV\,104 likely indicates a low X-ray luminosity, as we needed a much reduced $L_\mathrm{X}/L$ to match it.  

\FloatBarrier

\paragraph{Sk\,191 (B1.5\,Ia)}
For this star, we did not manage to obtain successful simultaneous fits for H$\alpha$ and H$\beta$, which both appear as quite developed P Cygni profiles. BSGs that display P Cygni/emission profiles in their Balmer series (at least H$\beta$) are considered hypergiants \citep[e.g.,][]{vanGenderen+1982,Lennon+1992,Clark+2012}, and therefore Sk\,191 could be classified as a BHG under this criterion. 

\begin{figure}
  \includegraphics[width=\columnwidth]{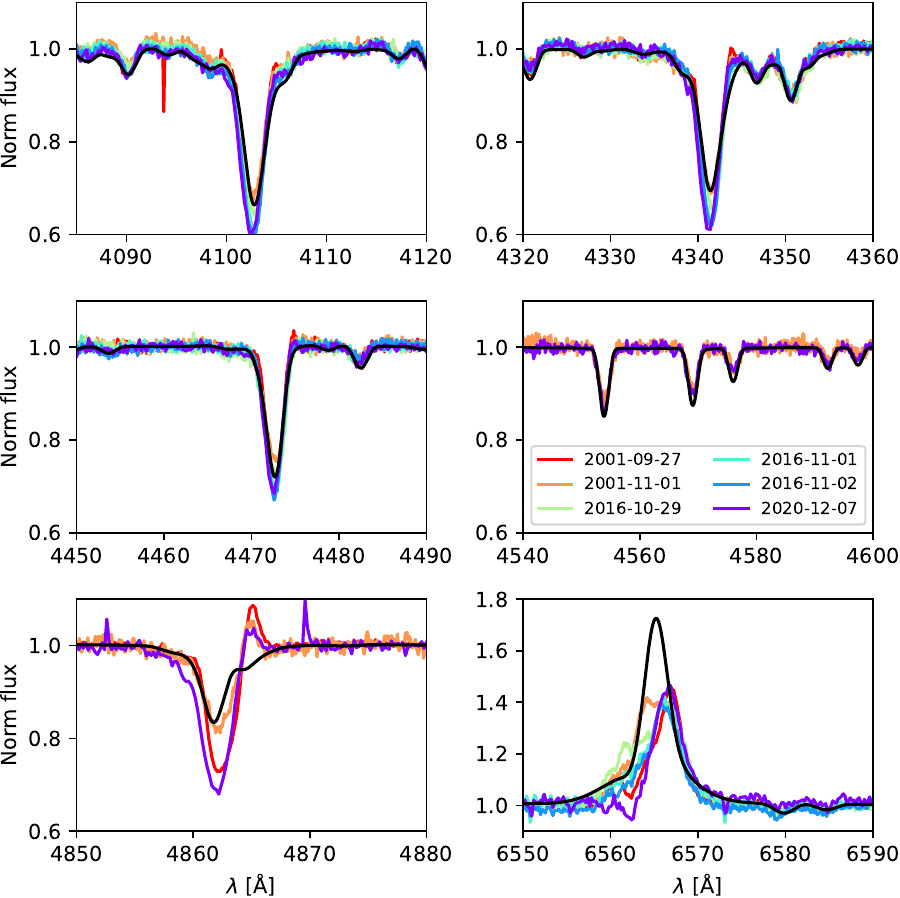}
  \caption{Optical time variability of Sk\,191. The figure shows the Balmer lines (upper and lower rows of plots) as well as \ion{He}{I}~$\lambda4471$ and \ion{Si}{III}~$\lambda$4552,67,78 in the middle row.}
  \label{fig:SK191_varop}
\end{figure}

Typically, when bluer Balmer lines present wind signatures, H$\alpha$ flux is considerably high, as is the case of AzV\,78. However, the H$\alpha$ profile of Sk\,191 seems rather weak as one would expect from its H$\beta$ -- the reason behind the difficulty for a simultaneous fit. The best compromise we obtained was by adopting a very high $\varv_{\mathrm{cl}} = 400$ and a very low amount of clumping aiming to limit the strength of H$\alpha$. Additionally, other lines of the Balmer series, as well as certain \ion{He}{I} lines, seem to display some wind signatures (either slightly displaced blueward or showing a P Cygni profile) -- see Fig.~\ref{fig:SK191_varop}. 

\FloatBarrier

\begin{figure}
  \includegraphics[width=\columnwidth]{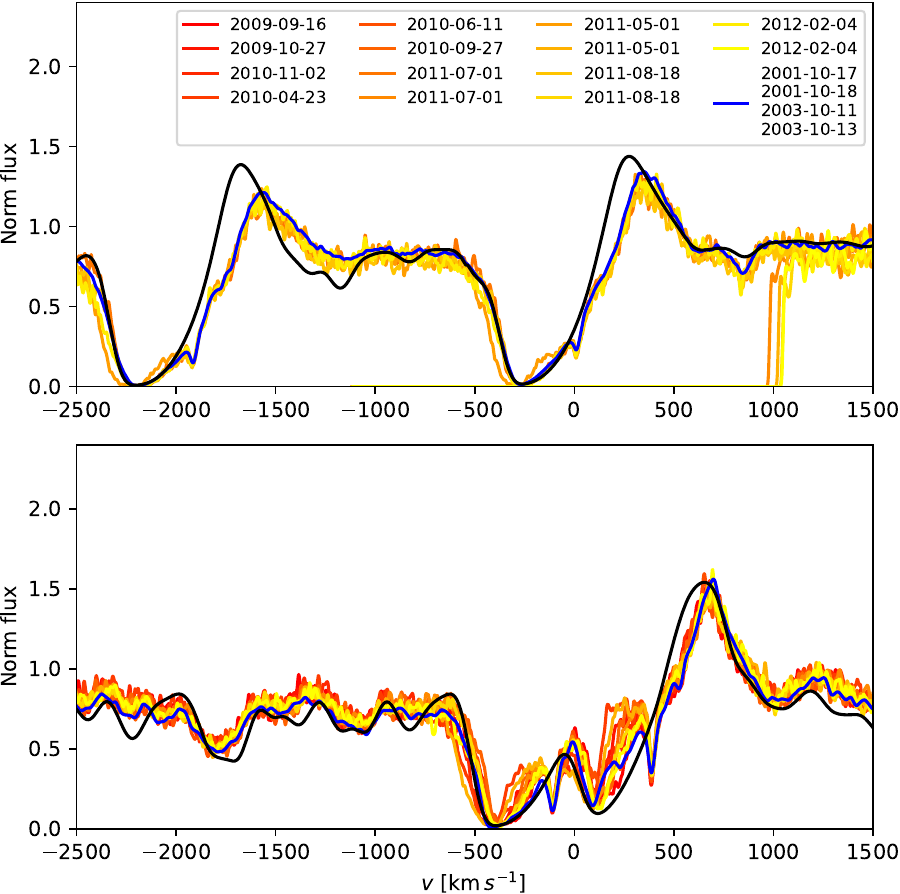}
  \caption{Time variability in the UV for Sk\,191. The upper and lower panels shows, respectively, \ion{Si}{IV}$\lambda1400$ and \ion{C}{IV}$\lambda1550$. The blue curve is the ULLYSES spectrum.}
  \label{fig:SK191_varuv}
\end{figure}

Interestingly, the UV spectrum was much easier to fit, and even normally problematic lines as \ion{Si}{IV}~$\lambda1400$ and \ion{Al}{III}~$\lambda1855$ were very well reproduced. The variability in the UV is not extensive (i.e., without changes to the overall morphology of the line), as it is possible to see in the \ion{Si}{IV}~$\lambda1400$ in Fig.~\ref{fig:SK191_varuv}. However, in the same figure, we can also notice that the width of the absorption trough of \ion{C}{IV}~$\lambda1550$ appears to vary about $\sim 100$~$\mathrm{km\,s^{-1}}$, which might also be indicative of variability in the wind terminal speed, or important ionization changes in the outer wind.

\FloatBarrier

\paragraph{AzV\,362 (B3\,Ia)}

The spectra in the literature of this star show ``typical'' well-developed P Cygni H$\alpha$ \citep{Trundle+2004, Evans+2004}. This indicates a quite variable behavior. In the UV, one does not observe strong wind lines, in principle incompatible with the strength of the H$\alpha$ -- which could be due to clumping, though. Such UV-H$\alpha$ incompatibility happens with AzV\,22 as well, albeit there is nothing particularly unusual about its line spectrum. Both AzV\,362 and AzV\,22 are the only cool BSGs in the sample with a higher degree of clumping.

\begin{figure}
  \includegraphics[width=\columnwidth]{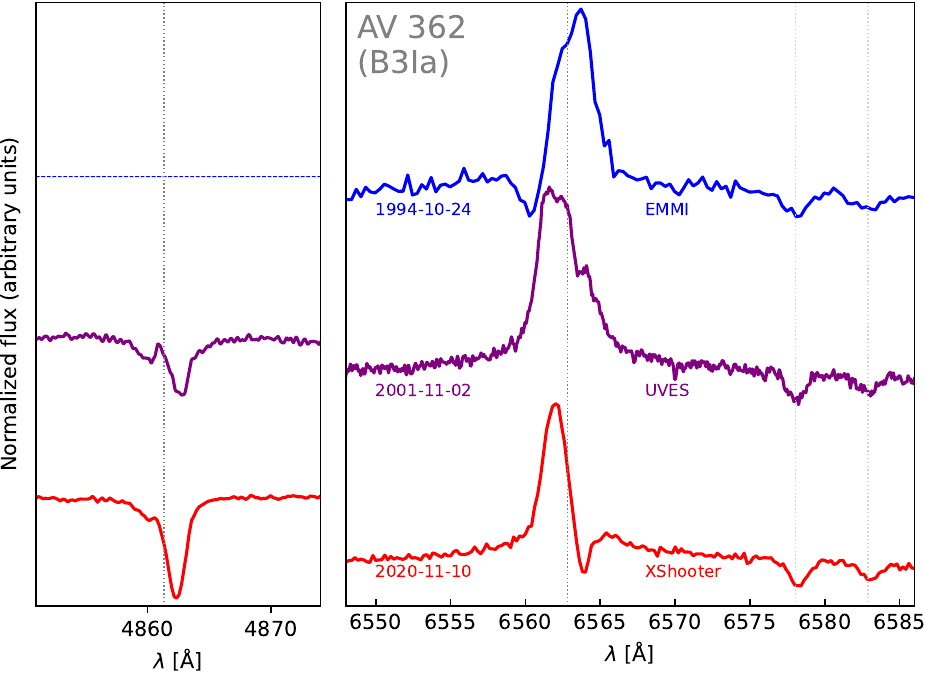}
  \caption{Time variation of H$\beta$ and H$\alpha$ of AzV\,362. Both lines seem to show a prominent blueshifted emission in 2001 and 2020. The H$\alpha$ in the recent X-Shooter observation even appears as a completely inverted P Cygni profile, which is also shown in Fig~\ref{fig:uv-op-spec}. Higher-order Balmer lines do not show significant variability. The \ion{C}{II} lines are shown for alignment purposes.}
  \label{fig:AV362-Hab}
\end{figure}

\begin{figure}
  \includegraphics[width=\columnwidth]{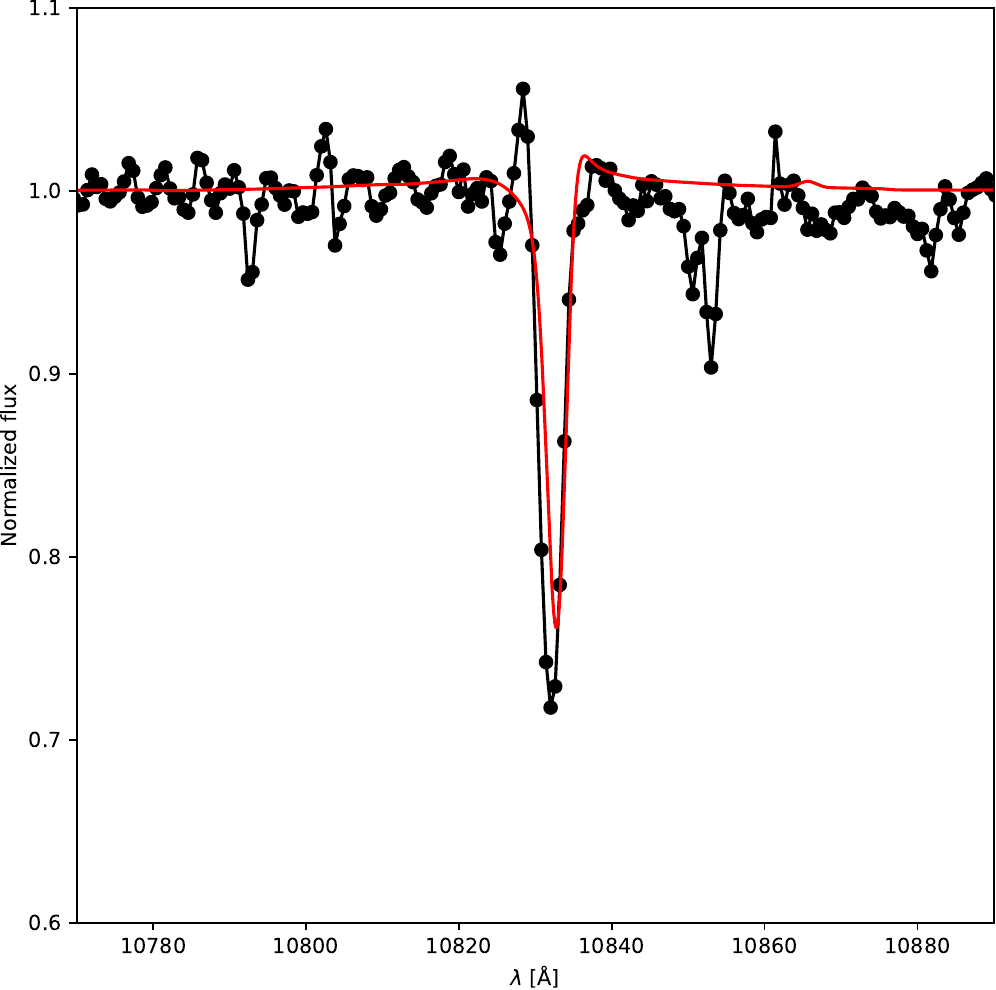}
  \caption{Observed (black line and dots) and model (red line) spectrum of AzV\,362 around \ion{He}{I}~$\lambda10830$.}
  \label{fig:AV362-ir}
\end{figure}

Looking at the infrared X-Shooter data, we can also notice a (possibly) inverted P Cygni in \ion{He}{I}~$\lambda10830$, as shown in Fig.~\ref{fig:AV362-ir}. Variability (and even momentarily unusual shapes) of H$\alpha$ among OBSGs is not something uncommon \citep[e.g.,][]{Martins+2015,Kraus+2015}, however, given the apparent persistence of it and the fact that it might be appearing in helium lines, it is possible that there is something special going on with this object.

\FloatBarrier

The flux of the FUSE spectrum for this star is considerably higher than the HST UV spectra (cf.\ Fig.\,F.3 in the appendix\,F available at Zenodo). This is likely due to an incorrect flux calibration of the FUSE range as other parts of the SED are well captured by the model.

%%%%%%%%%%%%%%%%%%%%%%%%%%%%%%%%%%%%%%%%%%%%%%%%%%%%%%%%%%%%%%%%%%%%%%%%%%%%%%%%%%%%%%%%%%%%

\section{Fitting diagnostic lines}

\begin{figure}
  \includegraphics[width=\columnwidth]{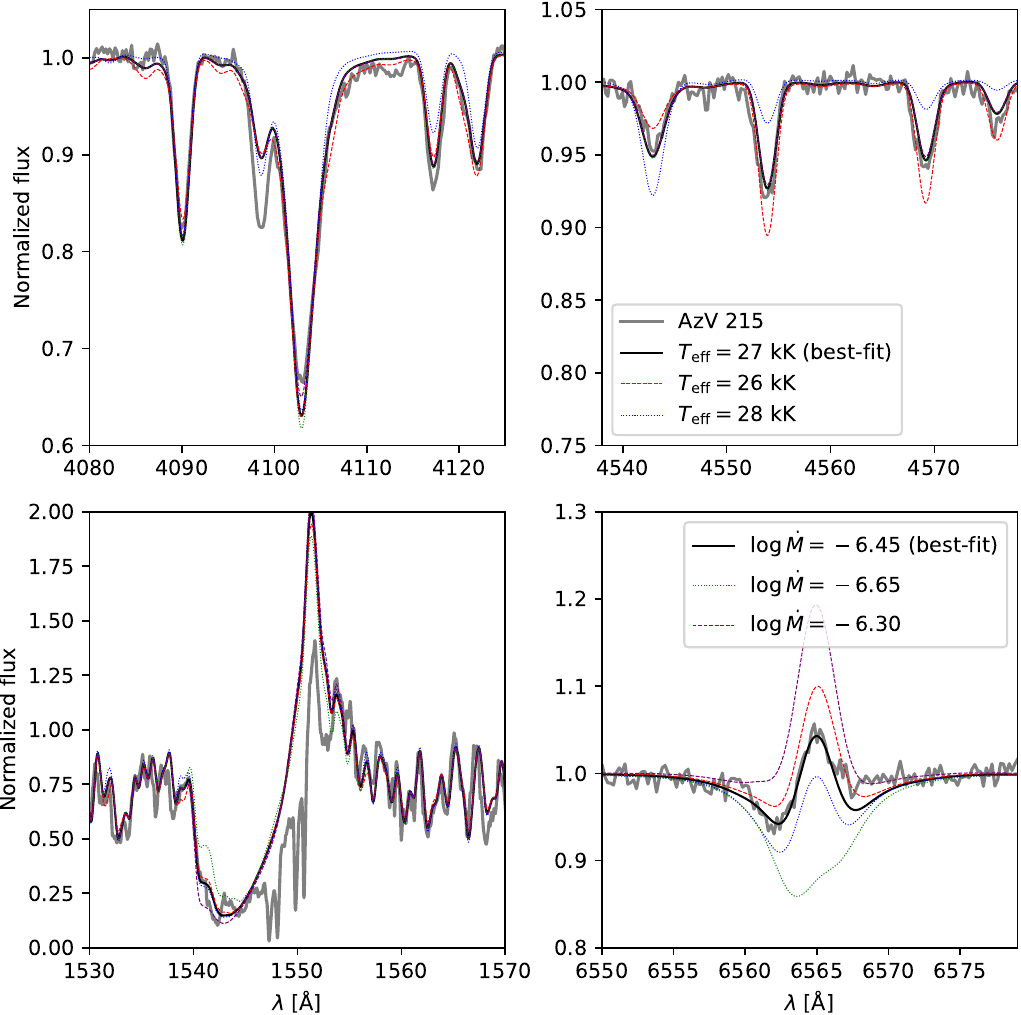}
  \caption{Illustration of the determination of parameters of AzV\,215. The solid gray lines are the observed spectra. The black filled line is the best-fit model. The blue dotted(red dashed) line represents a model with higher(lower) temperature. Likewise, the green dotted(purple dashed) line shows models with lower(higher) mass-loss rates.}
  \label{fig:diag_Teff_Md}
\end{figure}
\FloatBarrier

In  Fig.~\ref{fig:diag_Teff_Md}, we provide an illustrative example for the spectroscopic fitting process to determine the stellar and wind properties. Different stellar atmosphere models with varying parameters are created with Fig.~\ref{fig:diag_Teff_Md} showing the observed spectrum compared to models with different values for $T_\mathrm{eff}$ and $\dot{M}$. 
Guided by the knowledge on how different lines are affected by variations of the atmosphere properties, we obtain the model which marks the best compromise in fitting the many diagnostic lines for the different stellar and wind properties.

\begin{figure}
  \includegraphics[width=\columnwidth]{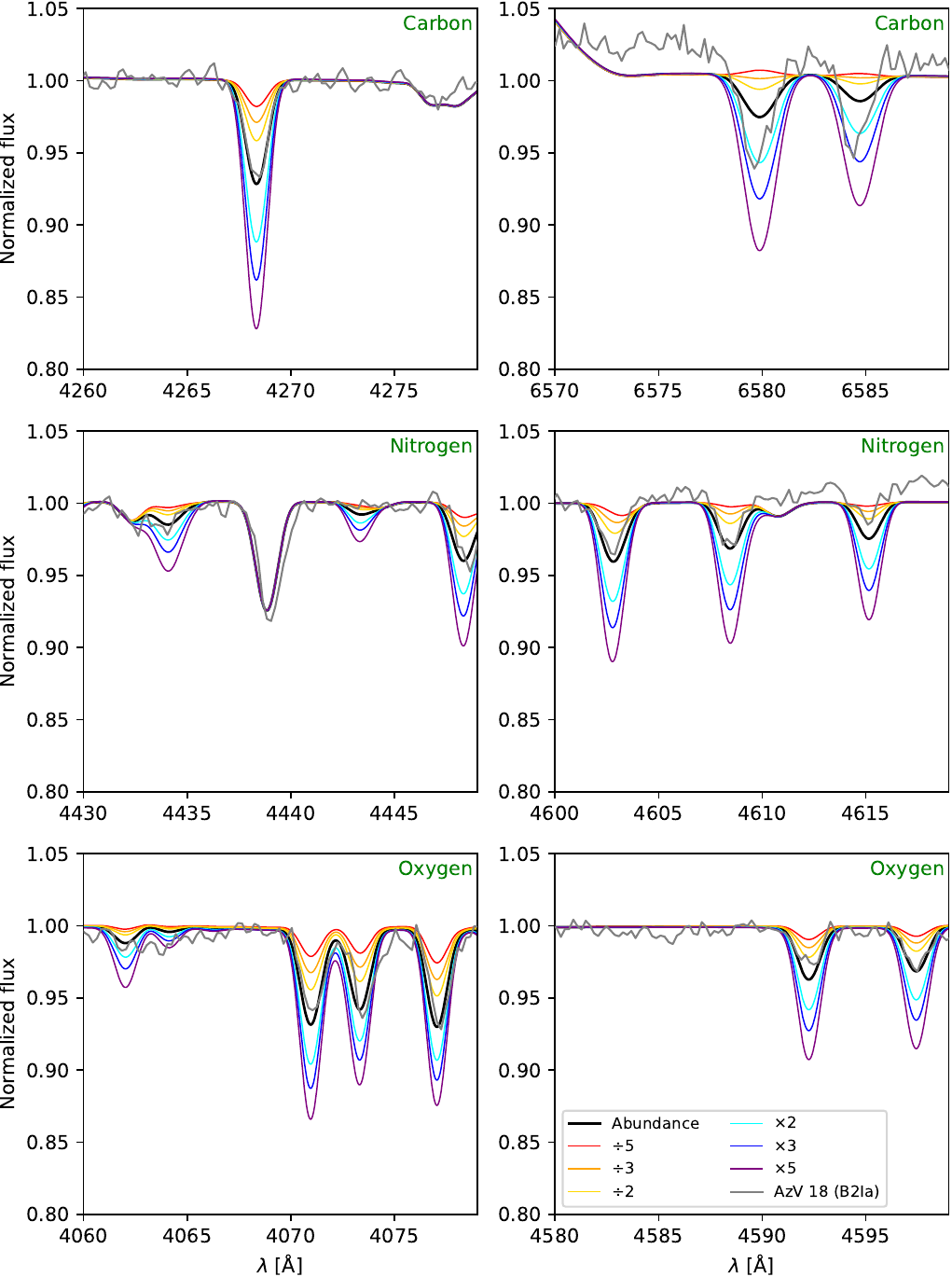}
  \caption{Illustration of the determination of the CNO abundance determination for AzV\,18.  The observed spectrum is the gray line. The thick black line is the model we found to have the best compromise between the diagnostic lines for each element. Each thin line represents the best-fit abundances for the respective element multiplied or divided by factors of 2, 3, and 5. This also guided us to adopt an uncertainty of 0.3~dex.}
  \label{fig:diag_CNO}
\end{figure}

Similarly to Fig.~\ref{fig:diag_Teff_Md}, in  Fig.~\ref{fig:diag_CNO} we illustrate the process of obtaining the surface abundance of CNO for one of our targets.

\FloatBarrier

\end{appendix}

\end{document}